\documentclass[]{PBCreport}

\usepackage{booktabs}
\usepackage{afterpage}
\usepackage[T1]{fontenc}
\usepackage{textcomp}
\usepackage{color, colortbl}
\usepackage{booktabs}
\usepackage{afterpage}
\usepackage{tabularx}
\usepackage{placeins}
\usepackage{siunitx}
\usepackage{hyperref}
\usepackage{graphicx,mathptmx,amsmath,amsfonts,amscd,amsthm,amssymb,eucal,psfrag,color,subfig,url,multirow,gensymb}
\usepackage{rotating}
\usepackage{soul}
\usepackage[switch]{lineno}
\usepackage[noadjust,compress]{cite}
\usepackage{url}
\usepackage{appendix}
\usepackage{xcolor,mdframed}

\hypersetup{
     colorlinks = true,
     linkcolor = blue,
     anchorcolor = blue,
     citecolor = blue,
     filecolor = blue,
     urlcolor = blue
     }

\definecolor{LightCyan}{rgb}{0.88,1,1}
\definecolor{LightYellow}{rgb}{1,0.97,0.9}

\documentlabel{CERN-PBC Report-2023-003}

\begin{document}

\title{Post-LS3 Experimental Options in ECN3}

\author{\parbox{\textwidth}{\small\it C.~Ahdida$^{1}$, G.~Arduini$^{*,1}$, K.~Balazs$^{1}$, H.~Bartosik$^{1}$, J.~Bernhard$^{1}$, A.~Boyarsky$^{2}$, J.~Brod$^{3}$, M.~Brugger$^{1}$, M.~Calviani$^{1}$, A.~Ceccucci$^{1}$, A.~Crivellin$^{4,5}$, G.~D'Ambrosio$^{6}$, G.~De~Lellis$^{6,7}$, B.~D\"obrich$^{8}$, M.~Fraser$^{1}$, R.~Franqueira~Ximenes$^{1}$, A.~Golutvin$^{9}$, M.~Gonzalez Alonso$^{10}$, E.~Goudzovski$^{11}$, J.-L.~Grenard$^{1}$, J.~Heeck$^{12}$, J.~Jaeckel$^{*,13}$, R.~Jacobsson$^{1}$, Y.~Kadi$^{1}$, F.~Kahlhoefer$^{\#,+,14}$, F.~Kling$^{15}$, M.~Koval$^{16}$, G.~Lanfranchi$^{+,17}$, C.~Lazzeroni$^{11}$, F.~Mahmoudi$^{1,18}$, D.~Marzocca$^{19}$, K.~Massri$^{1}$, M.~Moulson$^{17}$, S.~Neshatpour$^{6}$, J.~Osborne$^{1}$, M.~Pospelov$^{+,20,21}$, T.~Prebibaj$^{1}$,
T.~R.~Rabemananjara$^{22,23}$,
Ch.~Rembser$^{\#,1}$, J.~Rojo$^{22,23}$, A.~Rozanov$^{\#,24}$, G.~Ruggiero$^{25}$, G.~Rumolo$^{1}$, G.~Schnell$^{\&,26}$, M.~Schott$^{27}$, Y.~Soreq$^{28}$, T.~Spadaro$^{17}$, C.~Vall\'ee$^{*,24}$, T.~Zickler$^{1}$, J.~Zupan$^{3}$.
}}

\abstract{
The Experimental Cavern North 3 (ECN3) is an underground experimental cavern on the CERN Prévessin site. ECN3 currently hosts the NA62 experiment, with a physics programme devoted to rare kaon decays and searches of hidden particles approved until Long Shutdown 3 (LS3). Several options are proposed on the longer term in order to make best use of the worldwide unique potential of the high-intensity/high-energy proton beam extracted from the Super Proton Synchrotron (SPS) in ECN3. The current status of their study by the CERN Physics Beyond Colliders (PBC) Study Group is presented, including considerations on beam requirements and upgrades, detector R\&D and construction, schedules and cost, as well as physics potential within the CERN and worldwide landscape.     
}

\maketitle

\newpage

\section*{Affiliations}
\begin{small}
$^{1}$CERN, Geneva, Switzerland\\
$^{2}$Institute Lorentz, Leiden University, Niels Bohrweg 2, Leiden, NL-2333 CA, the Netherlands\\
$^{3}$Department of Physics, University of Cincinnati, Cincinnati, Ohio 45221,USA\\
$^{4}$Physik-Institut, Universit\"at Z\"urich, Winterthurerstrasse 190, CH–8057 Z\"urich, Switzerland\\
$^{5}$Paul Scherrer Institut, CH–5232 Villigen PSI, Switzerland\\
$^{6}$INFN-Sezione di Napoli, Complesso Universitario di Monte S. Angelo,
Via Cintia Edificio 6, 80126 Napoli, Italy\\
$^{7}$Universita` degli Studi di Napoli Federico II, I-80126 Napoli, Italy\\
$^{8}$Max-Planck-Institut  für Physik (Werner-Heisenberg-Institut), F\"ohringer Ring 6, 80805 M\"unchen, Germany\\
$^{9}$Blackett Laboratory, Imperial College London, Prince Consort Road, London, SW7 2AZ, UK\\
$^{10}$Department de Física Teòrica IFIC, Universitat de València-CSIC, Parc Científic, Paterna 46980, Valencia, Spain\\
$^{11}$School of Physics and Astronomy, University of Birmingham, Edgbaston, Birmingham, B15 2TT, United Kingdom\\
$^{12}$Department of Physics, University of Virginia,
Charlottesville, Virginia 22904-4714, USA\\
$^{13}${Institut f\"{u}r Theoretische Physik, Universit\"{a}t Heidelberg, Philosophenweg 16, 69120 Heidelberg, Germany}\\
$^{14}$Institute for Theoretical Particle Physics (TTP), Karlsruhe Institute of Technology (KIT), D-76131 Karls-
ruhe, Germany\\
$^{15}$Deutsches Elektronen-Synchrotron DESY, Notkestr. 85, 22607 Hamburg, Germany\\
$^{16}$Charles University, Prague, Czech Republic\\
$^{17}$INFN Laboratori Nazionali di Frascati, Frascati (Rome), Italy\\
$^{18}$Universit\'e de Lyon, Universit\'e Claude Bernard Lyon 1, CNRS/IN2P3, Institut de Physique des 2 Infinis de Lyon, UMR 5822, F-69622, Villeurbanne, France \\
$^{19}$INFN, Sezione di Trieste, SISSA, Via Bonomea 265, 34136, Trieste, Italy\\
$^{20}$ William I. Fine Theoretical Physics Institute, School of Physics and Astronomy, University of Minnesota, Minneapolis, MN 55455, USA\\
$^{21}$School of Physics and Astronomy, University of Minnesota, Minneapolis, MN 55455, USA\\
$^{22}${Nikhef Theory Group, Science Park 105, 1098 XG Amsterdam, The Netherlands}\\
$^{23}${Physics and Astronomy, Vrije Universiteit Amsterdam, NL-1081 HV Amsterdam, The Netherlands}\\
$^{24}$Aix Marseille Univ, CNRS/IN2P3, CPPM, Marseille, France\\
$^{25}$Faculty of Science and Technology, University of Lancaster, Lancaster, United Kingdom\\
$^{26}$Department of Physics \& EHU Quantum Center, University of the Basque Country UPV/EHU, 48080 Bilbao and IKERBASQUE, 48009 Bilbao, Spain\\
$^{27}$Johannes Gutenberg-Universit\"at Mainz, 55128 Mainz, Germany\\
$^{28}$Technion—Israel Institute of Technology, Haifa 32000, Israel\\

\vspace{0.8cm}
\noindent $^{*}$Physics Beyond Collider (PBC) coordinator\\
$^{\#}$PBC Beyond the Standard Model (BSM) Working Group convenor\\
$^{+}$PBC Feebly Interacting Particle Physics Centre (FPC) convenor\\
$^{\&}$PBC QCD Working Group convenor\\
\end{small}

\newpage
\section*{PBC Working Group contributors}
\label{sec:PBCContributors}

The studies benefited from contributions of the PBC physics working groups (\href{https://pbc.web.cern.ch/bsm}{BSM Working Group}, \href{https://pbc.web.cern.ch/fpc-mandate}{FIPs Physics Centre} and  \href{https://pbc.web.cern.ch/qcd}{QCD Working Group}) and of several PBC accelerator working groups including a dedicated ECN3 Beam Delivery Task Force (ECN3-TF)~\cite{bib:ECN3TFmandate,bib:ECN3TFmemo}.

\subparagraph{}
The contributing members of the accelerator working groups are listed below with the names of the conveners underlined.

{\parindent0pt \subparagraph{Accelerator complex capabilities:}
\underline{H. Bartosik}, T. Prebibaj, \underline{G. Rumolo}.}

{\parindent0pt\subparagraph{Beam Dump Facility:}
O. Aberle, C. Ahdida, P. Arrutia, K. Balazs, M. Calviani, Y. Dutheil, L.S. Esposito, R. Franqueira Ximenes, \underline{M. Fraser}, F. Galleazzi, S. Gilardoni, J.-L. Grenard, T. Griesemer, \underline{R. Jacobsson}, V. Kain, L. Krzempek, D. Lafarge, S. Marsh, J.M. Martin Ruiz, G. Mazzola, R.F. Mena Andrade, Y. Muttoni, A. Navascues Cornago, P. Ninin, J. Osborne, R. Ramjiawan, F. Sanchez Galan, P. Santos Diaz, F. Velotti, H. Vincke, P. Vojtyla.}

{\parindent0pt\subparagraph{Conventional Beams ECN3:}
C. Ahdida, D. Banerjee, A. Baratto Roldan, \underline{J. Bernhard}, \underline{M. Brugger} F. Butin, A. Ceccucci, N. Charitonidis, L.A. Dyks, L. Gatignon, J.-L. Grenard, Y. Kadi, L. Krzempek, G. Lanfranchi, C. Lazzeroni, K. Massri, M. Moulson. L.J. Nevay, E. Nowak, E.G. Parozzi, M. Van Dijk.}

{\parindent0pt\subparagraph{ECN3 Beam Delivery Task Force:} 
\underline{M. Brugger}, C. Ahdida, J. Bernhard, M. Calviani, Y. Dutheil, L.A. Dyks, L. S. Esposito, R. Folch, R. Franqueira Ximenes, \underline{M. Fraser}, J.-L. Grenard, Y. Kadi, E. Nowak, R. Ramjiawan, F. Sanchez-Galan, P. Schwarz, M. van Dijk, F. Velotti, C. Vendeuvre, H. Vincke, T. Zickler.}

\newpage

\tableofcontents

\newpage

\section*{Executive summary}

\addcontentsline{toc}{section}{Executive summary}

The PBC study group has supported the preparation of the proposals for future experiments in the CERN SPS North Area ECN3 experimental cavern beyond the currently approved programme, including their implementation and physics potential within the worldwide landscape.

\subsection*{Context}

There is strong and growing evidence from both particle physics and astrophysical observations for the existence of physics Beyond the Standard Model (BSM). Yet, so far it has evaded direct discovery in high energy colliders. 
This calls for novel experiments increasing the scope to search for new, low mass Feebly Interacting Particles (FIPs) as well as to indirectly probe the multi-TeV domain beyond direct LHC reach. High precision and high intensity are crucial tools in this endeavour.
In this context the CERN SPS complex provides a worldwide unique combination of high energy beams up to 400 GeV, high intensity and high duty cycle.

At CERN, completion of the CNGS neutrino beam program in 2012, together with injector upgrades performed for HL-LHC, leaves room for a new high-intensity facility as regards proton yield.
The best opportunity for such an implementation is the ECN3 underground experimental hall in the SPS North Area (NA), which was initially designed for high-intensity beams and currently hosts the NA62 experiment. NA62 data taking interleaves $K^+$ beam for $K^+$ rare decays measurements with Beam Dump (BD) mode for FIP searches. The program is approved until the LS3 shutdown scheduled from 2026 to 2028, and foresees to collect integrated intensities of $\approx10^{19}$ PoT and $\approx10^{18}$ PoT in the two modes, respectively.

Two main options are in competition in ECN3 beyond LS3. HIKE/SHADOWS combines an upgrade of NA62, HIKE, to perform higher precision measurements of rare kaon decays in two consecutive phases respectively devoted to $K^+$ and $K^0$ beams, with the possibility to take data in BD mode by closing a collimator, as is done by NA62, to look for FIPs. In the BD mode HIKE would be complemented by an off-axis detector, SHADOWS, to extend the acceptance at higher FIP masses and perform neutrino measurements. A possible longer term third phase optimized for the ultra-rare decay $K^0\to\pi^0\nu\bar{\nu}$ is not part of the current HIKE proposal and has not been considered in this study. Alternatively, BDF/SHiP is the implementation in ECN3 of the SHiP detector and the associated Beam Dump Facility (BDF). The latter was initially proposed as a new underground complex, and can be realized in ECN3 with a significant cost reduction. BDF/SHiP is designed as a state-of-the-art Beam Dump experiment with a dual spectrometer for searches of FIPs and neutrino measurements. It has been slightly downsized as compared to the former proposal to fit the ECN3 experimental hall, and brought closer to the proton beam dump to preserve the initial acceptance.

\subsection*{Beam and infrastructure upgrades}

HIKE/SHADOWS (resp. BDF/SHiP) request $\geq$~\SI{4.5}{\s} (resp.~$\geq$~\SI{1}{\s})-long proton spills with integrated intensities of up to 1.2 (resp.~4.0) $\times 10^{19}$ PoT/year.  New SPS operation modes have been designed to fulfill these needs in ECN3. They are compatible with the  delivery of more than $0.4 \times 10^{19}$ (resp.~$0.6 \times 10^{19}$) PoT/year to the other SPS experimental areas for the HIKE/SHADOWS (resp.~BDF/SHiP) scenario, which is comparable with the PoT delivered in recent years. The optimal operation mode for such high-intensity was found to consist in dedicated ECN3 spills which are directly transferred from the SPS slow extraction area to the target serving ECN3, and are characterized by significantly lower transfer losses as compared to the present operation mode.  The required proton beam line upgrades are the same for the two experimental options. They benefit from the already funded NA consolidation program, to which they add an extra material cost estimated to 14~MCHF with an uncertainty from 30 to 50~\%. 

The target serving ECN3 has to be fully rebuilt for both BDF/SHiP and HIKE/SHADOWS in order to stand the higher intensity and harsher radiation environment. The total target- and infrastructure-related costs are estimated to be in the range of 50~MCHF and similar for BDF/SHiP and HIKE/SHADOWS, though the design of the HIKE Phase 2 beamline is still ongoing including radiation protection and integration studies. The overall uncertainty for the cost estimate ranges from 30 to 50\%.

\subsection*{Experimental detectors}

The three considered detectors have similar global layouts consisting in a very low-pressure decay vessel followed by a spectrometer, with subdetector technologies adapted to the different operational constraints of the kaon and BD modes. In addition, SHiP and SHADOWS plan to host a small fine-grained dense detector with emulsions for FIP indirect detection and neutrino measurements. 

The HIKE detector will keep the NA62 structure and components with upgrades for each HIKE phase. Phase 1 primarily aims at a better timing resolution to stand the higher data taking rates, and at a better radiation hardness. This can benefit from HL-LHC-oriented R\&D (e.g. for silicon trackers) to match the stringent requirements. Phase 2 will adapt to the $K^0$ decay modes and associated background by removing some subdetectors and re-arranging others. The SHiP and SHADOWS detectors use well-established technologies with, however, harsher irradiation conditions for SHADOWS. Critical components of all projects are the magnet systems, especially those aimed to sweep the muon background out in BD mode. Final magnet designs will have to compromise between cost, electricity consumption, construction schedule and ability to achieve the very low background required by the experiments. The total material costs of the detectors are estimated to 27~MCHF (HIKE phase 1\&2 upgrades), 12~M\texteuro{} (SHADOWS) and 51~MCHF (SHiP), with uncertainties ranging from 10 to 30\%. 

\subsection*{Construction and operation schedules}

The preliminary beam upgrade schedule would allow the modifications required upstream of the target serving ECN3 to be implemented before the end of LS3 provided a timely decision. Because of resource competition with HL-LHC accelerator and detector upgrades, the ECN3-specific upgrades will extend beyond LS3 by at least one year.  This decoupling would allow other NA users to restart operation after LS3 while ECN3 components are installed and commissioned during Run~4. The three detector construction schedules are feasible but tight, especially for components still under R\&D, and will require timely decisions on subdetector options and funding to match the beam upgrades schedule.

Indicative operation schedules of BDF/SHiP and HIKE/SHADOWS options have been sketched. They span over more than 15 years of nominal data taking extending to the second half of the 2040s and they assume operation of the North Area will follow a pattern similar to the present one after the HL-LHC shutdown. HIKE/SHADOWS operation foresees 9 years shared between $K^+$ and BD mode and 6 years devoted to $K^0$ mode (BD operation in this mode is still under evaluation). The corresponding integrated intensities used as references to quantify the physics reach are: $6 \times 10^{20}$ PoT for BDF/SHiP; $5 \times 10^{19}$ PoT for HIKE/SHADOWS BD mode; $3.6 \times 10^{19}$ and $7.2 \times 10^{19}$ PoT for HIKE Phase 1 and Phase 2, respectively.  

\subsection*{Physics reach in worldwide landscape}

The main physics goals of the proposed projects include precision kaon physics, which is specific to HIKE, new neutrino measurements by SHiP and SHADOWS, and searches for FIPs by all three experiments. 

All planned measurements are based on rare processes and therefore highly sensitive to background. The dominant backgrounds which may affect the signals are random coincidences and DIS interactions of muons and neutrinos issued from the target area, as well as, for rare $K$ decays, contamination from the dominant $K$-decay channels. They were estimated for all projects with state-of-the-art detailed simulation tools and taking into account detector resolution. In addition, HIKE Phase 1 and SHADOWS benefit from extrapolations of NA62 real data in $K^+$ and BD modes, and SHiP has performed dedicated beam tests of muon production in a BDF target replica. The present results indicate that the most dangerous backgrounds will be kept under control for the targeted reference integrated intensities. The background estimations however require consolidation, especially for the $K^0$ beamline which is still under design. In case unexpected backgrounds show up in first real data, the long lifetime of the experiments should allow for detectors to be upgraded in order to mitigate them and ensure that backgrounds will in-fine not be the limiting factor of the measurements.

Kaon precision physics would extend the approved NA62 program with a $K^+$ integrated intensity higher by a factor $\approx$~4 and a novel study of rare $K^0$ decays. This gives access to hypothetical BSM high-mass states beyond the range directly accessible at the LHC, and to insights into the CKM matrix unitarity and Lepton Flavour Universality (LFU). Quantification of the agreement to the SM within BSM effective theories confirms the complementarity with B physics. $K^+$ precision physics at CERN is unique worldwide, and HIKE Phase 2 addresses $K^0$  channels that are complementary to the $K^0\to\pi^0\nu\bar{\nu}$ mode addressed in priority by KOTO at JPARC.

The SPS 400 GeV proton beam gives a worldwide unique possibility to efficiently search for FIPs in the MeV--GeV range up to the $b$ quark mass. HIKE has sensitivity to low-mass FIPs in the forward direction in BD mode, and (uniquely) to very-low mass FIPs from rare decays in Kaon mode. The addition of SHADOWS off-axis in BD mode extends the sensitivity to high-mass states, so that the HIKE/SHADOWS combination would significantly extend the exploration of FIPs within the worldwide landscape. The BDF/SHiP configuration is fully optimized for FIP searches in BD mode by providing sensitivity to low-mass FIPs produced forward, high-mass FIPs decaying at large angle, and scattering of invisible FIPs. It would provide ultimate sensitivity in the full mass range reachable at the SPS energy, in most cases beyond what would be achievable at CERN by the LHC proposed Forward Physics Facility~(FPF) and large angle FIP projects, as well as at FNAL by the DarkQuest Collaboration on the 120 GeV Main Injector.  

The highlight of neutrino studies planned by SHiP and SHADOWS would be the first quantitative measurement of $\tau$~neutrino and anti-neutrino interactions. 
SHADOWS may, however, be limited in $\nu_{\tau}$ statistics due to the lower neutrino flux in the off-axis position of its detector and its lower integrated intensity. SHiP on the other hand plans to measure several thousand of $\nu_{\tau}$ and $\bar{\nu_{\tau}}$ interactions, a sample which may be limited by systematic uncertainties rather than statistics. More studies are needed to quantify the projects' fundamental reach with neutrinos. Similar measurements are planned at the FPF though with a somewhat lower statistics than SHiP and in a different, complementary energy range.
 
All-in-all, a future high-intensity facility in ECN3 will have a unique impact in the worldwide landscape of the next decades. The physics criteria to select the experimental program will depend on the relative weights given to improvements in precision kaon physics, ultimate exploration of the FIPs territory in the SPS energy range, and novel neutrino measurements.

\newpage

\section{Introduction}

The PBC Study Group was initially mandated by the CERN Management to prepare the European Particle Physics Strategy Update~(EPPSU) for CERN projects other than high-energy frontier colliders. Following the EPPSU process, the PBC Study Group was confirmed on a permanent basis with an updated mandate~\cite{bib:PBC_mandate} taking into account the strategy recommendations. The Study Group is now in charge of supporting the proponents of new ideas to address the technical issues and physics motivation of the projects ahead of their external review by the CERN Scientific Committees and decision by the Management.     
The European Particle Physics Strategy Update has highlighted that {\it the quest for dark matter and the exploration of flavour and fundamental symmetries are crucial components of the search for new physics} and it has reaffirmed the importance of {\it a diverse programme that is complementary to the energy frontier}~\cite{Strategy:2019vxc, bib:EPPSU}.

The SPS North Experimental Area~(NA) is one of the major experimental facilities available at CERN and it is at the very heart of many present and proposed explorations for Beyond the Standard Model~(BSM) Physics. The area is presently ongoing an extensive consolidation campaign with major activities planned during the forthcoming LS3 (currently scheduled from 2026 to 2028) and the following LS4 under the NA Consolidation~(NA-CONS) Project. ECN3 is an underground cavern in the North Area suited for experiments requiring high-intensity.

ECN3 currently hosts the NA62 experiment \cite{bib:NA62} with an approved programme until LS3.
The following experimental proposals to be hosted in TCC8/ECN3\footnote{TCC8 is the Target Chamber Cavern upstream of ECN3.} 
have been studied within PBC: 
\begin{itemize}
    \item HIKE (High Intensity Kaon Experiment) proposing an extension of the current NA62 programme with charged kaons at higher intensity in a first phase and neutral kaons in a second phase. HIKE proposes phases~1~($K^+$) and 2~($K^0$) for approval in 2023. This programme will be complemented by the search for visible decays of Feebly-Interacting Particles (FIP) in Beam Dump (BD) mode on-axis~\cite{bib:HIKELOI_2022,bib:HIKE_PROPOSAL_2023};
    \item SHADOWS (Search for Hidden And Dark Objects With the SPS) to search for visible decays of FIPs and perform neutrino measurements by operating off-axis in parallel to HIKE BD mode~\cite{bib:SHADOWSLOI_2022,bib:SHADOWS_PROPOSAL_2023};
    \item BDF (Beam Dump Facility) and the associated SHiP (Search for Hidden Particles) experiment to search generically for Hidden Sector particles~\cite{Ahdida:2703984,Ahdida:2704147,bib:BDFSHIPLOI_2022,bib:BDFSHIP_PROPOSAL_2023} through both scattering and decay signatures. The detector system for scattering signatures is also suited for neutrino interaction physics, in particular exploring the tau neutrino.
\end{itemize}

Decisions should be taken well ahead of LS3 for a timely implementation of the chosen options and to profit of the potential synergies with the NA-CONS Project. The present document is aimed as an input to recommendations by the SPS and PS Experiments Committee (SPSC) and decision by the CERN Management. After a short reminder of the current ECN3 hall set-up and beam characteristics (Section~\ref{sec:current_status}), the main technical aspects and physics motivations of future options are presented (Section~\ref{sec:ExpProposals}). The technical issues of beam production, operation mode and detectors integration are summarized in Sections~\ref{sec:Operation} and~\ref{sec:ExtractionTransfer}. Technically-driven schedules and first cost estimates are given in Section~\ref{sec:ScheduleCost}. Finally, Section~\ref{sec:PhysPotential}
presents the physics reach of the various options within the CERN and worldwide physics landscape.

\section{Current status}
\label{sec:current_status}

\subsection{The North Experimental Area}

\begin{figure}[htbp]
  \centering
  \includegraphics[width=\linewidth]{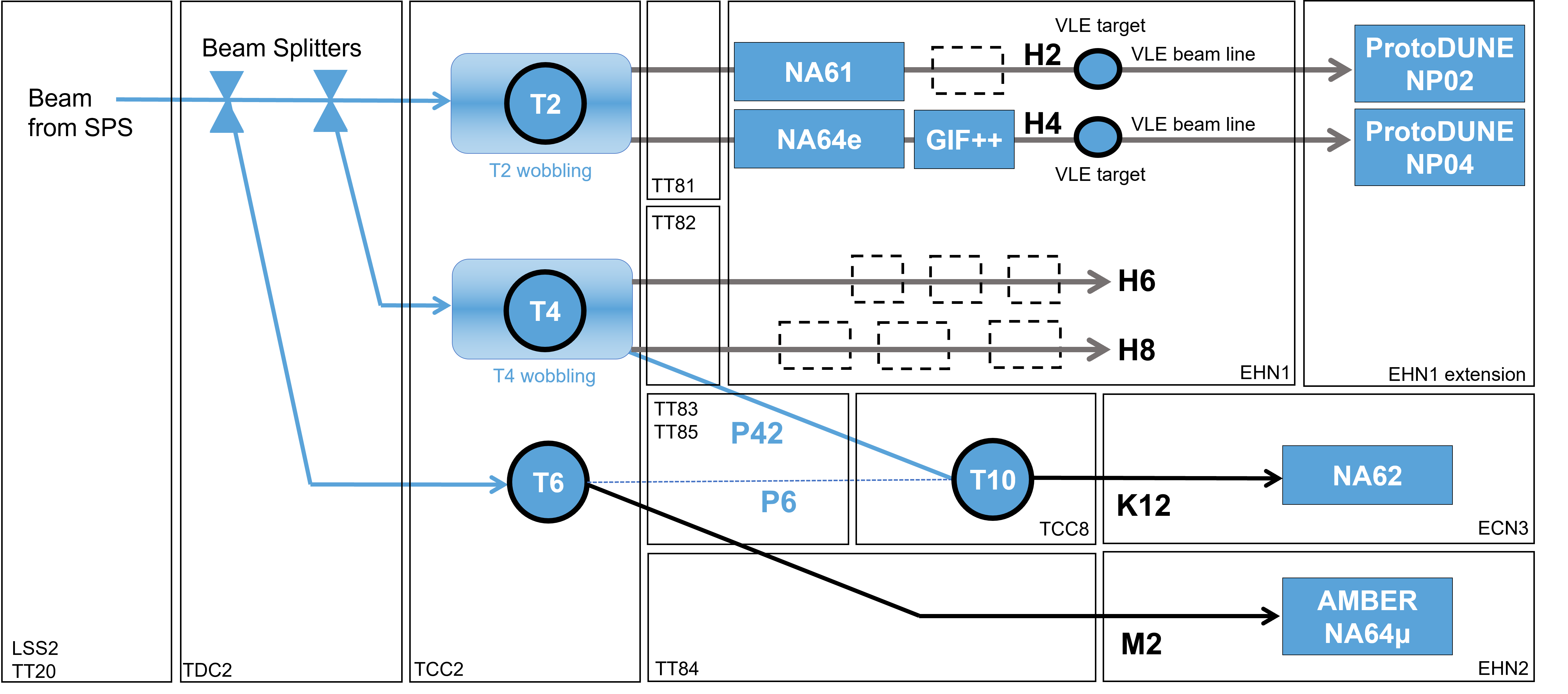}
  \caption{\small A schematic layout of the NA beamline and experiment complex as of 2023.}
  \label{fig:NA-layout}
\end{figure}

 The NA (Figure~\ref{fig:NA-layout}) is located on the CERN Pr\'evessin site. The three beryllium Targets T2, T4 and T6 in the TCC2 Target Hall (see Figure~\ref{fig:NA-layout}) are served by slow-extracted beams from the SPS via a dedicated transfer line (TT20).
NA comprises two surface halls~\cite{Banerjee:2774716}, EHN1 and EHN2, and an underground cavern, ECN3. 

EHN1 is the biggest surface hall at CERN 
and houses the H2, H4, H6, and H8 beamlines. The T2 target feeds the H2 and H4 beamlines, which are normally operated as versatile secondary or tertiary beams but may occasionally be configured as attenuated primary beams. The H4 beam is a particularly clean electron beam but can also serve its users with high-quality hadron and muon beams. The H6 and H8 beamlines are fed by secondary particles produced in the T4 target. These are versatile hadron and electron beams that can also provide low or medium intensity muon beams. The EHN1 beamlines are used for test-beam activities and currently host two physics experiments: the NA61
experiment~\cite{Gazdzicki:2029881,Abgrall:1642156} on the H2 beamline has a rich and varied physics programme with hadron and ion beams, and the NA64 experiment~\cite{Banerjee:2223648} on H4 performs a competitive dark photon search with high purity electron beams~\cite{Depero:2256460}. A future heavy ion experiment, NA60+~\cite{bib:NA60+LoI},  is also in discussion for implementation on H8~\cite{Gerbershagen:2806698}.

EHN2 is served by the M2 beamline~\cite{Doble:250676} from the T6 target. M2 provides a worldwide unique high-energy, high-intensity muon beam, and can also be operated as a high-intensity hadron beam. An option to operate it as a tertiary electron beam exists, but the rates are very low. EHN2 currently hosts the NA66/AMBER experiment~\cite{Abbon:1028264,Abbon:1950827} (successor of COMPASS), proposed to operate as a long-term QCD facility, and NA64$\mu$~\cite{Gninenko:2653581}, with similar objectives as NA64 in H4, but employing muon beams. M2 may also host MUonE~\cite{Abbiendi:2677471} and other projects in the future.

ECN3 is served by the K12 beamline derived from the T10 target: the primary protons not interacting in T4 are transported by the P42 beamline over almost $900$~metres to the T10 beryllium target located in the Target Hall TCC8. T10 initiates the K12 beamline which delivers a high-intensity mixed secondary hadron beam at 75~GeV/c with a ${\approx}$~6~\% kaon component to the NA62 experiment~\cite{CortinaGil:2257518} in ECN3.

\subsection{TCC8, ECN3 Experimental Cavern and the NA62 Experiment} 
 
\begin{figure}[htbp]
    \centering
    \includegraphics[width=\linewidth]{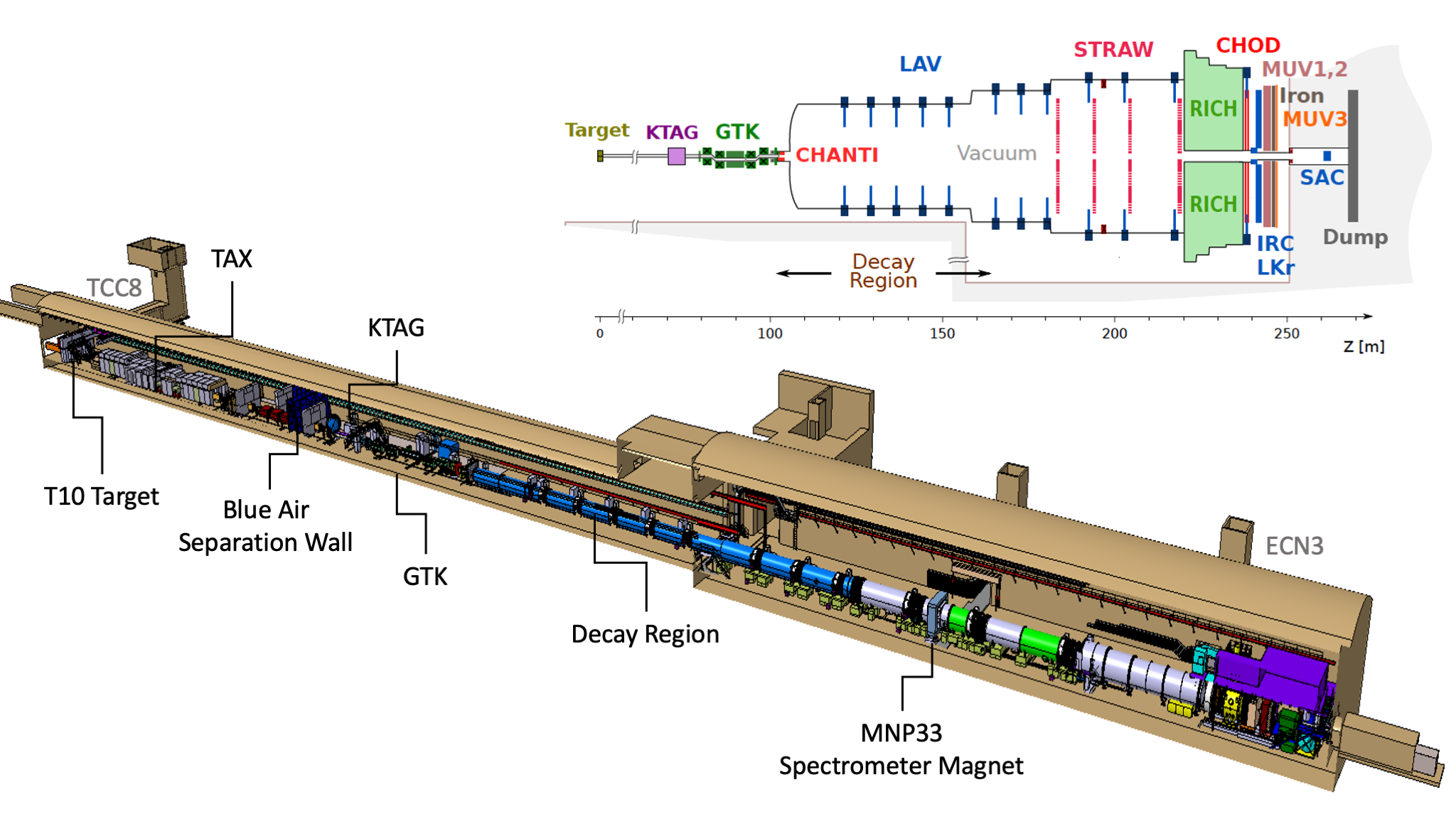}
    \caption{\small Overall layout of the current TCC8/ECN3 underground complex hosting the T10 target and TAX, the K12 beamline and the NA62 detector components.}
    \label{fig:TCC8-ECN3-layout}
 \end{figure}

 The current overall layout of the TCC8/ECN3 underground complex is shown in Figure~\ref{fig:TCC8-ECN3-layout} together with the K12 beam and the main detector components of NA62.
 
 TCC8 is split in two parts by an over-pressure double "Blue Wall" aimed to separate the air volumes of the target and detector/beamline areas during operation. It is followed by the ECN3 experimental hall.

The K12 mixed beam is produced by interaction of the primary protons with the~\SI{400}{\mm} long beryllium T10 target and focused onto a pair of dump collimators (TAX for "Target Attenuator eXperimental areas") made of massive copper and steel blocks. The beamline optics, and in particular a set of four strong dipoles surrounding the TAX ("first achromat"), ensure a selection of secondary particles at a momentum of~\SI{75}{\GeV}/c with a 1.1~\% RMS momentum resolution. Off-momentum and neutral particles are directly dumped into the TAX and positrons are filtered out with the help of a thin tungsten converter. The ${\approx}$~6~\% kaon component of the selected mixed hadron beam is tagged by the KTAG Cherenkov detector. After collimation and cleaning stages, the beam passes a second set of dipoles ("second achromat") that has been equipped with the fast Silicon strip detectors of the NA62~\SI{750}{\MHz} GigaTracker~(GTK), which measure the momentum, position and direction of each beam particle. A key component of the K12 beamline is the active muon sweeping system, to reduce the muon rate from hadron decays in the NA62 detector, consisting of several iron-filled dipole magnets and a toroid.

The NA62 experiment can also be operated in beam-dump mode. In this case the T10 target is moved out of the beam remotely and the full beam is dumped on the TAX collimators including primary and secondary particles. The muon sweeping system is left activated but with a modified configuration.

The NA62 experiment~\cite{bib:NA62} is currently mainly focusing on the core of its baseline programme devoted to the $K^+\to\pi^+\nu\bar{\nu}$ ultra-rare decay. 20 candidate events have been observed before LS2, in agreement with the Standard Model~(SM) expectation of 10 physics + 7 background events. The main goal of the approved programme until LS3 is 
to perform a $\mathcal{O}$(15-20\%) measurement of the $K^+\to\pi^+\nu\bar{\nu}$ branching ratio.
Other rare $K$ decays are being investigated in parallel. 
To perform the approved programme, NA62 has recently implemented detector upgrades allowing to operate at the nominal beam intensity of $3\times10^{12}$~protons per~\SI{4.8}{\s} long spill, with the goal to accumulate $10^{19}$~Protons on Target~(PoT) until LS3.
The approved programme also includes several months of data taking in beam dump mode to search for hidden particles up to
an integrated intensity of~$10^{18}$~PoT.  A sample of about $1.4\times 10^{17}$~PoT has already been collected in dump mode, and confirms that the expected combinatorial background is under control.

\subsection{Current operation mode and limitations}
\label{Ch3_RPaspects}

The number of protons that can be delivered to NA is primarily driven by the present performance of the SPS, which can accelerate more than $4\times10^{13}$~particles~(protons)~per pulse~(ppp) at an energy of~\SI{400}{\GeV}. In the present NA shared operation mode, more than $3.5\times10^{13}$~ppp can be routinely extracted from the SPS extraction Long Straight Section~(LSS)~2, transported via the TT20 transfer line and distributed to the three NA targets in TCC2 by means of two consecutive magnetic beam splitters located in the TDC2 area~(Figure~\ref{fig:NA-layout}) according to the user needs. The global transfer efficiency from SPS to the targets of 76~\% corresponds to a total of ${\approx}2.7\times10^{13}$~ppp impinging on the NA targets. Its measurement suffers from large uncertainties related to the calibration of the intensity monitors at the target stations and the above value should be considered as pessimistic.

An SPS cycle includes a 400 GeV flat-top~(FT), during which the slow extraction over~\SI{4.8}{\s} takes place, preceded by the injection plateau and acceleration ramp and followed by the magnet ramp down, for a total cycle duration of \SI{10.8}{\s}. The minimum repetition period is \SI{14.4}{\s}, limited by the maximum average power dissipation in the SPS main magnets ($\approx$~\SI{41}{\mega\watt})~\cite{Bartosik:2650722, Prebibaj:2848908}. 
The SPS cycle that serves the NA is part of a global "supercycle" with cycles serving other CERN users. LHC injection cycles are present only a few hours per day in average, so that the supercycle duration is primarily defined by non-LHC user needs. The typical NA duty cycle (NA spill length over supercycle length) is ${\approx}$20~\%. A typical number of ${\approx}$~3000 spill/day can be assumed, taking into account an SPS availability for physics of approximately 80~\%. For a typical 200 days of SPS operation within a year, the $2.7\times10^{13}$~ppp deliverable to the NA targets therefore corresponds to a maximum of $1.6\times10^{19}$~PoT/year. This estimate is based on the assumption that maximum intensity is reached from the start of the run and does not take into account running time with ions. Typically, both the accelerator and detectors require some time for the intensity ramp-up at the beginning of each yearly run.

NA beam operation poses several radiation protection~(RP) constraints that are already nowadays a challenge for operation and maintenance of accelerator components. Beside residual and prompt dose rate constraints, also activation of air, water and soil and radioactive waste production have to be considered.

Due to the nature of the slow extraction process and the need to serve multiple target stations simultaneously, significant activation of components occurs in LSS2, the TDC2 splitter area and the TCC2 area hosting the target stations. These areas were designed and built in the 1970s when RP regulations were less restrictive in comparison to today. Interventions in these areas are challenging due to the very high dose rates of some of the components and the lack of extensive remote handling and manipulation features. As a consequence, significantly long cool-down times might be needed before interventions. In addition, radiation damage to cables, considering the typical frequency of recabling campaigns, limits the annual integrated intensities to NA in shared mode to $\approx 1\times10^{19}$ PoT ~\cite{Bartosik:2650722,Li_IEFC_0302203} unless beam loss reduction measures are put in place.

The T2, T4, T6 and T10 NA targets have been partially renovated during LS1, and are designed to withstand the slow extraction (typical spill length of~\SI{4.8}{\s}) of a maximum proton intensity of at least~$1.5\times10^{13}$~ppp with a repetition period of~\SI{14.4}{\s}~\cite{EDMS1267311}. Operational experience with the T6~TAX indicates that similar conditions are acceptable for the downstream TAXs (deformation of the TAX holes or risk of local melting could occur for higher peak or average power deposition). The current nominal intensity of the P42 line is 5 times lower and amounts to $3.3\times10^{12}$~ppp. 
The corresponding intensity of the K12 selected~\SI{75}{\GeV}/c mixed beam is $2\times10^{9}$~ppp, for a spill duration of~\SI{4.8}{\s}, to match the specifications of the NA62 GTK. 
The operation of the K12 TAX in beam dump mode at the present intensity already puts the materials of the TAX blocks close or beyond their operational limits~\cite{EDMS2303290}. Operation at significantly higher intensities would therefore require a redesign of the overall target systems located in TCC8, including T10 target and K12 TAX~\cite{CBreport,EDMS2303297}.

In addition to high residual dose rates, prompt beam losses may also cause elevated dose rates in the areas of the NA that are accessible during beam operation. Recent studies~\cite{Ahdida1} have identified two critical locations above the P42 line where the current ECN3 beam operation provokes elevated prompt radiation levels close to or even exceeding the classification limit of the given area:
\begin{itemize}
    \item ramp on the Sal\`eve side of EHN1 where only $\approx$~1.2~m of soil is present between the P42 line and the ramp, in the following referred to as \textit{EHN1 ramp};
    \item bridge over a watercourse flowing above a section of the P42 line where only $\approx$1.2~m of soil is present, in the following referred to as \textit{ECN3 bridge}.
\end{itemize}
Tiny losses in the beamline elements below the \textit{EHN1 ramp} can produce the observed prompt radiation fields~\cite{Ahdida2, p42_radiation, Ahdida1} for the present-day beam parameters and therefore requiring a series of mitigation measures that were already, or are currently, being implemented~\cite{ECN3TFreport}. 

 P42 has an uninterrupted vacuum sector that spans from the T4 XTAX to the T10 target. Historically, the vacuum in P42 was achieved by means of turbomolecular pumps, however, these were moved to K12 and replaced by rotary pumps, as part of the preparation of the NA62 experiment, for financial reasons. The resulting pressure is now limited to $10^{-3}$~mbar and deemed adequate for proton transport today, but contributing to distributed losses and prompt radiation as vacuum levels are degrading due to ageing problems of the vacuum equipment. 

Access to ECN3 is not possible during beam operation, but can proceed immediately after beam stop downstream the Blue Wall between the TCC8 and ECN3 caverns. In the TCC8 target area upstream of the Blue Wall, a cool down period of \SI{30}{\minute} followed by an air flush during~\SI{90}{\minute} is needed after beam stop and before access.

\section{Post-LS3 experimental proposals}
\label{sec:ExpProposals}

\subsection{Overview of possible operational scenarios} \label{sec:OPScenarii}

From the existing experimental proposals two possible operational scenarios can be envisaged:

\begin{itemize}
    \item An extension of kaon physics and hidden sector exploration at higher intensity combining the HIKE~\cite{bib:HIKELOI_2022,bib:HIKE_PROPOSAL_2023} and SHADOWS~\cite{bib:SHADOWSLOI_2022,bib:SHADOWS_PROPOSAL_2023} projects. The continuation of high-intensity kaon experiments at CERN with HIKE provides a flavour probe into BSM physics. HIKE phase~1 would include an upgrade of the $K^+$ beam intensity, ultimately by a factor 4 (requiring up to $1.2\times10^{13}$~ppp on the T10 target over~$\geq$~\SI{4.5}{\s}), together with corresponding improvements of detector performances. During HIKE phase~2, a high intensity $K^0$ beam would be produced by up to $2\times10^{13}$~ppp on the T10 target and the detector configuration changed for $K^0$ decays, still keeping tracking devices, with the main goal of observing for the first time the ultra-rare decay $K^0_L\to\pi^0 l^+ l^-$ and performing a wide-range exploration of $K^0_L$ decays. Operation in BD mode at $2\times10^{13}$~ppp on the T10 TAX would alternate with kaon beam runs during HIKE phase-1. In order to maximize the reach of this extended BD operation, the SHADOWS decay spectrometer is proposed to be built off-axis downstream of the T10 TAX and to be operated in parallel to HIKE during BD runs. 

    \item Hidden Sector exploration with the implementation in ECN3 of the proposed SHiP detector and the associated Beam Dump Facility~\cite{bib:BDFSHIPLOI_2022,bib:BDFSHIP_PROPOSAL_2023}, formerly proposed on a new dedicated site in Pr\'evessin~\cite{Ahdida:2703984, Ahdida:2704147, Ahdida:2654870}. Following the EPPSU recommendations, the BDF proposal has been further optimized and other possible locations have been considered and compared, identifying ECN3 as the most suitable and cost-effective option~\cite{Aberle:2802785}. Fitting the SHiP detector within ECN3 requires a resizing of the detector components, and a shortening of the distance to the beam dump in order to preserve the signal acceptance. SHiP is proposed as a state-of-the art dual spectrometer, able to measure hypothetical hidden particles, both through their scattering in an instrumented high-density interaction target, and through their decays in a large acceptance decay spectrometer. The BDF implementation in ECN3 would correspond to a further increase of the proton beam intensity to $4\times10^{13}$~ppp over~$\geq$~\SI{1.0}{\s}. 
\end{itemize}

The experimental requirements are summarized in Table~\ref{tab:ExpReq} and the corresponding SPS/NA operation modes and proton sharing are discussed in Section~\ref{sec:Operation}.

\begin{table}[h!]
    \centering 
    \resizebox{\columnwidth}{!}{
    \begin{tabular}{lccccc}
    \toprule
    {} &   Intensity to TCC8 & Spill Length & PoT/nominal operation year & \# nominal operation years & Total PoT \\
    {} &    [$10^{13}$ p/spill] & [s] & [$10^{19}$]&  & [$10^{19}$]\\
    \midrule
    HIKE phase~1 ($K^+$)~\cite{bib:HIKELOI_2022}& 1.2 & $\geq 4.5$ & 0.72 &   5 & 3.6 \\
    HIKE phase~2 ($K^0$)~\cite{bib:HIKELOI_2022}  & 2.0 & $\geq 4.5$ &  1.2 &   6 & 7.2 \\
    HIKE/SHADOWS BD~\cite{bib:HIKELOI_2022,bib:SHADOWSLOI_2022}     & 2.0   &$\geq 4.5$ &  1.2 & 4  & 5\\
    BDF/SHiP~\cite{bib:BDFSHIPLOI_2022}     & 4.0 &  $\geq 1.0$ & 4 &   15 &  60\\
    \bottomrule
\end{tabular}
    } 
    \caption{\small Experimental beam requirements~\cite{userrequirements}. The different spill parameters required in the first two columns result from detector rate management and signal/background considerations for the different experiments. As an example the combinatorial background (see section \ref{sec:FIPs}) has an inverse quadratic dependence on the spill length for a given spill charge.}
    \label{tab:ExpReq}
\end{table}

\begin{figure}[h!]
    \centering
    \includegraphics[width=1.0\linewidth]{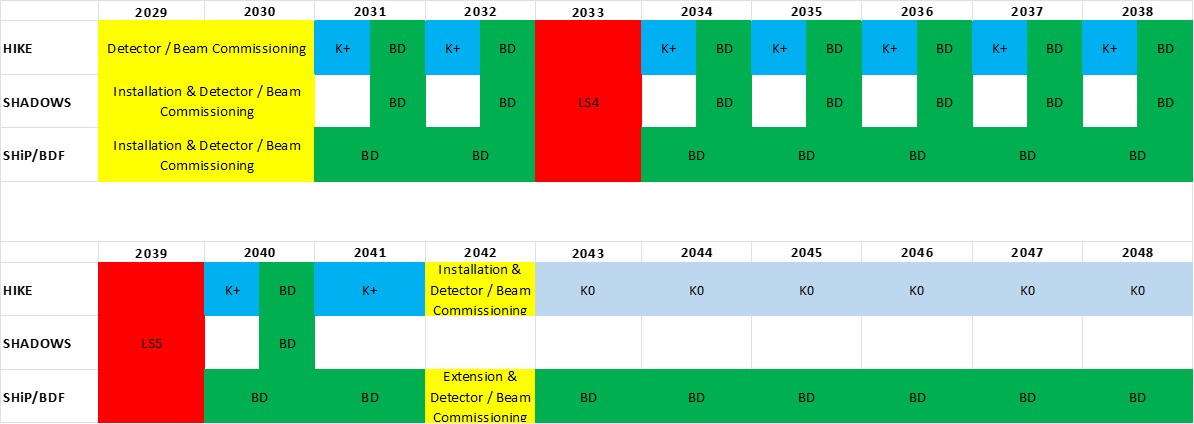}
    \caption{\small Indicative operation schedule for the proposed experiments.}
    \label{fig:indicativeschedule}
 \end{figure}

An indicative schedule based on the presently available long-term CERN Accelerator Complex schedule up to the end of High Luminosity-LHC~(HL-LHC)~\cite{bib:LongTermAccSchedule}, consistent with the above described operational scenarios and compatible with the requirements summarized in Table~\ref{tab:ExpReq}, is shown in Figure~\ref{fig:indicativeschedule}. The above schedule assumes:
\begin{itemize}
    \item ECN3 nominal operation starting in 2031;
    \item equal sharing of the operation time between $K^+$ and BD mode during the first 8 years of nominal operation for the HIKE/SHADOWS scenario;
    \item same calendar time span for both scenarios.
\end{itemize}

For both operational scenarios the experimental programme extends to the second half of the 2040s, well beyond the HL-LHC operation.
It is assumed that the operation of the North Area will follow a pattern similar to the present one also after the end of HL-LHC. The distribution and duration of LSs might change the experiments calendar-year duration. 

In the following, the experimental sensitivities of the projects are quantitatively estimated for the total numbers of PoT given in Table~\ref{tab:ExpReq} in compliance with the indicative schedule of  Figure~\ref{fig:indicativeschedule}.

\subsection{HIKE}

\subsubsection{Physics case}
The continuation of high-intensity kaon experiments at CERN with HIKE provides a unique probe into BSM physics, that can reach mass scales of~$\mathcal{O}$(100) TeV and gives access to a different, and in some cases higher, sensitivity to new physics than the $B$ and $D$ meson sectors~(see Section~\ref{subsec:flavor}). 
The primary goal of HIKE is to improve the accuracy of the kaon rare decay measurements, in order to match and possibly challenge the theory precision, to study and measure for the first time channels not yet observed, and to search with unprecedented sensitivity for kaon decays forbidden by the SM. HIKE can also address BD physics in a complementary mass range and phase space to other existing and planned experiments~(see Section~\ref{sec:FIPs}). A summary of HIKE sensitivity reach in the flavour sector is reported in Table~\ref{tab:hike-flavour}.


\begin{table}[htb]
{\small
\centering

\footnotesize
\begin{tabular}{lll}
\hline
$K^+\to\pi^+\nu\bar\nu$ & $\sigma_{\cal B}/{\cal B}\sim5\%$ & BSM physics, LFUV \\
$K^+\to\pi^+\ell^+\ell^-$ & Sub-\% precision on form-factors & LFUV \\
$K^+\to\pi^-\ell^+\ell^+$, $K^+\to\pi\mu e$ & Sensitivity ${\cal O}(10^{-13})$ & LFV / LNV \\
Semileptonic $K^+$ decays & $\sigma_{\cal B}/{\cal B}\sim0.1\%$ & $V_{us}$, CKM unitarity \\
$R_K={\cal B}(K^+\to e^+\nu)/{\cal B}(K^+\to\mu^+\nu)$ & $\sigma(R_K)/R_K\sim {\cal O}(0.1\%)$ & LFUV \\
Ancillary $K^+$ decays & \% -- \textperthousand & Chiral parameters (LECs) \\
(e.g. $K^+\to\pi^+\gamma\gamma$, $K^+\to\pi^+\pi^0 e^+e^-$) \\
\hline
$K_L\to\pi^0\ell^+\ell^-$ & $\sigma_{\cal B}/{\cal B} < 20\%$ & ${\rm Im}\lambda_t$ to 20\% precision, \\
&& BSM physics, LFUV \\
$K_L\to\mu^+\mu^-$ & $\sigma_{\cal B}/{\cal B}\sim 1\%$ & Ancillary for $K\to\mu\mu$ physics \\
$K_L\to\pi^0(\pi^0)\mu^\pm e^\mp$ & Sensitivity ${\cal O}(10^{-12})$ & LFV \\
Semileptonic $K_L$ decays & $\sigma_{\cal B}/{\cal B}\sim 0.1\%$ & $V_{us}$, CKM unitarity \\
Ancillary $K_L$ decays & \% -- \textperthousand & Chiral parameters (LECs), \\
(e.g. $K_L\to\gamma\gamma$, $K_L\to\pi^0\gamma\gamma$) & & SM $K_L\to\mu\mu$, $K_L\to\pi^0\ell^+\ell^-$ rates\\
\hline
\end{tabular}
\caption{\small Summary of HIKE sensitivity for flavour observables. The $K^+$ decay measurements will be made in Phase~1, and the $K_L$ decay measurements in Phase~2. The symbol $\cal B$ denotes the decay branching ratios. More details will be given in section~\ref{subsec:flavor}.}
\label{tab:hike-flavour}
}
\end{table}


Sensitivity projections in BD mode are produced assuming $5\times 10^{19}$ PoT. Operation at $2\times 10^{13}$ ppp for 4.8 s spills is assumed (although HIKE could accept a somehow higher beam intensity when running in BD mode). Parallel operation of SHADOWS with HIKE in BD mode increases acceptance at large angle and improves searches for large-mass hidden particles such as Heavy Neutral Leptons (HNLs), light Dark Scalars and Axion Like Particles~(ALPs) with respect to HIKE alone.

\subsubsection{Experiment description}
The HIKE programme~\cite{bib:HIKELOI_2022,bib:HIKE_PROPOSAL_2023} uses shared detectors and infrastructure to address flavour physics both with charged and neutral kaon beams: a charged kaon phase and  a neutral kaon phase with tracking are put forward for SPSC review in 2023.

\begin{figure}
\centering
\includegraphics[width=0.95\textwidth]{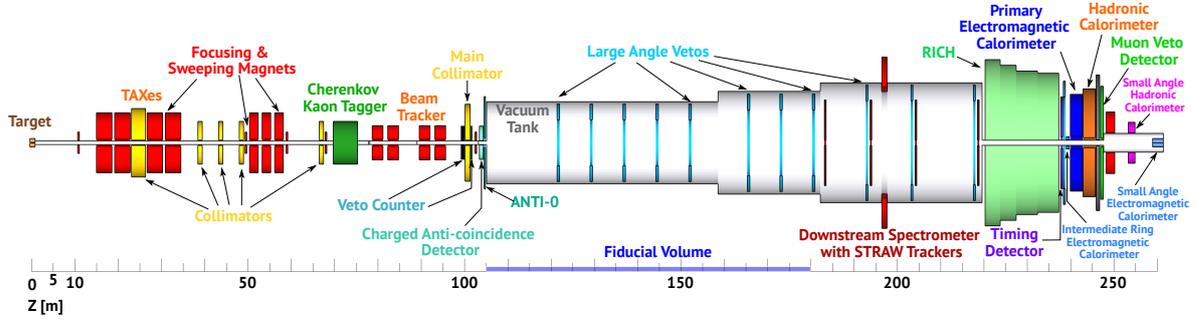}
\caption{\small Beamline and detector layout for HIKE Phase 1 ($K^+$), with an aspect ratio of 1:10.}
\label{fig:hike_phase1}
\end{figure}
The setup and detectors in the charged kaon phase (Phase 1) will be optimised for the precision detection to $\mathcal{O}$(5\%) of the branching ratio of $K^+ \rightarrow \pi^+ \nu \bar\nu$. While the conceptual layout is based on the successful one of NA62, new detectors will replace those of NA62 with the goal of improving the performance and sustaining higher rates; prime examples are the beam tracker and the tracking spectrometer. The detector configuration for Phase 1 is illustrated in Figure~\ref{fig:hike_phase1}.

Thanks to the relatively compact detector, the neutral beam plus tracking phase (Phase 2) allows for a 90 m long fiducial decay volume to be accommodated in the present ECN3 experimental hall, with no major civil engineering work.
\begin{figure}
\centering
\includegraphics[width=0.95\textwidth]{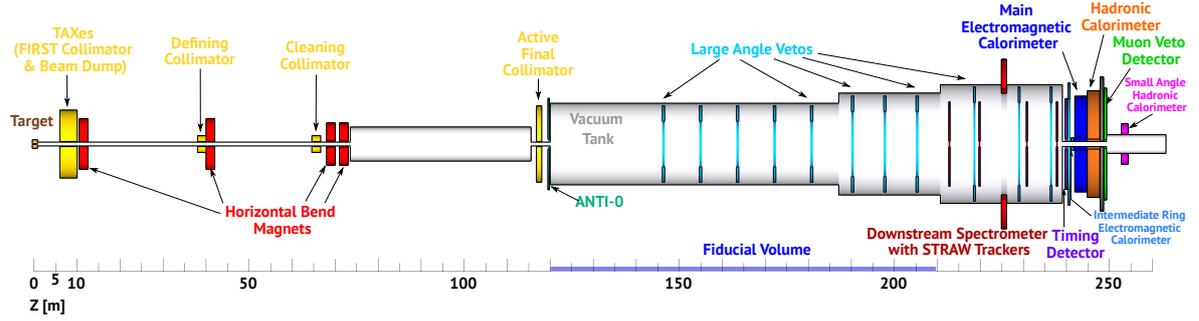}
\caption{\small Beamline and detector layout for HIKE Phase 2 ($K_L$), with an aspect ratio of 1:10.}
\label{fig:hike_phase2}
\end{figure}
This phase will use an experimental setup with minimal modifications with respect to the charged kaon phase, but important modifications will have to be implemented in the K12 beamline. The beam tracker, kaon-identification, pion-identification detectors will be removed; the main tracking spectrometer will be shortened, and central holes of the chambers will be realigned on the neutral beam axis; the Large Angle Veto detectors will be moved and possibly reduced in number and the small angle calorimeters will be moved.
The detector configuration for Phase 2 is illustrated in Figure~\ref{fig:hike_phase2}.
Many of the same requirements arise in the design of the electromagnetic calorimeter~(ECAL) for the $K^+$ and $K_L$ phases. 
A design for a fast calorimeter with excellent photon detection efficiency and energy resolution to be used in all phases of the HIKE programme is therefore preferable and chosen as the baseline. However, the LKr calorimeter remains a valuable option for the early commissioning and data taking phases.
The foreseen evolution of the detector configuration is summarized in Table~\ref{tab:hike-detectors}.
The efficiency of the trigger and data acquisition system of NA62 is affected by increasing intensity; besides, a hardware triggered approach as currently used by NA62 is intrinsically prone to limitations. For these reasons, a trigger-less approach is foreseen for HIKE, where data filtering is mostly at high-level-trigger level.

\begin{table}[h!]
\centering

\begin{tabular}{lccl}
\hline
Detector & Phase~1 & Phase~2 & Comment \\
\hline
Cherenkov tagger & upgraded & removed & faster photo-detectors\\
Beam tracker & replaced & removed & 3D-trenched or monolithic silicon sensor \\
Upstream veto detectors & replaced & kept & SciFi \\
Large-angle vetos & replaced & kept & lead/scintillator tiles\\
Downstream spectrometer & replaced & kept & STRAW (ultra-thin straws)\\
Pion identification (RICH) & upgraded & removed & faster photo-detectors\\
Main EM calorimeter & replaced & kept & fine-sampling shashlyk \\
Timing detector & upgraded & kept & higher granularity\\
Hadronic calorimeter & replaced & kept & high-granularity sampling\\
Muon detector & upgraded & kept & higher granularity\\
Small-angle calorimeters & replaced & kept & oriented high-Z crystals\\
HASC & upgraded & kept & larger coverage\\
\hline
\end{tabular}
\caption{\small Detector configurations for HIKE Phases 1 and 2. The evolutions are indicated versus the current NA62 configuration for Phase 1, and versus the upgraded Phase 1 configuration for Phase 2.}
\label{tab:hike-detectors}
\end{table}

HIKE BD operation will build upon the experience accumulated in NA62 with BD data taking, 
where the proton beam is made to interact in the T10 TAX. A similar procedure will be possible in HIKE, which will be able to switch between kaon and dump mode during an 8-hours SPS Machine Development~(MD) time slot.

The reach for the various channels assumes: $2\times 10^{13}$ kaon decays in decay volume per year 
for the $K^+$ beam and $3.8\times 10^{13}$ kaon decays in decay volume per year 
for the $K^0_L$ beam plus tracking.
A uniformly distributed intensity over $\geq$\SI{4.5}{\s} spill is essential in all phases, to optimally collect high statistics while effectively managing detector rates and spurious intensity effects.

\subsubsection{Present status, required R\&D}
Details of specific technologies envisaged for detectors and readout systems can be found in~\cite{bib:HIKELOI_2022,bib:HIKE_PROPOSAL_2023}. In brief, state-of-the-art technologies considered to push the time resolution, granularity and rate performances are:
\begin{itemize}
    \item Beam tracker, based on the TimeSpot sensor and  Application Specific Integrated Circuit~(ASIC) technology, or new monolithic silicon sensors. Sensors with the desired performances exist already. Related ASICs are being developed, in synergy with other high-energy experiments happening on a similar timescale.
    \item Main tracker based on ultra-thin Straws. A prototype is being developed already.
    \item Electromagnetic~(EM) calorimeter, a fine-sampling shashlyk based on PANDA~(antiProton ANihilation at DArmstadt) forward EM calorimeter.
    \item Small-angle EM calorimeter based on a compact Cherenkov calorimeter with oriented high-Z crystals. Test beam results already indicate feasibility.
    \item Photon detectors for kaon and pion identification detectors, based on Micro-Channel Plate-Photo Multipliers~(MCP-PMTs). These devices already satisfy the requirements but are susceptible to aging. Aging tests with modified Atomic Layer Deposition~(ALD) prototypes are ongoing.
    \item Large-angle photon vetoes, based on lead/scintillator tiles with Wavelength Shifting~(WLS) read-out by Silicon Photo-Multipliers~(SiPMs). The technology is well established.
    \item Hadron calorimeter, based on a high-granularity sampling calorimeter.
    \item Timing planes and charged particle vetoes, based on scintillating tiles readout by SiPMs.
    \item Veto counter based on Scintillating Fibre~(SciFi) technology, as that used in LHCb.
\end{itemize}
In summary, all the mentioned technologies are established, at least as proof of principle, and several are synergetic to detector developments for High-Luminosity LHC~(HL-LHC) experiments.

\subsection{SHADOWS}

\subsubsection{Physics case}

SHADOWS aims to perform a comprehensive search for FIPs (discussed in detail in Section~\ref{sec:FIPs}) in the range from the MeV scale to a few GeV. It aims at exploiting the upgraded 400 GeV proton beam line P42, slowly extracted from the SPS, by running {\it off-axis} concurrently with the proposed HIKE experiment. 

In the MeV-GeV range, the strongest bounds on the interaction strength of new light particles with SM particles exist up to the kaon mass; above this mass the bounds weaken significantly. SHADOWS can take an important step forward into this still poorly explored territory and has significant discovery potential for FIPs if they have a mass between the kaon and the beauty mass. If no signal is found, SHADOWS will push the limits on their couplings with SM particles by up to two orders of magnitude, depending on the model and scenario, opening new directions in model building.

SHADOWS is meant to expand HIKE's capability to search for FIPs from kaon decays and in BD mode, by enhancing the sensitivity for FIPs coming from charm and beauty hadron decays. The combined system SHADOWS+HIKE can span the still uncovered parameter space of many well motivated FIP models, {\it below} and {\it above} the kaon mass, with a competitive sensitivity in the international landscape. 
Since theoretically there is no uniquely preferred mass range for FIPs, the capability of spanning {below} and {above} the kaon mass, ranging from a few MeV up to the $b$ mass, is paramount.

The {\it off-axis} position allows SHADOWS to be less impacted by backgrounds (especially neutrinos) 
than an {\it on-axis} setup, and to be placed close to the FIP production point.

With the NaNu subdetector, SHADOWS also aims to study neutrino physics (in particular $\tau$ neutrinos) in a phase space complementary to the one explored at SND and FASER experiments, currently running at the LHC.
The capability of the NaNu subdetector to search for light DM is currently being studied.

\subsubsection{Experiment description}
\label{sec:SHADOWS_exp}
The SHADOWS detector \cite{bib:SHADOWSLOI_2022, bib:SHADOWS_PROPOSAL_2023} requirements are defined by the characteristics of FIPs produced in the interactions of the 400~GeV/c proton beam with a dump. At these energies, FIPs with masses above the kaon mass are mostly produced in the decays of charmed and beauty hadrons and in proton bremsstrahlung and/or Primakoff effect occurring in the dump.
At the SPS centre-of-mass energy ($\sqrt{s} \approx 28$~GeV) the heavy hadrons are produced with a relatively small boost so that FIPs emerging from their decays have a large polar angle and can be detected by an off-axis detector. 
The distance of the detector with respect to the impinging point of the proton beam onto the dump is a compromise between the maximisation of FIP flux in acceptance (that requires short distances) and the maximisation of the probability that the FIP decays before reaching the detector (that requires long distances).
The optimal distance varies as a function of the FIP model and benchmark. The current compromise, also taking into account beam background and irradiation, corresponds to an off-axis distance of the decay vessel lateral wall to the beamline of \SI{1.45}{\m} , and to a longitudinal distance of the decay vessel upstream window from the upstream face of the TAX dump of \SI{15}{\m}. The background lateral veto wall remains to be integrated in the layout.

The SHADOWS detector must be able to reconstruct and identify most of the visible final states of FIP decays while simultaneously reducing the background to a level of less than 1 event in the whole data set.

To this aim a standard spectrometer with excellent tracking and timing performance, and an efficient veto system and some particle identification capability is required. The spectrometer will be made of:

\begin{itemize}

  \item {\it A magnetic muon sweeping system} based on magnetised iron blocks (MIB) in front and aside the decay volume to sweep away from the detector acceptance the muons emerging from the dump.  
     
    \item {\it An efficient veto system} able to tag the residual muon flux surviving the MIB system before it enters the decay volume. This system is made of two components, an {\it upstream veto} and a {\it lateral veto}, made of 2 active layers instrumented with micro-megas, to veto muons entering from the front-face and the side close to the beam line of the decay vessel. The exact configuration of the lateral veto is still being optimized. The sensitivity simulations presented later assume a baseline lateral veto instrumentation along the full length of the decay volume. 

    \item {\it A Tracking System} able to reconstruct with high accuracy the mass, the decay vertex and the impact parameter with respect to the impact point of the beam on the dump for FIP decays with at least two charged tracks in the final state. The requirements are:  i) a vertex resolution of $\approx {\mathcal{O}}(1)$~cm in the transverse plane over a volume length of $\approx$~20~m;  ii) an impact parameter resolution  of $\mathcal{O}$(cm) for FIP decays into two charged tracks when the total momentum is extrapolated backward at the impact point of the beam on the dump. Two technologies for the tracking stations are currently under scrutiny, the scintillating fibre option and the NA62-like straw tubes option, which is currently the baseline. 

   \item {\it The dipole magnet: }
    Two designs of the dipole magnet providing a bending power of about \SI{0.9}{\tesla\meter} are being considered: i) a normal-conducting~(NC) option, designed in order to have a power consumption of \SI{287}{\kilo\watt}, i.e. 10 times lower than that of the NA62 dipole magnet for the same bending power; ii) a superconducting~(SC) option. The NC option is the current baseline.

   \item {\it a Timing Detector} with $\approx 100$~ps time resolution in order to reduce any combinatorial background (and in particular the muon one, see Sec.~\ref{sec:FIPs}) by requiring the tracks to be coincident in time. The tracks of combinatorial background events are indeed intrinsically out-of-time with respect to each other as they have origin times spread over the 4.8~sec duration of a typical P42 proton spill.
   The timing layer will be made of scintillating bars of \SI{1}{\cm} thickness with SiPM readout.  
   
   \item {\it An Electromagnetic Calorimeter} able to reconstruct the energy with a mild resolution of $\sigma(E)/E \approx 10-15\%/\sqrt{E(\text{GeV)}}$, a time resolution of few ns and some pointing capability in order to reconstruct the mass of fully neutral decays such as $ALP \to \gamma \gamma$. Two options are currently under study: the SplitCal option and a StripCal option, based on scintillating strips. The StripCal option looks very compelling and represents to date our baseline.
   
    \item {\it A Muon Detector} to positively identify muons with timing capabilities to reinforce the rejection of the combinatorial muon background in combination with the timing detector. The muon detector will be based on scintillating tiles with direct SiPM readout. This technology allows a compact, efficient and cost-effective detector to be built. The measured time resolution per station is ${\mathcal{O}}$(250) ps.
    
\end{itemize}

The baseline solution to reduce the background of inelastic interactions of neutrinos with the air of the decay volume (Section~\ref{sec:FIPs}) will be to put the decay volume in a mild ($\approx$~1~mbar) vacuum. A compelling alternative is a decay volume made of a balloon filled with Helium, to be studied for the Technical Design Report~(TDR).

\begin{figure}[!t]
    \centering
    \includegraphics[width=\linewidth]{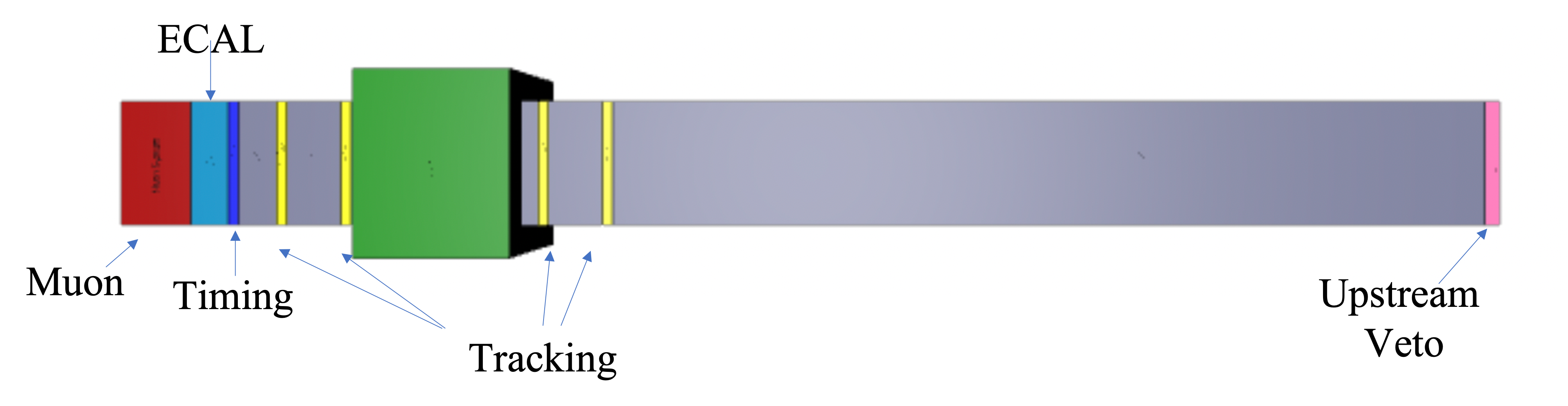}
   \caption{\small Schematic layout of the SHADOWS spectrometer.}
    \label{fig:shadows}
\end{figure}

\begin{figure}[!t]
    \centering
    \includegraphics[width=\linewidth]{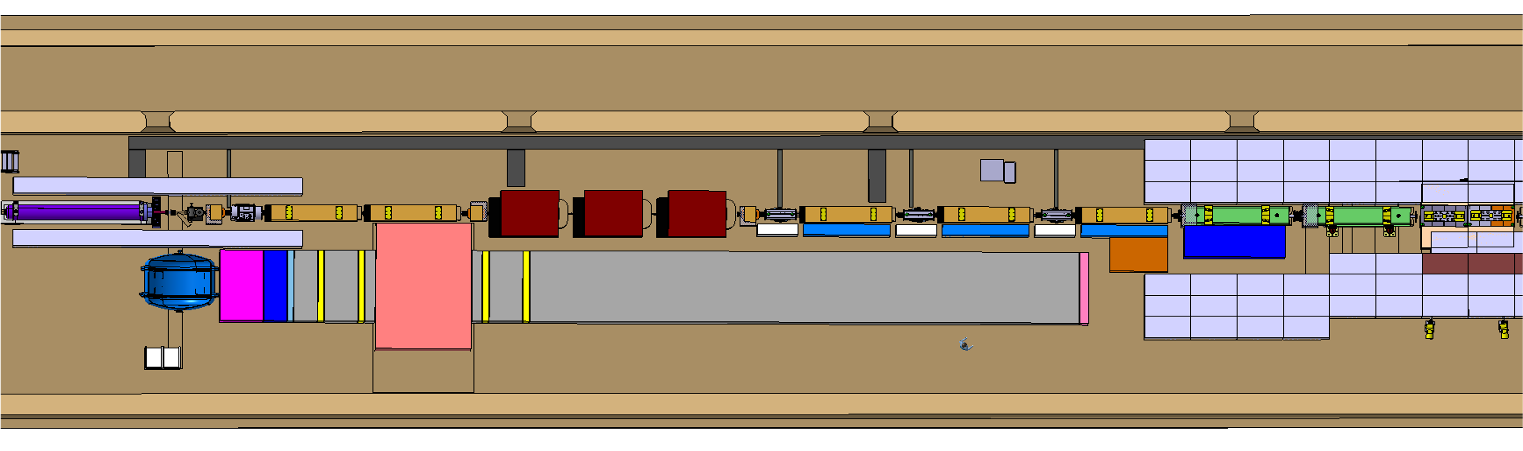}
   \caption{\small Schematic layout of the SHADOWS spectrometer integrated in the experimental area~\cite{EDMS_Integration_Shadows_Hike_Phase1}. The side wall of the reserved space for the decay vessel and detector volume is \SI{1.45}{\m} from the beamline axis. The reserved space includes 100~mm for cables and services on each side. The longitudinal distance of the decay vessel upstream window from the upstream face of the TAX dump is \SI{15}{\m}. The background lateral veto wall between the decay volume and the beamline is not shown since its configuration is still being optimized. It will have to be included in the space reserved for the detector to keep sufficient space for access to the beamline.}
    \label{fig:shadows_in_area}
\end{figure}

 The baseline layout of the spectrometer with the in-vacuum decay vessel is shown in Figure~\ref{fig:shadows}. 
 The spectrometer integrated in the experimental area close to the dump is shown in Figure~\ref{fig:shadows_in_area}.

\begin{figure}[t!]
\includegraphics[width=0.42\textwidth]{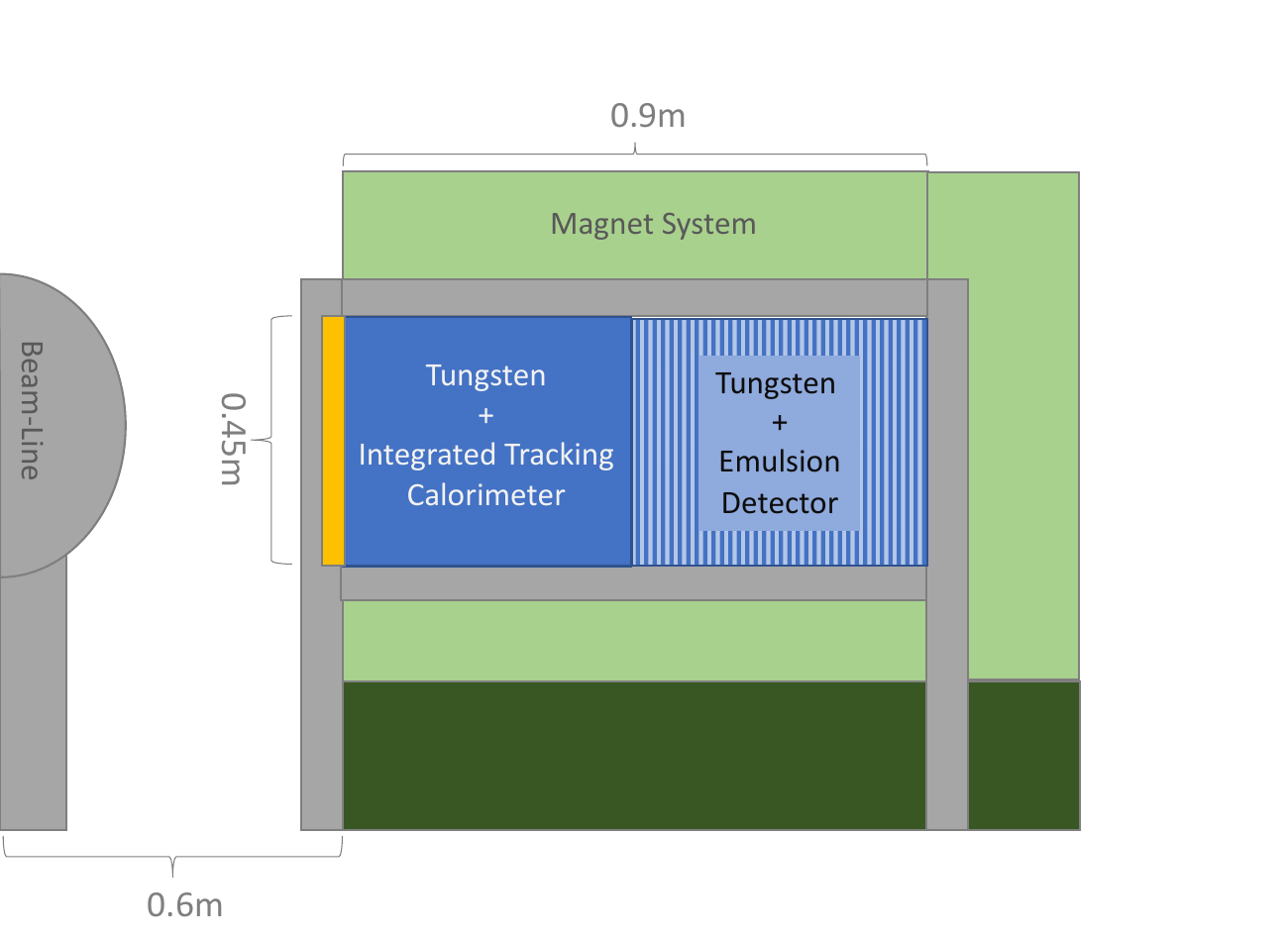}%
    \includegraphics[width=0.58\textwidth]{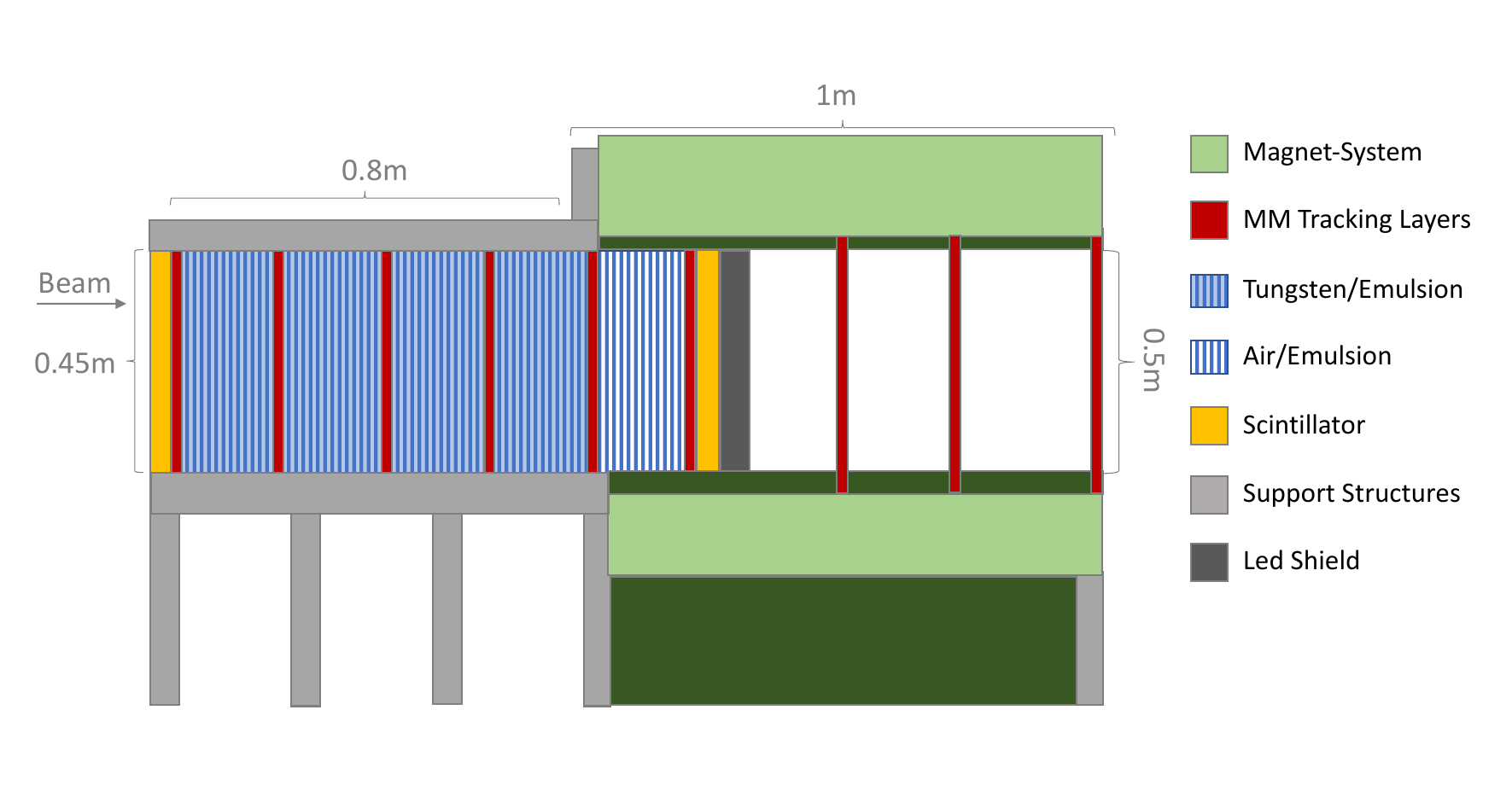}%
\caption{\small Baseline implementation of the NaNu subdetector: transverse view (left) showing the emulsion-based detector as well as the active detector based in micromegas and scintilator technologies; longitudinal view (right) showing the magnet and tracking stations downstream of the detectors.}
\label{fig:FrontView}
\end{figure}

Directly downstream the main SHADOWS detector, a specific neutrino sub-detector system called NaNu (NorthArea NeUtrino Experiment, \cite{Neuhaus:2022hji}) will be positioned approximately \SI{50}{\m} downstream from the beam dump and \SI{0.6}{\m} off-axis. The baseline concept involves two main detector components: the "active-detector" and the "emulsion-detector," both having dimensions of $45 \times 45 \times 100$~cm$^3$. These detectors will be partially located inside an existing dipole magnet at CERN with gap dimensions of $50 \times 100 \times 100$~cm$^3$ and a magnetic field strength of 1.4~T generated by a current of 2500~A. The transverse plane of the NaNu subdetector, facing the interaction point, has a total size of $45 \times 90$~cm$^2$. The active detector component, positioned close to the beamline, is a calorimeter system that utilizes passive tungsten plates and plastic scintillators with tracking capabilities using Micromegas chambers. Its purpose is to study muon neutrino interactions. The emulsion detector consists of tungsten plates interleaved with emulsion films and is designed to study interactions involving tau and electron neutrinos. The combined passive material in both systems amounts to a total mass of approximately 2.4 tons in each detector component. Following the detectors, there is a spectrometer for measuring the momentum of muons, utilizing a \SI{1.5}{\tesla} magnetic field over a length of \SI{1}{\m}. Depending on funding availability and the feasibility of reducing muon background, it is possible to replace the active detector component with a second emulsion-based detector design, which could increase the expected number of tau neutrino interactions by up to a factor of five. The schematic layout of the baseline NaNu version is shown in Figure~\ref{fig:FrontView}.

The SHADOWS detector description with its sub-detector options is fully documented in the SHADOWS Proposal \cite{bib:SHADOWS_PROPOSAL_2023} including still open issues.

\subsubsection{Present status, required R\&D}
To a large extent, prototypes or even full-size detectors based on the technologies proposed for SHADOWS have already been built or operated. Hence, in most cases the R\&D is meant to further optimize the design of an already well established and known technology, rather than proving that a given technology is suitable for the task.

\begin{itemize}

\item {\it Upstream and Lateral Vetoes:} The measured performance of micromegas prototypes are: i) a few mm spatial resolution, ii)  MHz/cm$^2$ rate capability; iii) 10-20 ns time resolution; iv) >95~\% single layer efficiency. A dedicated prototype for SHADOWS is currently being built. 

\item {\it Tracking system:} Detectors in operation (SciFi Tracker in LHCb and Straw tracker in NA62) guarantee the reliability of the two technologies in consideration. A thorough R\&D is expected to happen in the coming years.

\item {\it Timing layer:} The scintillating material will be chosen from what is commercially available. The scintillating bars will then be read out at both ends with commercially available SiPMs, with SiPMs mounted on front-end~(FE) electronics Printed Circuit Boards~(PCBs) derived from those developed and produced for other projects such as the ATLAS Phase-II ITk Strip upgrade. 

\item {\it ECAL:}  A SplitCal prototype has already been built and successfully operated in the context of the R\&D for the SHiP detector~\cite{Bonivento:2018eqn}. The StripCal option is currently being studied. 
A dedicated R\&D is foreseen to happen in the coming years.

\item {\it Muon Detector: } A thorough R\&D has already been performed in the past 2 years within the AIDA-Innova European Grant. 
Two SHADOWS full-size prototypes have been built and used in June 2023 to measure the off-axis muon flux in the ECN3 cavern. 

\item {\it TDAQ: } The TDAQ system will be as much as possible in common with HIKE. This will allow to design a high-performance and cost-effective system and share with HIKE expertise and person-power. 

\end{itemize}

\subsection{BDF/SHiP}

\subsubsection{Physics case}
\label{sec:SHiP_physics}

BDF/SHiP is a state-of-the-art experimental setup designed to perform a generic search for FIPs with maximal sensitivity in a region of mass and coupling that is only accessible with a dedicated beam-dump configuration. The physics programme includes searches through both decay and scattering signatures. 

The beam parameters listed in Table~\ref{tab:ExpReq} for BDF/SHiP give SHiP access, {\it annually}, to~$\approx 2\times10^{17}$~charmed hadrons, $\approx1.4\times10^{13}$~beauty hadrons, $\approx 2\times 10^{15}$ tau leptons, and ${\mathcal{O}}(10^{20})$ photons above~\SI{100}{\MeV} within the acceptance of the detectors. 

The overall detector concept provides sensitivity to as many final states as possible~\cite{detectorEPJC2022}, including both fully and partially reconstructed final states, to ensure model-independent searches. Sensitivity to decay modes with neutrinos enable SHiP to explore for instance HNLs with enhanced $U_\tau$ – coupling and neutralinos.

The BDF/SHiP physics programme was explored in detail in a dedicated physics book in 2015~\cite{2016SHiPPhysicsCase} prepared by a large collaboration of theorists. It has been further elaborated over the years and was part of the comprehensive coverage of the field in the EPPSU 2020 Physics Briefing Book~\cite{bib:EPPSU}. Beyond the exploration of FIPs, BDF/SHiP is also particularly suitable for a rich program of tau-neutrino physics and measurements of neutrino-induced charm production. More details on both these aspects can be found in section~\ref{sec:PhysPotential}. It has also been shown that the BDF/SHiP target system can give unique access to a high-intensity neutron spectrum~\cite{Ahdida:2703984} that is not easily accessible at spallation facilities. This makes it possible to implement a user platform~\cite{PhysRevAccelBeams.25.103001} for studying neutron-induced reactions on short-lived isotopes that is relevant for nuclear and astrophysics~\cite{https://doi.org/10.48550/arxiv.2209.04443}, as well as for material testing~\cite{Harden_2019}, and radiation-to-electronics (R2E) studies.

The BDF/SHiP physics performance is anchored in an optimised acceptance to all FIP production mechanisms~\cite{bondarenko2023optimal} accessible with the~\SI{400}{\GeV} protons, combined with a highly efficient background suppression. The background suppression relies on a set of critical components:
\begin{itemize}
\item target of high density material with short interaction length to suppress weak decays of pions and kaons to muons and neutrinos,
\item iron hadron stopper to absorb hadrons and electromagnetic radiation produced in the dump,
\item magnetic muon shield starting with magnetisation of the hadron stopper and followed by free-standing magnets to deflect the muons produced in the dump ($\approx10^{11}$ per spill), away
from the detector acceptance,
\end{itemize}
and in particular for the search for FIP decays:
\begin{itemize}
\item background taggers fully surrounding the decay volume, both upstream and on all sides, to protect against residual muons leaking through the shield, and against hadrons from muon and neutrino deep-inelastic scattering~(DIS) interactions, as well as cosmics,
\item  vacuum in the decay volume to suppress in particular neutrino DIS.
\end{itemize}

The detector systems provide further suppression by the reconstructed quantities in terms of fiducial volume, track quality, vertex quality, impact parameter at the dump target, timing, and particle identification. The designed redundancy in the background suppression allows for a common, very simple and robust event selection with these quantities, and to measure background components by relaxing criteria. The selection has been demonstrated through full simulation to be entirely inclusive with respect to different types of long-lived particle decays. This ensures maximum sensitivity in the FIP searches, while remaining generic to new models that may be proposed in the future.

In addition to improving present constraints on many models by several orders of magnitude, the SHiP decay spectrometer allows distinguishing between different models, and, in a large part of the parameter space, measure parameters that are relevant for model building and cosmology. At the limit of sensitivity of other experiments, BDF/SHiP expects $\mathcal{O}$(100--1000) events throughout the mass range. These features make BDF/SHiP a unique direct-discovery tool for FIPs.  Moreover, together with the direct search for Light Dark Matter~(LDM), and neutrino physics, BDF/SHiP represents a wide-scope general-purpose beam-dump experiment.

\subsubsection{Experiment description}
\label{sec:SHiP_exp}

A detailed description of the detector, the design and the detector performance from measurements with prototypes in test beam have been reported in Refs.~\cite{Ahdida:2704147,Ahdida:2654870,detectorEPJC2022, bib:BDFSHIPLOI_2022} (complete list of dedicated reports in~\cite{bib:BDFSHIPLOI_2022}) and updated in~\cite{bib:BDFSHIP_PROPOSAL_2023}. Below is a summary of the most relevant features of the SHiP detector. 

\begin{figure*}
    \centering
    \includegraphics[width=0.99\linewidth]{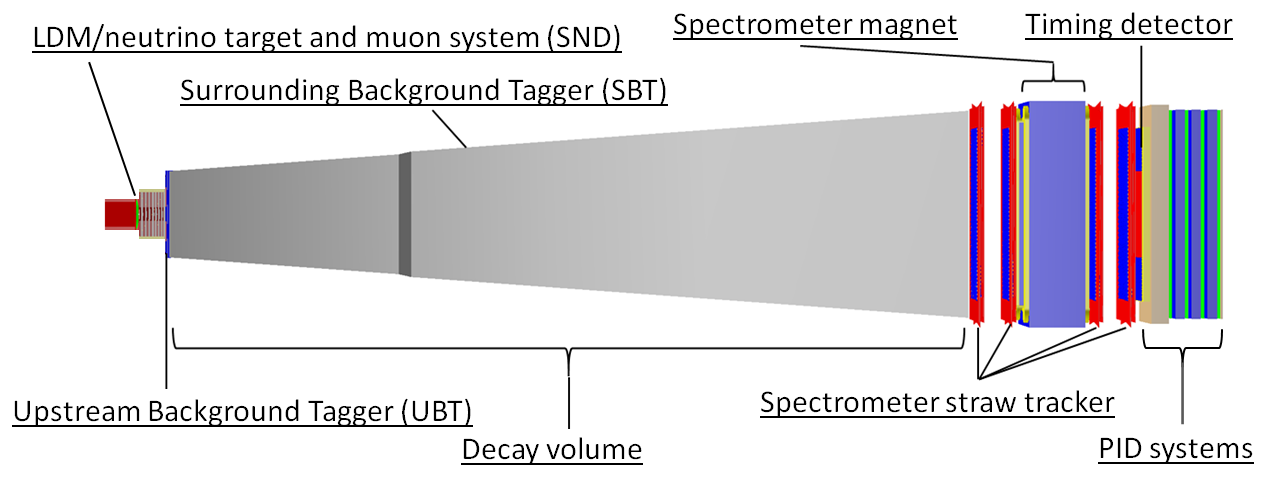}
   \caption{\small Schematic layout of the Scattering and Neutrino Detector~(SND) and Hidden Sector Decay Spectrometer~(HSDS).}
    \label{fig:HSDS}
\end{figure*}

The SHiP experiment is composed of a muon shield and dual system of complementary apparatuses, shown in Figure~\ref{fig:HSDS}. The upstream system, the Scattering and Neutrino Detector~(SND), is designed for the search for LDM scattering and for neutrino physics. 
The downstream system, the Hidden Sector Decay Search~(HSDS) detector is designed to reconstruct the decay vertices of FIPs, measuring invariant mass and providing particle identification of the decay products.

The revision of SHiP from the original Comprehensive Design Study~(CDS)~\cite{Ahdida:2704147} in ECN4 to the smaller ECN3 experimental hall has required reducing the lateral dimensions of the HSDS spectrometer.  The aperture of the spectrometer has been reduced from~\SI{5}{\m} width and~\SI{10}{\m} height to $4\times$\SI{6}{\square\meter}, consequently  also leading to a reduction of the decay volume and the particle identification systems in height and width. The lengths of the decay volume and the HSDS detector systems remain unchanged. This work has been accompanied by an effort to shorten the muon shield. The aim of bringing the experiment closer to the proton beam dump is to preserve the signal acceptance for all physics modes, production and scattering/decay kinematics convolved together, with a detector that is also decreased in cost. The first studies of the experimental layout for ECN3, as described in the LoI~\cite{bib:BDFSHIPLOI_2022}, continued focusing on a muon shield entirely based on NC magnets. The studies led to a fully developed NC alternative with a $\approx$~\SI{5}{\m} shorter muon shield, and a ~\SI{3}{\m} shorter configuration of the SND, and acceptable background rates and sensitivity. First explorations with SC technologies were done during the CDS phase and have continued in the context of ECN3~\cite{bib:BDFSHIP_PROPOSAL_2023} with the help of external expertise. These studies have concluded on an optimised hybrid muon shield in which the first section is based on SC technology, and the second alternate-field section is based on NC technology. This has made it possible to further reduce the overall length of the muon shield by $\approx$~\SI{5}{\m}, and to implement the SND with a length of about~\SI{6}{\m}. Given that the investigations of the SC magnets are promising, and that they lead to on overall reduction in size of the NC section of the muon shield, the experimental layout, physics performance, and cost have been evaluated with the hybrid muon shield~\cite{bib:BDFSHIP_PROPOSAL_2023}, shown in Figures~\ref{fig:SHiP} and~\ref{fig:SHiP_geometry}.

The SND detector consists of a LDM/neutrino target with vertexing capability incorporated in the form of tungsten plates alternated with emulsion films and fast electronic detector planes. The SND target system is followed by a muon spectrometer that is designed to identify and determine charge and momentum of muons produced in the $\nu_{\tau}$ interactions at high efficiency.

The electronic detector planes in the SND target region are based on scintillating fibres. The configuration allows reconstructing the shower produced by the recoil electron in LDM scattering to determine the initial particle angle and energy. In addition, the micro-metric accuracy of the nuclear emulsion provides topological discrimination of LDM interactions against neutrino-induced background events. For the neutrino physics programme, the emulsion technique is crucial to detect tau leptons and charmed hadrons by disentangling their production and decay vertices with the help of the sub-micrometric position and milliradian angular resolution. 

\begin{figure}[t!]
\begin{center}
\includegraphics[width=0.55\textwidth]{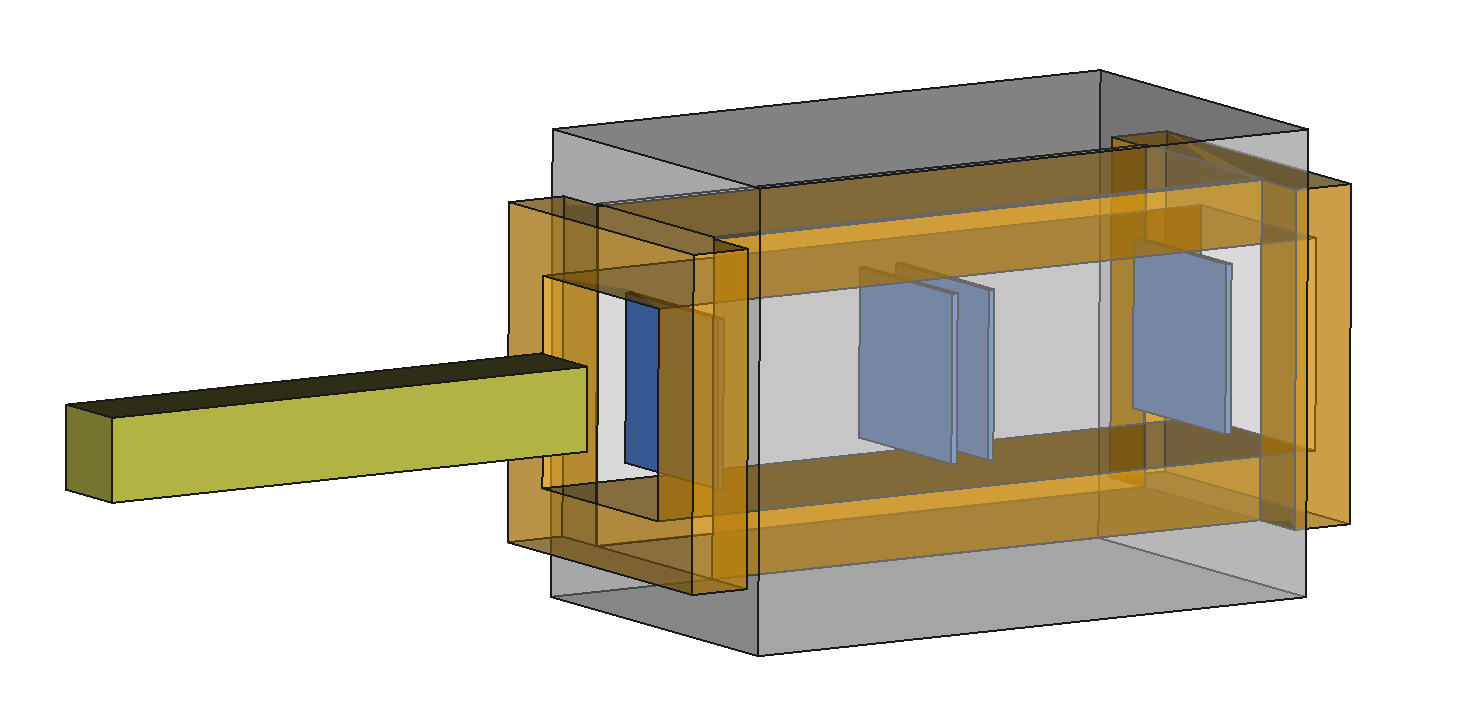}
\caption{\small
Conceptual layout of the Scattering and Neutrino Detector (SND) in the SHiP apparatus. 
\label{fig:SNDSHiP}}
\end{center}
\end{figure}

With respect to the CDS design, the magnet around the LDM/neutrino target has been removed, leading to a loss of the charge determination in the hadronic modes of the $\nu_\tau$ interactions. Instead the magnetised muon spectrometer distinguishes between $\nu_\tau$ and $\overline{\nu}_\tau$ in the golden mode $\tau\rightarrow\mu\nu_{\tau}\overline{\nu}_{\mu}$. Without the magnet around the target, the momentum of charged pions and kaons is measured through the detection of their multiple Coulomb scattering in the target~\cite{KODAMA2007192}. Neutral pions are also detected in the emulsion films and their energy measured. 
The detector is designed to observe all three neutrino flavours and perform searches for new particles through the scattering with the electrons and the nucleons of the SND target. The LDM/neutrino target and vertex detector are implemented as walls of emulsion cloud chamber (ECC) technology. Each wall consists of alternating layers of nuclear emulsion films, acting as the micrometric precision tracking stations, interleaved with tungsten layers, acting as the high-density passive layers of the target. The role of the target tracker between the walls is to provide the time stamp of the interactions located in the ECCs and to connect muon tracks between the target and the muon spectrometer. A conceptual layout of the SND detector is shown in Figure~\ref{fig:SNDSHiP}.

With the shorter distance to the proton target, the same yield of tau neutrinos as in the CDS design may be achieved with a $\approx$\SI{3}{\tonne} and $0.4\times0.4$\,m$^2$ LDM/neutrino target (\SI{8}{\tonne} in~\cite{Ahdida:2704147}), thus reducing the required surface of emulsion films to 145\,m$^2$. 

Immediately downstream of the SND, the HSDS detector measures both fully reconstructable decays of FIPs as well as partially reconstructable decays with neutrinos in the final state in a~\SI{50}{\m} long decay volume of a pyramidal frustum shape that is delineated by the deflected beam-induced muon flux.

The HSDS decay volume is followed by a spectrometer. The main element of the spectrometer is the spectrometer tracker, designed to accurately reconstruct the decay vertex, the mass, and the impact parameter of the reconstructed FIP trajectory at the proton target. The initial design of the magnet was based on a NC coil~\cite{Ahdida:2704147, HSspectrometer_magnet, Bajas:2839592}. In order to significantly reduce the power consumption, the CDS phase included a study of a new type of superconductor-based design~\cite{Bajas:2839592}. An R\&D programme with the goal of developing a demonstrator is currently underway at CERN with the involvement of a SHiP institute. 

A particle identification system, including an electromagnetic and a hadronic calorimeter, provide particle identification, which is essential in discriminating between the very wide range of models with FIPs, but also in providing information for background rejection. The electromagnetic calorimeter is a scintillator/lead sampling calorimeter, consisting of two parts of 3 and 17 radiation lengths ($X_0$), respectively, which
are mechanically separated in the longitudinal direction. Each part is equipped with a high spatial resolution layer in order to precisely measure the shower axes and allow reconstructing the vertex of ALP$\rightarrow\gamma\gamma$ decays and the invariant mass. Measurements of shower profiles with a prototype in test beam show that, with a few mm transverse shower-position resolution in the high-precision layers, an angular resolution of the order of a few mrad is achievable. The longitudinal segmentation of the calorimeter also improves the electron/hadron separation. 

 Background from neutrinos interacting within the decay volume is eliminated by maintaining the decay volume at a pressure of $\approx$~\SI{1}{\milli\bar}. The decay volume wall is instrumented upstream and on all sides by a system of high-efficiency background taggers in order to provide regional and temporal veto against muon and neutrino interactions in the vessel walls and against particles entering the volume from outside, including cosmics. The taggers covering the surrounding walls (SBT) are based on a liquid scintillator system segmented in cells, resulting in an efficiency of >99\%, and $\approx$\,ns time resolution. The tagger on the upstream vessel wall~(UBT) is based on three six-layer Multigap Resistive Plate Chambers~(MRPC), each with $\approx$50\,ps resolution, 98\% efficiency and spatial resolution of a few millimetres. A dedicated timing detector is located between the last spectrometer tracker plane and the calorimeters to provide a measure of time coincidence in order to reject combinatorial backgrounds. It is based on scintillating bars and has a time resolution of about 85\,ps. Due to the criticality of the veto systems and the timing detector, they have been through several test-beam campaigns, including measurements with large-scale prototypes.

The SHiP physics performance has been evaluated with 15 years of nominal operation, i.e. $6\times 10^{20}$ PoT. It has been verified that this is compatible with the zero-background strategy and the constraints from technical/radiation point-of-view in the current accelerator complex, as well as in the implementation of BDF~(see Section~\ref{sec:BDFintegration}).

\subsubsection{Present status, required R\&D}

The work packages for the BDF and the SHiP TDR studies, including the associated resource requirements, were discussed in the CDS reports~\cite{Ahdida:2703984,Ahdida:2704147}. The work packages are built on the understanding of the designs developed in the extensive joint studies performed during the six years of the Technical Proposal and CDS phases, which concentrated a large part of the effort on tuning the design of the components to maximise the signal acceptance and minimise the background.  All critical components of the facility have been studied, analysed and in some cases prototyped. The target system as one of the most challenging components has been through a first validation in a beam test in which the operating conditions of the real target were reproduced~\cite{PhysRevAccelBeams.22.113001,PhysRevAccelBeams.22.123001,Kershaw_2018}. All the SHiP sub-detectors have undergone at least a first level of prototyping and measurements with the prototypes in test beam~\cite{Ahdida:2654870, Ahdida:2704147}. In particular, the MRPC technology for the UBT, the liquid scintillator technology for the SBT, and the scintillating bar technology for the timing detector have had larger-scale prototypes in test beam. The beam tests have revealed the main technological challenges to be addressed during the TDR phase. With this information at hand, all major subsystems of the SHiP detector have been through conceptual design reviews, with the focus on outlining the work up to the TDR. The SND@LHC experiment~\cite{SNDLHC:2022ihg}, currently installed and operating in TI18 of the LHC, is a successful demonstration of the detector concept first developed for the OPERA experiment~\cite{OPERA_experiment}, and then improved within SHiP for an environment with a significantly higher rate of background. Collaboration with SND@LHC is established to pursue the development of the SND detector for BDF/SHiP, and most importantly, the studies towards an upgraded SND@LHC can make significant contributions to the LDM/neutrino programme at BDF/SHiP.

The principal technological challenges for the experiment lie in the further development of the muon shield, the decay volume and the spectrometer magnet, and involve mechanics and the full-size production. It is of high interest to develop the SC options for the muon shield with the potential to enhance the physics reach, and for the main spectrometer magnet with the aim to reduce the power consumption and the operational costs. The integration of the SBT and the HSDS spectrometer tracker is associated with important design challenges that must be addressed early in the TDR phase.

\section{Operation and proton sharing}
\label{sec:Operation}

The compatibility and possible proton sharing scenarios between the proposed future experiments in ECN3 and other NA experiments have been studied, also considering the parallel operation of the LHC, AWAKE, HiRadMat and MD sessions~\cite{Bartosik:2650722, Prebibaj:2848908}.

\subsection{Operation mode}

For the future proton sharing scenarios, operational periods with and without dedicated ion physics have been considered. Scenarios with dedicated SPS cycles for ECN3 users (\textit{dedicated ECN3 spills}) as well as scenarios with a concurrent beam delivery to the TCC2 and TCC8 targets (\textit{shared spills}) have been studied. Different flat top lengths have been analysed taking into account realistic supercycle compositions while respecting the SPS limits on power dissipation in the magnets. The intensities considered are based on operationally achieved values during the past operation of the SPS while an operational efficiency of 80\% (consistent with the expectations after the ongoing NA-CONS) has been assumed.

\begin{figure}[h!]
     \centering
         \includegraphics[width=0.7\linewidth]{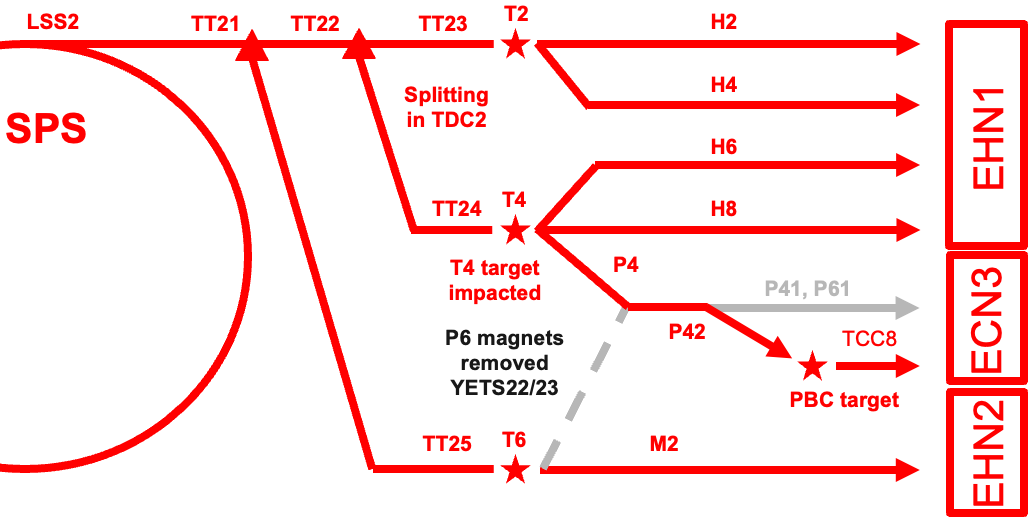}

         \includegraphics[width=0.7\textwidth]{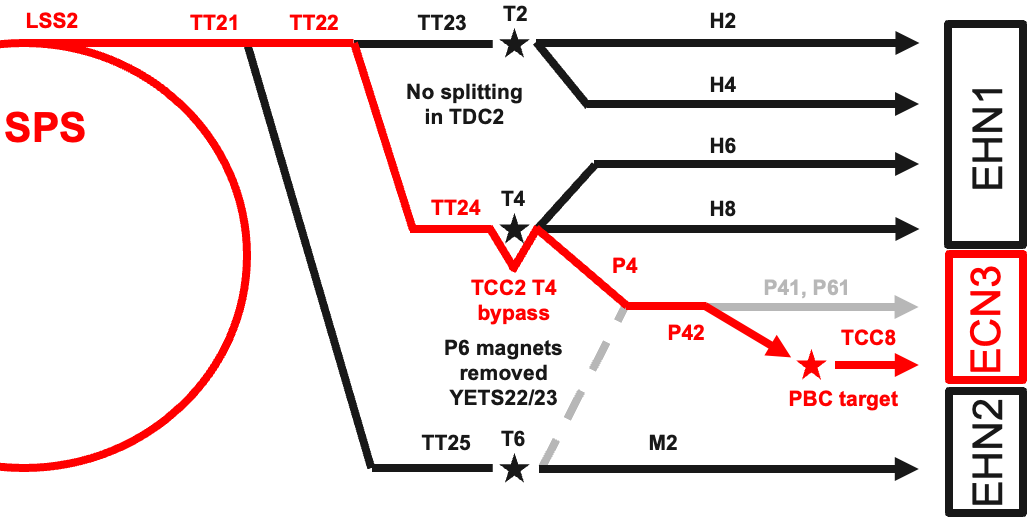}
        \caption{\small Schematic diagram of the two ECN3 beam delivery scenarios considered: T4 target in beam (top) and T4 target bypassed (bottom).}
        \label{fig:scenarios1}
\end{figure}

Presently, the TCC2 targets are served  simultaneously with \textit{shared spills} by splitting the extracted beam, transported via the TT20 transfer line, by means of the two splitter magnets in TDC2 (see Section~\ref{sec:current_status} and Figures~\ref{fig:NA-layout}~and~\ref{fig:scenarios1}--top). The corresponding transmission efficiencies (used to determine the amount of PoT on the TCC2 targets) are listed in the first column (\textit{shared spills}/TCC2) of Table~\ref{table:Transmissions}. 
In this mode of operation, the T10 target in TCC8, serving ECN3, receives the non-interacting fraction of the beam delivered to T4, which then is transferred to TCC8 via the P4/P42 lines. The remaining fraction of the beam interacting on T4 serves the H6 and H8 secondary lines. The transmission efficiencies to TCC8 are listed in the second column (\textit{shared spills}/TCC8--T4 in beam) of Table~\ref{table:Transmissions}.
\begin{table}[h!]
\tiny
    \centering 
    \resizebox{\columnwidth}{!}{
    \begin{tabular}{lccc}
    \toprule
    {} &   \textit{shared spills}/TCC2 & \textit{shared spills}/TCC8--T4 in beam & \textit{dedicated ECN3 spills}/TCC8--T4 bypassed\\
    \midrule
    Extraction &   0.98 &   0.98 & 0.98 \\
    TT20       &   0.99 &   0.99 & 0.99 \\
    Splitting  &   0.95 &   0.95 & 1.0 \\
    T4/TAX         &      - &   \textcolor{black}{0.78-0.94} & 0.98\\
    P42        &      - &   \textcolor{black}{0.97} &  0.99\\
    Total      &  0.922 &   0.697-0.840 &      0.941 \\
    \bottomrule
\end{tabular}
    } 
    \caption{\small Assumed transmission coefficients from SPS extraction to the TCC2 and TCC8 targets.}
    \label{table:Transmissions}
\end{table}

A new mode of operation with \textit{dedicated ECN3 spills} can be conceived where beam is transported through TT20 and TCC2 and delivered exclusively to TCC8. This scenario assumes that the primary beam can be cleanly transported without splitting in TT20 to the T4 target station bypassing the target with a trajectory bump~(see Figure~\ref{fig:scenarios1}--bottom). No other NA experiment will receive beam when a \textit{dedicated ECN3 spill} is delivered. The corresponding transmission efficiencies are listed in the third column of Table~\ref{table:Transmissions}.

Operation with \textit{dedicated ECN3 spills} is characterized by significantly lower beam losses at the splitters and at the T4 target as compared to the operation with \textit{shared spills} (see Table~\ref{table:Transmissions}), and therefore implies lower prompt and induced radiation as well as a reduction of the overall muon background in the NA.
Moreover, as dedicated cycles would be played outside of the shared cycles during which the other NA experiments are taking data, no adverse effect of ECN3 High-Intensity~(HI) operation on the backgrounds for the EHN1 and EHN2 experiments is expected. The only exception might be emulsion experiments, which currently are not planned. 
The RP studies carried out to-date and the various mitigation measures identified conclude that HI operation of ECN3 with dedicated ECN3 cycles is expected to be compliant with the CERN RP code~\cite{ECN3TFreport}.

In addition, operation with super-cycles delivering \textit{shared spills} for EHN1 and EHN2 and dedicated high-intensity cycles for ECN3 remains compatible with the present T4 target and TCC2 TAX design. Therefore, upgrading them is not required, provided that the appropriate machine protection measures are put in place. Recent studies have confirmed the assumed transmission efficiency through the T4 target station and TAX (see Table~\ref{table:Transmissions} - fourth row --- T4/TAX) for the~\textit{dedicated ECN3 spill} (third column) while indicating lower values for the~\textit{shared spills} with T4 in beam~(second column)~\cite{bib:Mazzola:2865692}. 

\textbf{For the above reasons the delivery of the required ECN3 intensity with~\textit{dedicated ECN3 spills} is preferred}.

\subsection{Proton sharing}

The SPS operation has been studied for ECN3 high-intensity~\cite{Prebibaj:2848908} by optimising SPS supercycles delivering both \textit{shared spills} for EHN1 and EHN2 and \textit{dedicated ECN3 spills} considering the operational scenarios presented in Section~\ref{sec:OPScenarii}.

\begin{figure}[ht!]
    \centering
    \includegraphics[width=0.49\linewidth]{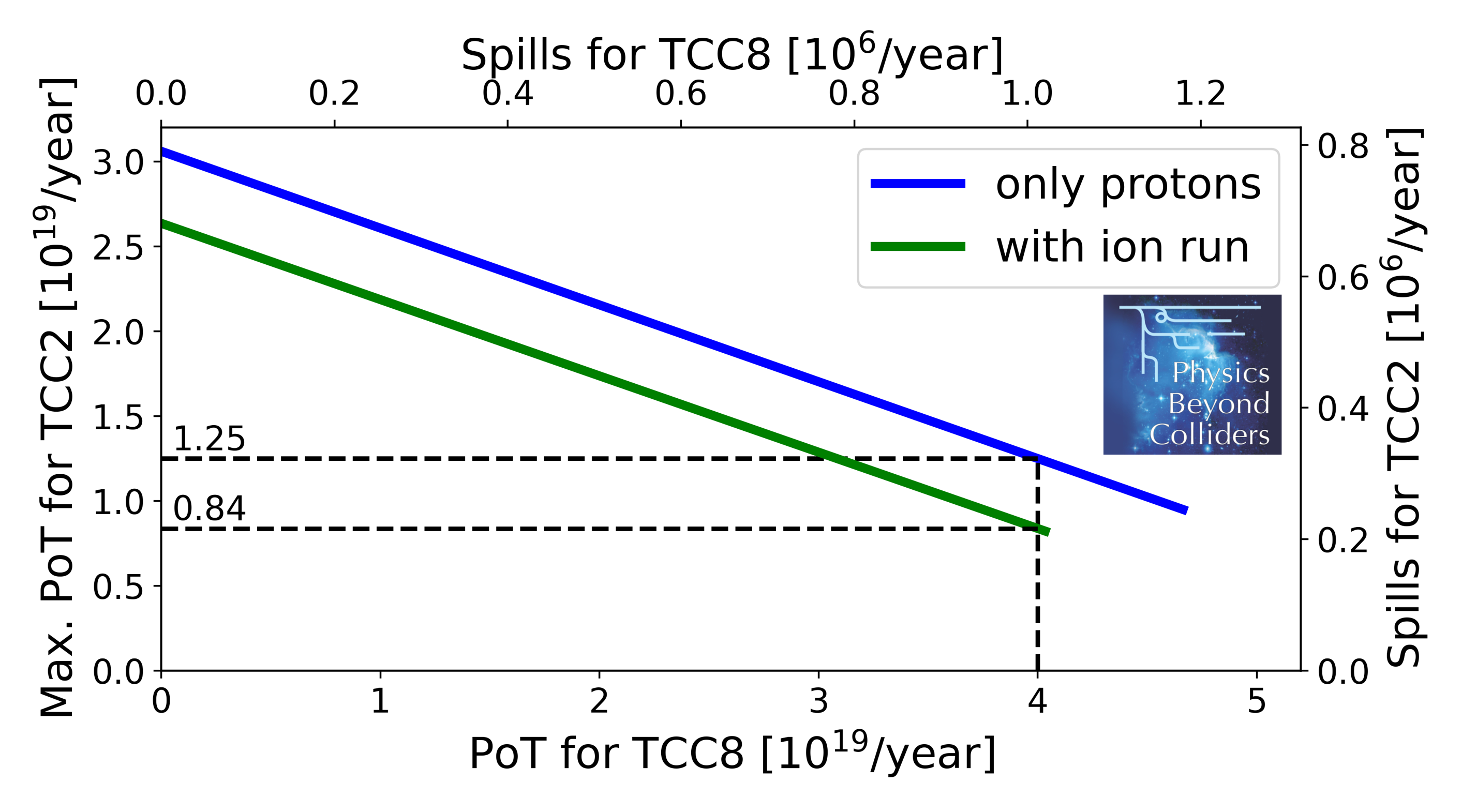}
    \includegraphics[
    trim=0 0 0 0, clip, 
    width=0.49\textwidth]{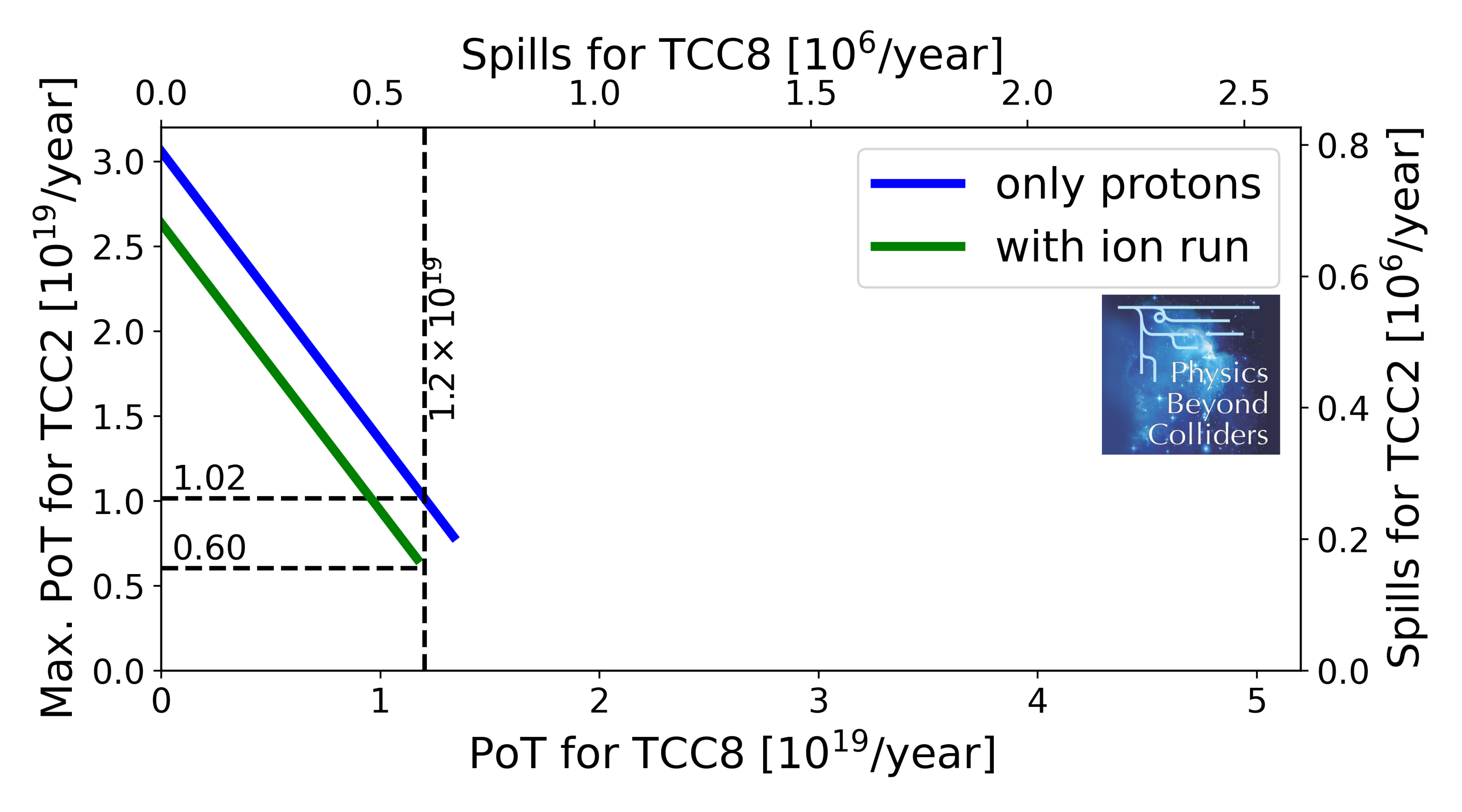}
    \caption{\small Future proton sharing scenarios with (green) and without (blue) ion operation for SPS operation with supercycles containing: \SI{4.8}{\s}-long \textit{shared spills} and \SI{1.2}{\s}-long \textit{dedicated ECN3 spills} (left); \SI{4.8}{\s}-long \textit{shared spills} and \SI{4.8}{\s}-long \textit{dedicated ECN3 spills} (right). The \textit{dedicated ECN3 spills} are delivered on the TCC8 target while the \textit{shared spills} impinge the TCC2 targets to feed the other NA users in EHN1 and EHN2. An intensity of $4.2\times 10^{13}$ ppp before SPS extraction has been assumed for the \textit{shared spills}, while intensities of $4.2\times 10^{13}$ ppp (left) and $2.1\times 10^{13}$ ppp (right) before extraction from the SPS have been assumed for the \textit{dedicated ECN3 spills} according to the requirements listed in Table~\ref{tab:ExpReq} for BDF/SHiP and HIKE/SHADOWS operational scenarios, respectively. The transmission efficiencies in the third column of Table~\ref{table:Transmissions} have been considered.}
    \label{fig:ECN3_D}
\end{figure}

Figure~\ref{fig:ECN3_D} shows that the experimental requirements for HIKE/SHADOWS (BDF/SHiP) can be met with a dedicated beam delivery  
while providing $\approx1\times10^{19}$~PoT/year ($\approx1.2\times10^{19}$~PoT/year) to the other NA experiments, provided no ion run takes place. Similarly, $\approx0.6\times10^{19}$~PoT/year ($\approx0.8\times10^{19}$~PoT/year) can be delivered in case an ion run (1 month) is included.
The integrated intensity to the other NA experiments is maximised by assuming the acceleration of $4.2\times10^{13}$ ppp on the \textit{shared spill} cycles with a \SI{4.8}{\s} FT. For some existing NA users this might be problematic due to rate limitations. A careful 
scheduling of rate-limited NA experiments exploiting longer cycles with a FT of \SI{9.6}{\s} would help to optimise beam delivery and to alleviate this problem. The study demonstrates that $\approx0.7\times10^{19}$~PoT/year ($\approx0.8\times10^{19}$~PoT/year) can be delivered to other NA users with \SI{9.6}{\s}-long \textit{shared spills} interleaved with \textit{dedicated ECN3 spills} for HIKE/SHADOWS (BDF/SHiP), provided no ion run takes place. In case an ion run is included in the operational year, $\approx0.4\times10^{19}$~PoT/year ($\approx0.6\times10^{19}$~PoT/year) can be delivered to other NA users. An additional optimization would consist in increasing the intensity of \textit{dedicated ECN3 spills} beyond $2.1\times 10^{13}$ ppp and correspondingly increasing the FT duration at constant extracted current for the HIKE/SHADOWS mode of operation.

Finally, it should be stressed that the PoT numbers would be reduced in case of more frequent LHC fillings, as compared to today's operation, during the HL-LHC era.

The energy consumption of the SPS main magnets and the NA magnets depends on the super-cycle composition. These elements are among the main contributors to the overall SPS and NA energy consumption during beam operation, representing more than 40~\% and almost 15~\% of the total SPS+NA consumption, respectively. Supplying beam to a HI facility in ECN3 will not change the power consumption significantly with respect to recent years. For 2022 the total energy consumption of the SPS main magnets was $\approx$~\SI{170}{\giga\watt\hour} and the estimated difference for all the ECN3 beam delivery scenarios considered (with~\SI{1.2}{\s} to~\SI{9.6}{\s} FT) is small and not larger than $\approx$~10~\%, see~\cite{Prebibaj:2848908}. 

\section{Required modifications and integration}

\begin{figure}[h!]
     \centering
     \includegraphics[width=\textwidth]
     {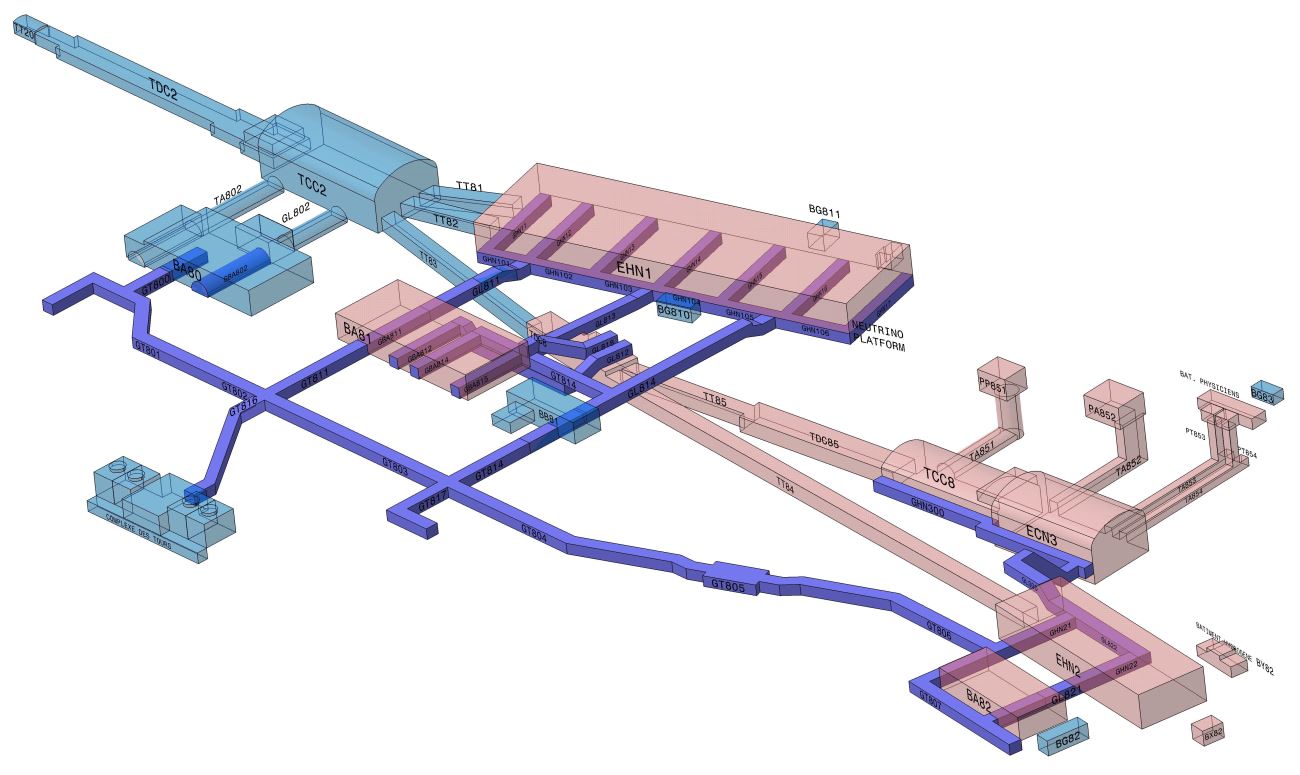}
     \caption{\small Overview of the facilities affected by the NA-CONS Project Phase~1~(light blue) and Phase~2~(light red). Technical galleries will be also subject of consolidation and are also indicated (dark blue/purple).}
     \label{fig:NA-CONS}
\end{figure}

ECN3 HI operation requires modifications of existing facilities. The extent of these modifications and the integration of the new experiments and the associated facilities are analyzed in this Section~(for more details see~\cite{ECN3TFreport}). The compatibility and synergy with the activities ongoing or planned within the NA-CONS Project are addressed. The NA-CONS project consists of two phases: 
\begin{itemize}
    \item Phase~1: 2022--2028 (up to end LS3), prioritising the primary beam areas TT20, TDC2, TCC2 and the initial section of the NA Transfer Tunnels.
    
    \item Phase~2: 2026--2034 (up to end LS4), completing the consolidation of the secondary beam areas.
\end{itemize}
The areas affected by NA-CONS are schematically shown in Figure~\ref{fig:NA-CONS}.

NA-CONS is expected to guarantee reliable operation in the North Area up to the end of 2040s provided regular maintenance is performed, such as the regular replacement of irradiated cables when needed. In that respect, the dedicated mode of operation and the upgrades described in this Section will reduce the radiological impact of HI ECN3 operation to a level comparable or lower than the present mode of operation for which NA-CONS has been conceived.

\label{sec:ExtractionTransfer}

\subsection{SPS extraction}

The consolidation of the electrostatic septa is already planned and funded as part of the Accelerator Consolidation (ACC-CONS) Project during LS3 and ready for Run~4, with a far longer-term R\&D objective to replace the septa with systems employing crystal technology. 

 At least a factor 4 reduction of the beam losses is needed to implement the proposed ECN3 HI upgrade without impacting the present day radiological situation in LSS2. R\&D on the LS3 timeline is focused on beam loss reduction techniques that significantly improve the efficiency of the present electrostatic slow extraction system~\cite{fraser:ipac17-mopik045,fraser:ipac19-wepmp031,Balhan:2668989}. In particular, the development of a low-density version of the septa tanks, of an anode with improved straightness and of thin crystals to `shadow' the septum blade is ongoing with PBC support.

The required extraction beam loss reduction factor can be achieved with the crystal shadowing technique developed at CERN~\cite{PhysRevAccelBeams.22.093502}. Up to a factor of 2 has already been demonstrated at the SPS with beam tests of prototype local and non-local shadowing systems installed in LSS2 and LSS4. The phase-space folding technique~\cite{PhysRevAccelBeams.22.123501} can be combined with the crystal shadowing technique to boost the loss reduction close to a factor 4, although it cannot be combined effectively in the shared mode of operation because the larger emittance of the folded beam will increase beam losses at the TT20 splitters~\cite{ArrutiaSota:2749240}.

\subsection{TT20, P4 and P42 Transfer Lines}
\label{sec:transfer}

The modifications required for the primary (TT20) and secondary (P4, P42) transfer lines for HI ECN3 operation with dedicated ECN3 cycles are independent of the experiment that could be installed in ECN3.

Recent studies of the current TT20 optics have revealed deviations of the measured  optics from the model~\cite{velotti-optics1,velotti-optics2}. Studies continue in 2023 to solve this issue. 
As discussed in Section~\ref{sec:Operation}, a new optics in TT20, rematched to provide a dedicated beam to ECN3 by transmitting it unsplit through the two TT20 splitters~\cite{tt20distribution} will have to be used and a vertical bump at the T4 target station will have to be implemented for the dedicated ECN3 cycles (see Figure~\ref{fig:scenarios1}~-~bottom). The largest T4 TAX collimator opening of~\SI{40}{\mm}$\times$\SI{20}{\mm} will be used to accommodate the large beam divergence at the T4 target. With this configuration the unsplit beam should be transported through TT20/TDC2/TCC2 to TCC8 without losses for the \textit{ECN3 dedicated spills}.

The front-end of the T4 target is composed of multiple \SI{2}{\mm} thick beryllium plates of different lengths (between 40  and \SI{500}{\mm}) arranged one on top of another with a vertical separation of \SI{40}{\mm}. This geometry provides the opportunity to bump the beam vertically between the target plates. 
With the installation of one additional vertical dipole magnet upstream of the T4 target, a closed solution for a trajectory bump can be found in combination with two other magnets already existing in the beamline for trajectory correction. A prototype system with a non-laminated magnet and spare power converter has been installed during the Year-End Technical Stop (YETS)~2022--2023~\cite{ecrBYPASS} and it has allowed initial tests with beam and the proof-of-principle of this mode of operation. In the operational configuration, the prototype magnet will need to be replaced by a magnet with a laminated yoke and a new power converter to allow Pulse-to-Pulse Mode~(PPM) operation. As a back-up solution for the magnetic bypass option, actuating the T4 target's head between cycles is being investigated. 

The MTN magnets in the wobbling system of T4 (that allow for a momentum selection of the secondary beam produced in the T4 target for H6/H8) cannot be operated in PPM. They can be kept powered at constant current and the beam transported into P42 to TCC8 on \textit{dedicated ECN3 spill} cycles, whilst still providing beam to H6/H8 on \textit{shared spill} cycles. The fraction of beam that does not interact with the T4 target during \textit{shared spills} will still enter P42, as it does today for NA62. During Run~4 it might not be possible to optimize the transport of this beam, as for the \textit{dedicated ECN3 spills}, because some of the power converters of the P42 line will not be operable in PPM. To ease the situation, the beam entering P42 during \textit{shared spills} can be reduced in intensity by reducing the primary beam intensity and increasing the T4 target length (up to \SI{500}{\mm}, according to the H6/H8 experimental programme). A new absorber located in P42 could then be used in case of unacceptable beam losses in P42 or experimental background in ECN3. A new laminated vertical dipole magnet, to be installed at the end of the P4 line, will direct the beam on the absorber~\cite{p42_dump} during \textit{shared spills}.

\subsubsection{Magnets and Power Converters}
\label{subsub:magnetsPC}
\begin{figure}[htb!]
     \centering
         \includegraphics[width=1.0\linewidth]{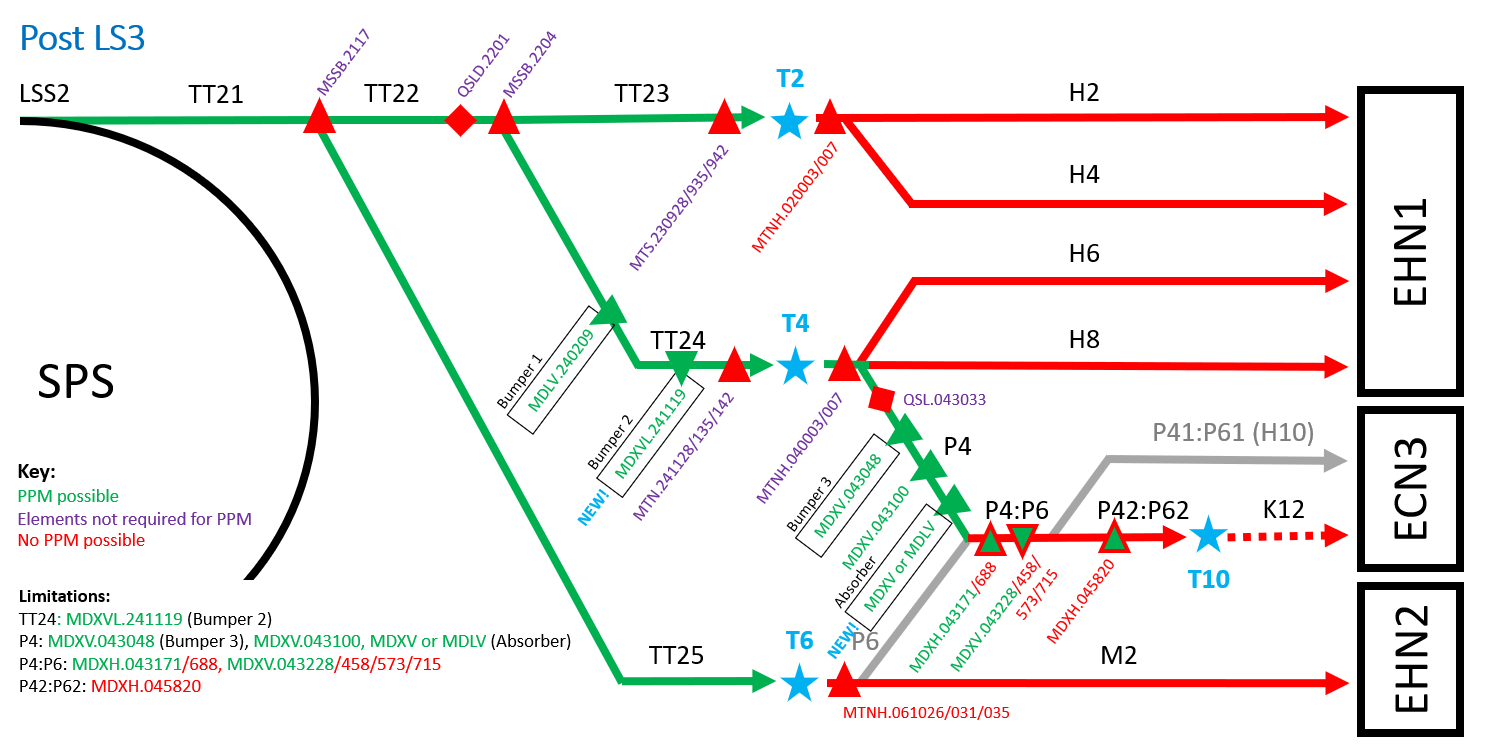}
         \includegraphics[width=1.0\textwidth]{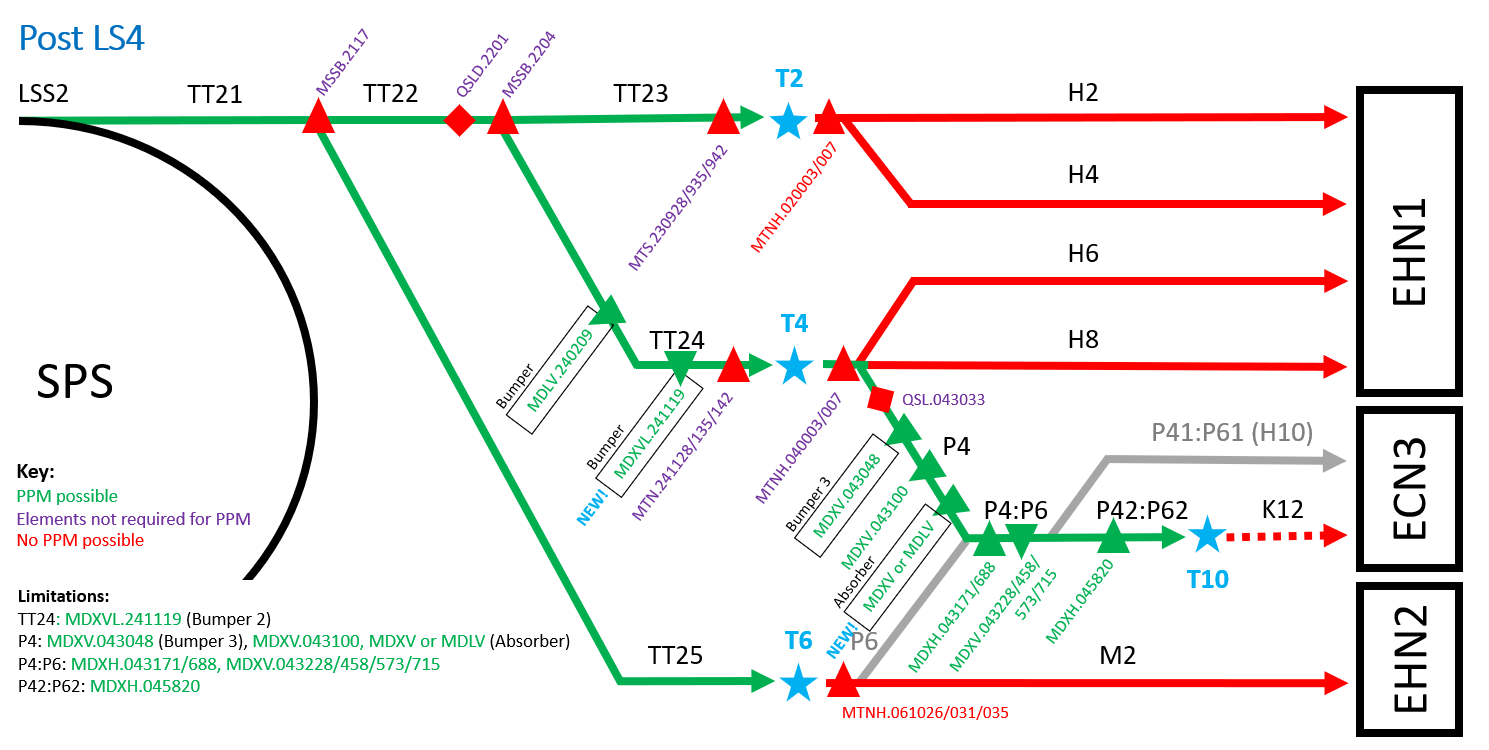}
        \caption{\small PPM compatibility of TT20, P4 and P42 lines after LS3 (top) and after LS4 (bottom). The two new magnets are indicated. Red label: magnet and/or power converter cannot be operated in PPM, green label: magnet and power converter can be operated in PPM, violet label: magnets and/or power converter not required to operate in PPM mode.}
        \label{fig:EPCLS3LS4}
\end{figure}

Two new laminated vertical bumper magnets (bumper MDXVL.24119 and the absorber magnet of MDXV or MDLV type in Figure~\ref{fig:EPCLS3LS4}) with the corresponding power converters, DC cables and Warm magnets Interlock Controller~(WIC) will be required and the non-laminated magnet MDXV.043048 and other eight MDX corrector magnets will have to be replaced by a laminated version.
During Run~4 not all power converters downstream of the T4 target will be capable of operating in PPM as shown in Figure~\ref{fig:EPCLS3LS4}~--~top, only those undergoing consolidation in LS3 will be. After LS4, when NA-CONS Phase~2 for power converters will be completed, all magnets and power converters in P42 will be PPM-compatible and \textit{dedicated ECN3 spills} and \textit{shared spills} could be optimised independently (Figure~\ref{fig:EPCLS3LS4}~--~bottom). 

The operation with~\SI{1.2}{\s}-long~\textit{dedicated ECN3 spills} might have an impact on the specifications of the electrical infrastructure while the correspondingly larger number of cycles might entail more frequent maintenance of the power converters. The impact on cost and maintenance budget of the above two items is being estimated~\cite{edmsNA-CONSpowering} but it is expected to be small.

\subsubsection{Beam instrumentation}
Efficient operation of the NA primary and secondary beam lines will require detailed optics models and accurate measurements of the beam characteristics to benchmark them as well as precise beam position and intensity measurements to minimize losses. Consolidation of the beam instrumentation is already part of the NA-CONS scope. High-intensity operation for ECN3 will require additional upgrades~\cite{ECN3TFreport,slawgBI,eabireport}:

\begin{itemize}
    \item 4 beam profile Secondary Emission Monitors (SEM) (Beam SEM Grids~---~BSGs) planned as part of NA-CONS--Phase 1 have been installed in P42 during YETS~2022--2023~\cite{ecrBSG} to conduct optics studies during the 2023 run. In order to operate, these monitors require vacuum pressures of at least $\approx10^{-4} - 10^{-5}$~mbar which are not presently reached in the P42 line (see Section~\ref{Ch3_RPaspects}). Due to the tight schedule for the installation, preventing to upgrade the P42 vacuum system to achieve the above vacuum levels, four small sectors around each of the BSG have been isolated under the required vacuum conditions and separated by thin (\SI{100}{\micro\m}) Aluminium windows from the rest of the vacuum line.  Additional BSGs will have to be installed in TT20 in view of the high-intensity operation;
    \item 13 new Beam Loss Monitors (BLM) have been installed~\cite{ecrBLM} as part of NA-CONS to instrument critical locations including the \textit{EHN1 ramp} and \textit{ECN3 bridge} and to permit optimisation of prompt beam losses. Additional BLMs will be installed in TDC2 and TCC2 upstream of the targets as part of NA-CONS~Phase~1;
    \item a passive optical fibre dosimeter covered by NA-CONS~Phase~1 will be installed at selected locations to measure integrated beam losses outside the coverage of the BLM system;
    \item following the experience gained during 2021--2022 operation the Target Beam Instrumentation (TBI) will be upgraded and it will include beam profile monitors based on BSGs~\cite{tbi};
    \item the installation of consolidated SEM for beam intensity measurements (BSI) is included in the present NA-CONS~Phase~1 scope;
    \item High-bandwidth spill monitoring to guide the optimization of the spill uniformity and reduce event pile-up is also included in NA-CONS.
\end{itemize}

\subsubsection{Vacuum System}
The consolidation of pumping units, main gate valves connecting the pumps to the vacuum system, vacuum gauges and the corresponding cabling and electrical sockets in the primary and secondary transfer lines is already planned as part of NA-CONS~Phase~1. The same applies to all vacuum chambers, bellows and windows exhibiting any signs of damage or deterioration (in particular in TCC2).

The present vacuum level in TT20 is not expected to limit performance. The scope of the NA-CONS~Phase~1 consolidation for the P42 beamline will have to be extended to achieve a vacuum level of at least $10^{-4}$~mbar all along the line without windows. Studies are underway to compute the effect of vacuum pressure on radiation levels and preliminary results indicate that the above target average vacuum pressure is sufficient.

\subsubsection{Beam intercepting devices}
The majority of the beam intercepting devices in the transfer lines are already being considered in Phase~1 of NA-CONS to increase reliability and address a series of operational issues encountered during operation in 2021--2022. These include~\cite{ECN3TFreport}:
\begin{itemize}
\item The TT20 Target External Dump~(TED), which is moved along the beam trajectory when required to prevent beam transport to the downstream part of TT20. The new TT20 TED design will be compatible with increased intensity per cycle ($> 4\times10^{13}$ ppp) and an appropriate duty cycle consistent with the operational scenarios described in Section~\ref{sec:Operation}. Cooling of the assembly will be optimised with sustainability in mind, while the core, the shielding and translation system will be designed considering  best practices and adaptation to the foreseen dumped intensities.
\item The TT20 Target Beam Stopper Extraction~(TBSE) stopper, providing a redundant safety element in case of access to the TDC2/TCC2 area. It will undergo a consolidation of its translation system, while keeping the same absorber.
\item The TT20 Target Collimator Splitter Copper~(TCSC), protecting each of the two TT20 splitter magnets, will intercept the beam during \textit{shared spills} only. Therefore, no significant losses and activation increase is expected as a result of the high intensity operation in ECN3. However, the TCSCs will remain among the most radioactive components in TT20 and following the operational experience in 2021--2022 design improvements including a low-activation tank with improved handling, new support tables (to allow more accurate alignment while allowing easier remote exchange of the assembly) and an improved water cooling system with quick connections to permit the possibility of installing marble shielding~\cite{bid_marco,tcsc_vincke} have been recommended to take place as part of NA-CONS~Phase~1. These upgrades will reduce the dose to personnel intervening in the area, but they will not reduce the splitting inefficiency at the origin of the beam losses estimated to be $\sim 3\%$ per splitter~\cite{ArrutiaSota:2749240,yann_sps}. Crystal set-ups installed upstream of the collimators and aligned in volume-reflection or channeling mode (\cite{ArrutiaSota:2749240,mvra} could offer a reduction of the splitting inefficiency by a factor between 2 and 5 and should be further studied and implemented as part of a general campaign of loss reduction for the operation of EHN1 and EHN2.
\item The T4 Target, and the corresponding Target Beam Instrumentation Upstream (TBIU) and Downstream (TBID) of it, will not require any modification for the mode of operation considered in Section~\ref{sec:Operation}, though a re-design of all the TCC2 target stations to guarantee an isostatic positioning of the TBIU and the TBID and of the target box itself have been requested to be implemented as part of NA-CONS~Phase~1. The beryllium plates will not require any specific upgrade~\cite{t4-tar}, provided that adequate beam interlocks are put in place preventing impact of high intensity beams, which would permanently damage them.
\item The TAX collimators are suffering from repeated reliability issues linked to their support tables reaching their end of life. The supporting table of 7 devices (including one spare) are included in NA-CONS~Phase~1 to address the reliability issues. The T6 TAX on P62 can be postponed to NA-CONS~Phase~2 because the line is presently not in use. The T10 TAX could also be removed from NA-CONS pending a decision on the physics programme to be conducted in ECN3. The dedicated mode of operation for ECN3 does not require {\it a priori} a modification of the beam absorbing elements, provided adequate beam interlocks preventing more than a single high intensity extraction to intercept the absorbing material are put in place. Two or more extractions of the dedicated beam at $4\times10^{13}$ ppp would risk melting the copper in the second block of TAX if impacting directly~\cite{t4-xtax,t10-xtax}. In fact, even shared beam at $2\times10^{13}$~ppp could damage the blocks in that case.
\end{itemize}

 As mentioned earlier, a new absorber will be installed in the P42 beamline to intercept the proton beam not interacting with the T4 target during \textit{shared spills}. The latter would be an internal dump under vacuum, with an aperture large enough to allow the beams during \textit{dedicated ECN3 spills} to pass through. An available spare of an earlier version of the SPS internal beam dump (TIDVG4)~\cite{p42_tidvg} could be used for that purpose.

\subsubsection{Survey and Alignment}
The alignment and smoothing of the NA primary and secondary lines is foreseen as part of the NA-CONS project. The connection of TT20 through the T4-TAX system to P4/P42 in TCC2 and the P42 beamline are of interest for ECN3 operation.

The work in TCC2 can only take place during LS3 because of the high radiation levels. A permanent survey network will be installed in TCC2 as part of NA-CONS~Phase~1 to ease the measurement in the area and reduce radiation to the personnel. The P42 transfer line has been surveyed and smoothed already in YETS~2022--2023. This activity is limited by the activation of certain collimators in the TT83 tunnel (see Figure~\ref{fig:NA-CONS}) and additional verifications will have to take place in LS3.

NA-CONS includes the update of survey instrumentation and measurement methods and in particular the target station consolidation that will ease the measurement of the equipment position.

\subsubsection{Radiation protection}

In addition to the studies carried-out to assess the origin of the observed and expected prompt radiation levels and to identify appropriate mitigation measures (see Section~\ref{Ch3_RPaspects}),
further studies were performed to investigate accidental beam loss scenarios along the shallow transfer tunnels (TT83 and TT85---see Figure~\ref{fig:NA-CONS}) housing the P4/P42 beamline~\cite{Ahdida1, Nowak}. The loss of an entire NA62 spill at the current nominal intensity would create a maximum dose of $\approx$~\SI{300}{\micro\sievert}/spill at the \textit{EHN1 ramp}, which is acceptable (below the limit of \SI{1}{\milli\sievert}) if there are no visitors in the area and provided the beam is interlocked after 1 spill. Presently, an RP monitoring system is installed with an interlock capability. When scaling to the higher intensities given in Table~\ref{tab:ExpReq} the limit would be exceeded and the following two mitigation measures should be implemented: 
\begin{itemize}
    \item halt the extraction and dump the beam in the SPS using an interlock input to the Beam Interlock System (BIS) from the BLM system and selected power converters;
    \item increase the effectiveness of the shielding at the ramp~\cite{Nowak}. Replacing the concrete shielding by iron yields a factor 50 reduction in the prompt dose, which would be sufficient to stay well below the \SI{1}{\milli\sievert} limit in case of accidental beam loss. This measure and other possible actions are presently being studied.
\end{itemize}

The situation at the \textit{ECN3 bridge} is similar with $\approx$~\SI{50}{\micro\sievert}/spill reached with the uncontrolled beam loss of the nominal NA62 intensity~\cite{Nowak}. A reduction of more than an order of magnitude in the prompt dose rates can be achieved with moderate improvements of the shielding at the bridge~\cite{Nowak} and civil engineering studies for such a solution have been launched.

\subsubsection{Machine protection system}

The machine protection architecture foreseen as part of the NA-CONS project is compatible with a dedicated ECN3 beam delivery scenario~\cite{mpp1,mpp2,mpp3,mpp4,mpp5}. The BIS is modular and distributed across the NA primary and secondary beamlines. It can be easily adapted to the needs of future beam transfer and target systems. A detailed study on the required machine protection inputs is needed for the HI facility in ECN3 in 2023. The technical specifications are presently being written and new interlocking requirements are now being worked out. The protection of the primary beamlines would exploit signals provided by several pieces of equipment. These include power converters’ current monitoring, WIC, BLM systems, vacuum valves, beam intercepting devices, transfer line elements and the access system. The BIS will have to decode which cycle-type is being played and it will allow SPS slow beam extraction only if safe conditions are met. The system has a reaction time ($\approx$~\SI{100}{\micro\s}) well below the spill length to avoid accidental damage to equipment. The deployment of the new BIS is foreseen as baseline in NA-CONS~Phase~1 and during LS3, however, there will be a transition period where modern interlocks will coexist with old and software interlocks because the consolidation of power converters in the auxiliary surface buildings BA81 and BA82 currently is not planned to happen until LS4 (see Figure~\ref{fig:NA-CONS}).

\subsubsection{Timing and controls}

In comparison to the non-PPM NA operation today, the introduction of a dedicated NA user in ECN3 will bring with it the concept of ECN3 user (USER) and ECN3 destination (DEST), not only for the relevant magnets and power converters, but also for the machine protection system and other systems that need to understand the cycle-type (\textit{dedicated ECN3 spills} or \textit{shared spills}) being played, including the NA users and experiments themselves. The distribution of timing signals to the NA is part of NA-CONS but the individual NA user requirements will need to be followed-up carefully to ensure that post-LS3 operation is compatible with a dedicated cycle and NA user in ECN3.

\subsubsection{Other TDC2/TCC2 infrastructure}

Although a dedicated beam delivery mode to ECN3 relaxes the need for significant upgrades in TDC2 and TCC2 during LS3, some targeted but significant consolidation is still required in the zone. In addition to the items already discussed in this Section, the following activities should be completed during LS3 to improve the future reliability of the NA in Run~4:
\begin{itemize}
    \item replacement and rerouting of DC and signal cables;
    \item replacement and rerouting of water cooling hoses and connections;
    \item deployment of higher performing fire detection and protection system with corresponding compartmentalisation and smoke extraction.
\end{itemize}

\subsection{TCC8/ECN3}
\label{sec:Integration}

The instantaneous and integrated beam intensities requested by BDF/SHiP and HIKE/SHADOWS both require the installation of a new target complex, associated cooling and ventilation systems, and shielding in TCC8.

Based on the experience of fixed-target operation at CERN and considering the best practices in the international community, as well as the need to comply with today's radiation protection and radiation safety regulations, the target systems of a new facility will have more stringent design requirements than currently operating facilities~(see also~\cite{bdf_cb_wg_marco_claudia}). Studies executed during 2021--2022~\cite{Aberle:2802785} proved that a high-power target station could achieve compliance with these criteria, provided that an appropriate shielding configuration as well as specific design requirements are implemented. Recovery of at least 100--\SI{120}{\cubic\meter} of passive cast iron blocks from facilities such as the CERN Neutrinos to Gran Sasso~(CNGS) hadron absorber and/or from the old PS neutrino facility in TT7 is being investigated in order to optimize costs and enhance sustainability~\cite{bib:EDMSGrenard}.

Three options with helium, nitrogen, or alternatively vacuum have been considered for the target vessel that should ensure an inert atmosphere to prevent corrosion and reduce residual gas activation within the target shielding. These need detailed investigations, together with the design of the proximity shielding and services. 

Optimisation concerning design, integration, handling and manipulation is also being sought in order to allow reasonable maintenance of highly radioactive devices according to the ALARA~(As Low As Reasonably Achievable) principle and in particular the possible replacement of the magnets installed between the target and the TAX (for the HIKE/SHADOWS configuration) and/or the target (for both configurations) during the lifetime of the HI facility, expected to span over at least 15 years of nominal operation.
While the concepts around the handling of the target and the target complex are well developed, the different components involved and the remote handling techniques also require detailed design and prototyping. Nevertheless, no showstopper has been identified so far.

The facility design with its shielding and infrastructure has been optimized to be compliant with CERN’s RP code~\cite{SafetyCodeF} regarding dose to the personnel and members of the public. The optimization considers the operational scenarios described in Section~\ref{sec:OPScenarii}. 
It takes into account the prompt and residual radiation, air activation and the environmental impact. Also, soil activation and transfer of activation products to groundwater has been considered in the shielding design. However, due to lack of information about the local groundwater transport, very conservative constraints on the activity concentration of longer-lived leachable radionuclides in the soil ($^3$H, $^{22}$Na) have been applied. A hydro-geological study is underway, and it will provide the information needed to relax the above constraints and to further reduce the required shielding.

A preliminary civil engineering study~\cite{CE} has been carried out on the required modifications to the existing infrastructure.
The installation of the new target complex with the associated shielding requires civil engineering works in TCC8. The existing floor will be lowered locally and a dynamically confined area with nuclear-grade ventilation will be created with fire resistant walls, separating the target area from the ECN3 hall and the rest of TCC8. The extent of the civil engineering work might be reduced once the results of the hydro-geological study above-mentioned are available.

A new service surface building will be constructed with an area of approximately~\SI{500}{\square\meter} to house all the dedicated services needed for the target complex sub-systems independently of the experiment, including an area for target system preparation as well as for the handling, repair and waste packaging of spent targets and the various beam intercepting devices. Additional~\SI{200}{\square\meter} will be available for the installation of power converters. The local electrical installation would require the construction of a concrete platform to support the transformers measuring about 12$\times$\SI{8}{\square\meter} for HIKE/SHADOWS or 12$\times$\SI{4}{\square\meter} for BDF/SHiP.

Access separation of TCC8 with respect to the rest of the NA will be needed to allow for work on the target and experiment installation during Run~4 whilst beam operation continues in the rest of the NA. Potentially, new fire doors will have to be installed with an impact on the compartmentalisation and on the fire detection scheme. A Fire-Induced Radiological Integrated Assessment~(FIRIA) analysis of the new target complex and compartmentalisation study must be conducted. New buildings and shafts will have to be equipped with fire detection as well. The recently renovated EHN2-BA82 control unit can be scaled to protect a larger perimeter. The access control system will have to be implemented according to the new premises and related restrictions (target building, target area, shafts, new service building for power converters and cooling station). The safety aspects in TCC8 and ECN3 will need a detailed and experiment-specific study, to be carried out in the TDR phase.

The EHN2 and ECN3 magnets are powered from the BA82 surface building. The consolidation of BA82 is foreseen only in phase 2 of NA-CONS during LS4 and its anticipation to LS3 is not possible, as emerged during the NA-CONS Cost, Schedule and Scope Review~(CSSR)~\cite{bib:NA-CONS_CSSR}.  Instead, the installation of the converters for ECN3 could be foreseen in the new service building planned for ancillary equipment for the target systems in TCC8. The installation work could be performed after LS3 without impacting the operation of M2 and the consolidation of EHN2 and BA82 could take place during NA-CONS~phase~2 as planned.

It is expected that the cooling and ventilation capacity available after the consolidation of the cooling towers as part of NA-CONS~phase~1 will be sufficient for the ECN3 upgrade. The new service building hosting the power converters for the experimental magnets will require dedicated ancillary cooling and ventilation equipment, including pumps, control racks and heat exchangers for the demineralized water. In addition, a corresponding local electrical infrastructure will have to be deployed.  

An important logistical support will be required all along the process of equipment decommissioning and, if needed, decontamination in the area with a particular care for materials like target, absorbers and highly activated equipment. Waste packaging and disposal will have to be organised accordingly. In the same way, transport and handling support will be needed for the installation of the target complex and the experimental equipment. An upgrade of the crane in TCC8 will be required to improve its movement system and remote handling capability. 

The impact on other services such as cryogenics, gas distribution and Information Technology~(IT) infrastructure will need iterating with the specific experiments.

\subsubsection{HIKE/SHADOWS}\label{sec:IntegrationHIKESHADOWS}

\begin{figure}[t]
    \centering
    \includegraphics[width=1.0\linewidth]{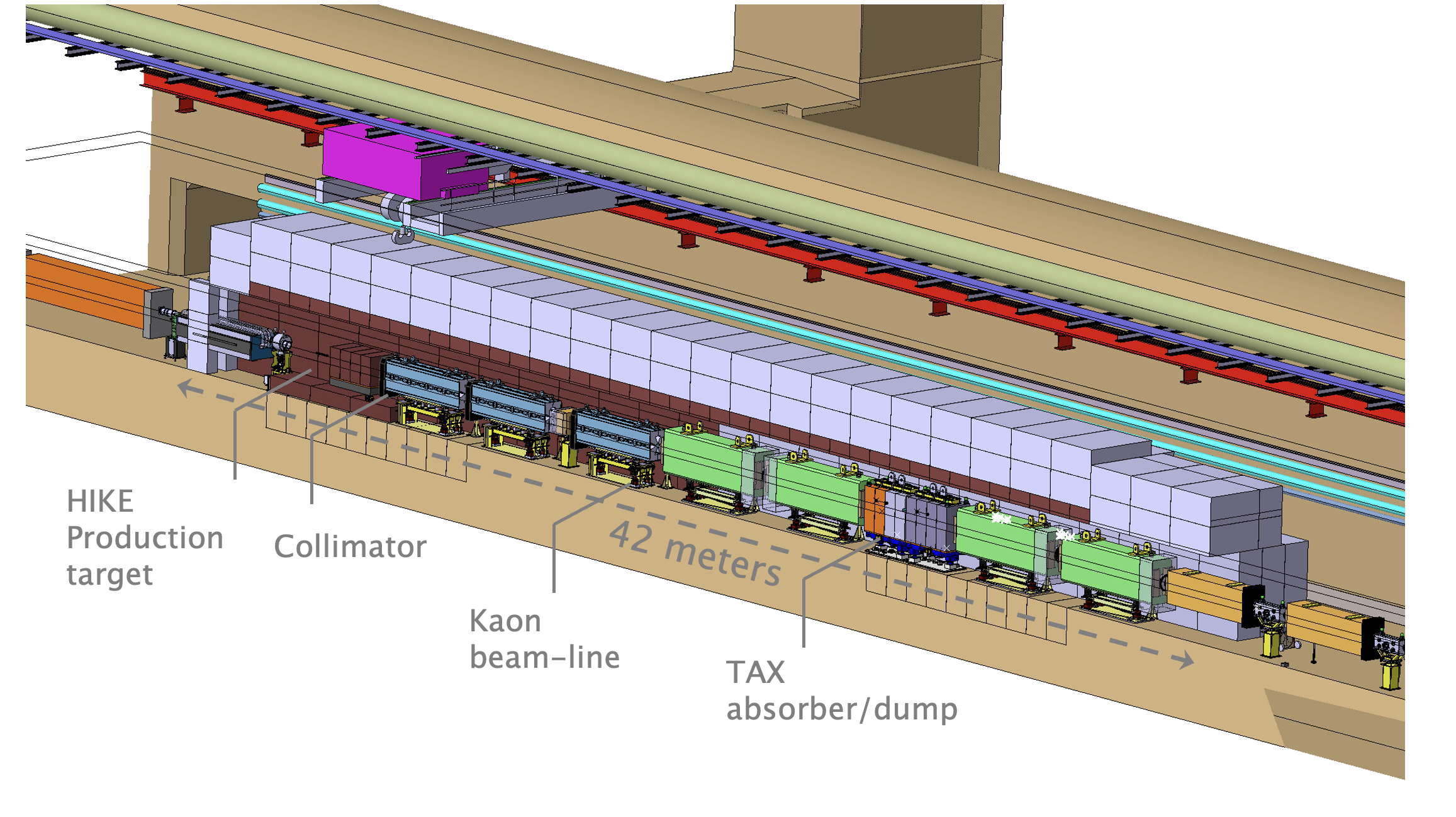}
    \caption{\small Overview of the HIKE Phase~1 / SHADOWS Target and TAX systems in TCC8/ECN3~\cite{EDMS_Integration_Shadows_Hike_Phase1}. Integration of Phase~2 equipment is not yet available, as beamline studies are still ongoing.}
    \label{fig:HIKE}
\end{figure}

For the HIKE proposal, a~\SI{100}{\kW}-class target complex based on radiation-cooled graphite or He-gas cooled beryllium, similar to the CNGS configuration, is proposed, at the place of the current T10/TAX target system, which will be completely dismantled. The physics requirements, resulting from the kaon beam, demand a significant shielding improvement with respect to the current NA62 target system~(see Figure~\ref{fig:HIKE}). Despite the lower requested number of PoT with respect to BDF/SHiP, the nature of the kaon production, its selection and secondaries in-flight decay, coupled with the need of dumping the remaining proton beam and hadrons, results in a target system that is stretched over a length of $\approx$~\SI{27}{\m} with several equipment in the secondary beamline that requires access (hence without hermetic shielding); this requires a significant amount of shielding to contain the radiation (more than~\SI{300}{\cubic\metre} of cast iron and~\SI{600}{\cubic\metre} of concrete).

An ad-hoc TAX system with a Cu-Fe sandwich configuration, upgraded cooling and maintenance/handling capabilities will replace the existing one. Full remote handling of the various components is also a pre-requisite to be compatible with ALARA requirements.

The integration of the layout considered for HIKE Phase~1 and SHADOWS has been conceptually validated~\cite{EDMS_Integration_Shadows_Hike_Phase1}. It must be noted that the design of the SHADOWS background lateral veto wall between the decay volume and the beamline is still ongoing and this will have to be integrated in the space presently reserved for the experiment shown in Figure~\ref{fig:shadows_in_area}. Access capabilities to the equipment between target and TAX will require further optimization since, due to the increased shielding and the expected residual dose rate, it is expected that maintenance operation would be relatively complex. Space availability in TCC8 would still have to be thoroughly evaluated as the integration of SHADOWS and the K12 beamline, particularly between the target and TAX for the latter, progresses.

A set of radiation protection studies were conducted based on extensive FLUKA Monte Carlo simulations~\cite{fluka:website, Ahdida:2806210, BATTISTONI201510} to optimize the facility and its shielding design~\cite{RP-TN-HIKE-SHADOWS}. 
The studies were performed for HIKE Phase~1 and HIKE/SHADOWS BD mode, however not yet for HIKE Phase~2 in view of the ongoing beamline design. 
The optimized shielding for HIKE Phase~1 and the HIKE/ SHADOWS BD mode allows reducing the soil activation to comply with the given design limits and the residual radiation in the target and experimental areas guaranteeing access for interventions. It further aims at containing the air activation and reducing the environmental impact from its releases to respect CERN's dose objective of~\SI{10}{\micro\sievert}/year for members of the public. The shielding decreases the prompt radiation above-ground. While for the area above TCC8 and ECN3 the shielding is sufficient to comply with a Non-Designated Area~(NDA), the area downstream of ECN3 must be reinforced with additional~\SI{4}{\m} of soil over an extended area and additional iron shielding in TCC8/ECN3 is required. This allows not only to meet the ambient dose equivalent limit of an NDA within the CERN fence, but also the limit at the CERN fence as well as the above-mentioned dose objective for members of the public.

HI operation in kaon mode will induce significant radiation dose to the coils of the magnets installed between the target and the TAX. Dedicated shielding will be required to avoid the necessity of frequent magnet replacements~\cite{Esposito_PBC_ACC_2203203}. An optimisation of the TCX, the fixed collimator mask downstream of the target encasement, is being considered. Due to the high prompt radiation levels (specifically high energy hadrons and neutrons, order of magnitudes higher than tolerated by commercial electronics according to CERN's Radiation Hardness Assurance criteria), radiation-tolerant electronics will have to be used in proximity of the SHADOWS detector and dedicated alcoves with iron shielding will have to be built for the electronics in TCC8~\cite{Esposito_PBC_ACC_2203203}. It is expected that for HIKE Phase~1 the front-end of the K12 beamline will have to be rebuilt. In particular, the magnets between target and TAX will have to be replaced with new magnets adapted for full remote handling that are busbar-powered to avoid manual cable connection.

The floor in the TCC8 cavern needs to be reinforced with iron shielding in the critical areas of HIKE/SHADOWS to prevent the soil activation going above the given design limits. Under the target the required excavation for the installation of the iron blocks is~\SI{6.5}{\m} long,~\SI{3}{\m} wide and~\SI{0.8}{\m} deep, while under the TAX the floor will be excavated on a~\SI{6}{\m} long area in three steps with a total depth of~\SI{1.35}{\m} and the width varying between 2 and~\SI{6}{\m}. Additionally, in the area between the target and the TAX the floor will be also lowered by \SI{0.5}{\m} over a~\SI{8.4}{\m} length and by~\SI{0.7}{\m} over a~\SI{8}{\m} length. Due to the size of the required modification, the slab will be excavated to the full depth and a new reinforced foundation slab will be built to maintain the structural stability of the tunnel. In addition, for the installation of SHADOWS a~\SI{3.5}{\m} long,~\SI{5.5}{\m} wide and~\SI{0.5}{\m} deep trench will be excavated under the spectrometer magnet.

SHADOWS requires a new power converter for the spectrometer, three for the MIBs and one for the NaNu magnet. In case the existing MNP33 magnet will be replaced by a new NC or SC spectrometer, one new converter will be required for HIKE instead of the two currently existing ones. The power converters will be installed in the target service building.

The present conceptual HIKE Phase~2 layout~\cite{HIKE_Phase2_Concept_Beam} implies a major rework of the K12 beamline in the transition from HIKE Phase 1 to Phase 2 during an LS. It includes the removal of the highly activated TCX and magnets between target and TAX, the removal of the K12 beamline itself, and the installation of an in-vacuum high-power beam dump few meters downstream the production target to reduce the muon background stemming from decays of secondary hadrons produced in the target. Optionally, the possibility to put the TAX absorber entirely under vacuum is considered to reduce background from kaon regeneration at the vacuum windows surrounding the TAX and the air in the TAX holes. This may imply a significant change in the shielding configuration and beamline system integration. 

\begin{figure}[h]
    \centering
    \includegraphics[width=1.0\linewidth]{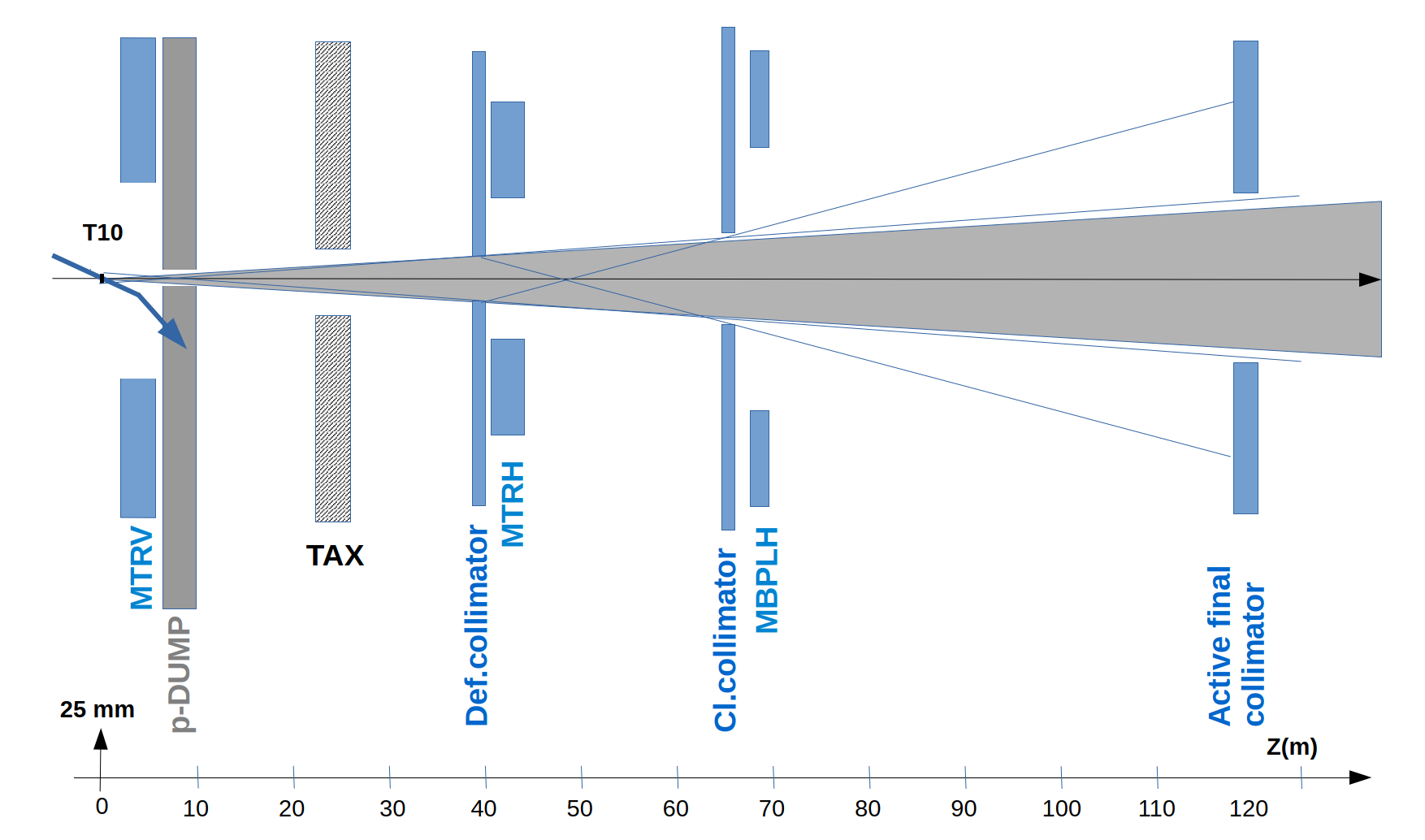}
    \caption{\small Conceptual overview on the HIKE Phase 2 neutral beamline.}
    \label{fig:HikePh2beamconcept}
\end{figure}

The new neutral beamline will consist mainly of three collimation and sweeping magnet stages as depicted in Figure~\ref{fig:HikePh2beamconcept}. The defining collimator is located at $1/3$ of the distance to the final collimator and defines the beam angular acceptance of $\pm$\SI{0.4}{\milli\radian}, matching the size of the central bore in the proposed HIKE calorimeter. A cleaning collimator stops debris from scatterings in the jaws of the defining collimator, and a final collimator stops scattering products from the cleaning collimator. Charged background from inelastic interactions at the collimators is reduced further by introducing strong sweeping magnets with apertures larger than the beam acceptance. The active final collimator is part of the experiment and defines the start of the fiducial volume of HIKE. The beamline between TAX and experiment is required to be under vacuum.

The HIKE Phase~2 layout has not been validated yet, either from a radiation protection or from a system integration point of view as the beamline design is still ongoing. Moreover, it is important to stress that the Phase~2 services will have to be available already during the construction period of Phase~1, as the dose rate at the end of Phase~1 is expected to be very significant. Ongoing studies seem to indicate that there is no need to keep the TAX absorber under vacuum while optimization of the spot size at the target is being considered to reduce the power density requirements on the proton dump~\cite{HIKE_Phase2_Concept_Beam}. 

In addition, minor modifications to the P42 beamline will be needed, i.e., re-alignment of the last three dipole magnets, if the experiment decides to run at a production angle larger than~\SI{2.4}{\milli\radian}. The target concept would allow to increase the angle up to 8~mrad.

\subsubsection{BDF/SHiP}
\label{sec:BDFintegration}

The design of BDF and the technology studies, including prototyping, have been documented in detail in the CDS report and other documents~\cite{Ahdida_2019, Ahdida:2703984} (complete list of dedicated reports in~\cite{bib:BDFSHIPLOI_2022}). The implementation in the existing TCC8 and ECN3 reuses the designs developed for the original proposal. Only the most relevant aspects for the implementation in TCC8/ECN3 are reported below.

The present T10 production target in TCC8 would be removed along with the entire K12 beamline and the corresponding magnet power converters as well as all the shielding assemblies. The latter will be reused for the target systems. At the upstream end of TCC8, the magnets of the BDF dilution system would be installed along with a vacuum chamber spanning the length of TCC8 towards the BDF/SHiP proton target, with the $\approx$~\SI{130}{\m} drift distance exploited to increase the beam size and develop the dilution pattern on the target's front face.

\begin{figure}[t]
    \centering
    \includegraphics[width=1.0\linewidth]{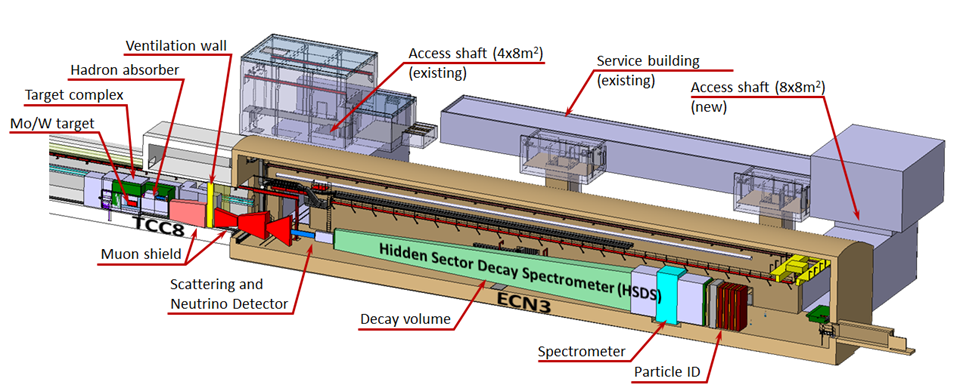}
    \caption{\small Overview of the BDF/SHiP experimental setup in the SPS TCC8/ECN3 beam facility.}
    \label{fig:SHiP}
\end{figure}

\begin{figure}[t]
    \centering
    \includegraphics[width=0.9\linewidth]{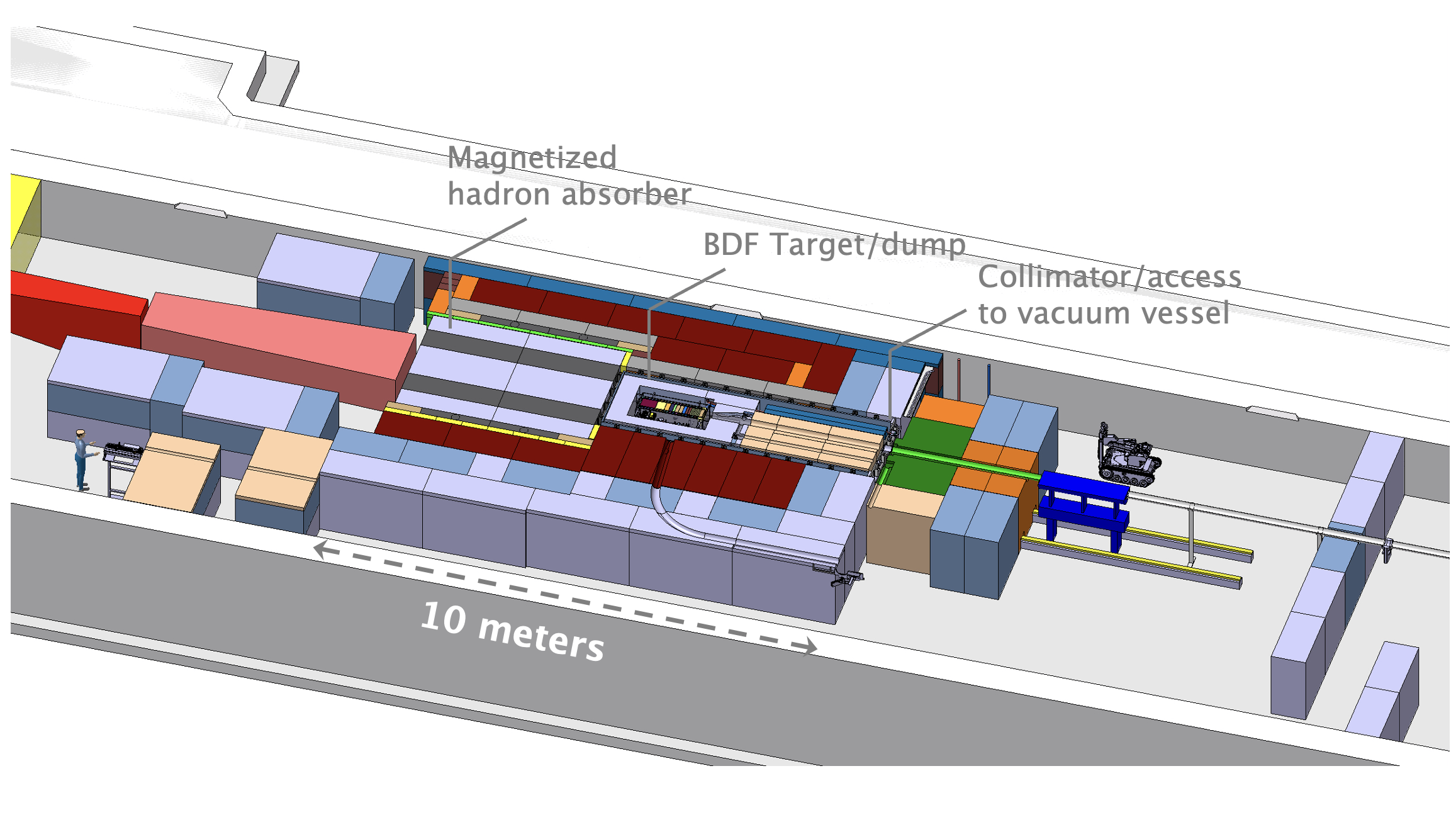}
    \caption{\small Preliminary design of the BDF-SHiP target area implemented in ECN3.}
    \label{fig:BDF_target_area}
\end{figure}

The layout of BDF/SHiP at the end of TCC8 and throughout ECN3 is shown in Figure~\ref{fig:SHiP}~\cite{bib:BDF_SHIP_ECN3_integration}. The setup consists of the high-density~\SI{300}{\kW}-class proton target, effectively acting as a beam dump and absorber, followed by a magnetised hadron absorber and a magnetic muon shield immediately downstream. The shield deflects the muons produced in the beam dump in order to reduce the flux in the detector acceptance to an acceptable level. The hadron absorber is an integral part of the overall shielding complex that is completely surrounding and sealing the target system. Together they form a compact and free-standing target complex, shown in Figure~\ref{fig:BDF_target_area}.

The target complex design draws from the experience gained during the CDS phase~\cite{Kershaw_2018,Ahdida:2703984}.
Significant simplification and reduction in shielding has been made possible thanks to the use of an already operational underground area and thanks to the depth of TCC8 with respect to the surface. The handling of the target systems may be carried out by the existing crane in TCC8 (after the upgrade of its movement system and remote handling capability), taking inspiration from the recently developed design of the new SPS beam dump~\cite{Pianese:IPAC2018-WEPMG004} and developments during 2023. This has led to a revision of the shielding and the system handling in ECN3 to cope with the space and access constraints, while fully respecting the constraints from radiation protection, equipment maintenance and operation.

In order to maximise the production of heavy flavoured hadrons and photons, and at the same time provide the cleanest possible background environment by suppressing decays of pions and kaons decaying to muons and neutrinos, the target should be long and made from a combination of materials with the highest possible atomic mass and atomic number, and be optimised for maximum density with a minimum of space taken by internal cooling. The corresponding target system developed during the CDS phase~\cite{PhysRevAccelBeams.22.113001,PhysRevAccelBeams.22.123001} requires no modifications with respect to the implementation in ECN3. The baseline design is still composed of blocks of titanium-zirconium-doped molybdenum alloy~(TZM), cladded by a tantalum-alloy, in the core of the proton shower, followed by blocks of tantalum-cladded pure tungsten. The blocks are interleaved with a minimum number of~\SI{5}{\mm} gaps for cooling, resulting in a total length of twelve interaction lengths. In order to cope with the~\SI{350}{\kilo\watt} average beam power, a bunker configuration with cooled stainless steel shielding, passive cast iron blocks (\SI{180}{\cubic\meter}), as well as concrete and marble shielding is foreseen (for a total volume of $\approx$\SI{360}{\cubic\meter}). A pit (\SI{4}{\m} long, \SI{4}{\m} wide and \SI{1}{\m} deep) will be excavated under the target station to embed part of the shielding and some of the services.

\begin{figure}[!t]
\centering 
\includegraphics[width=.99\textwidth]{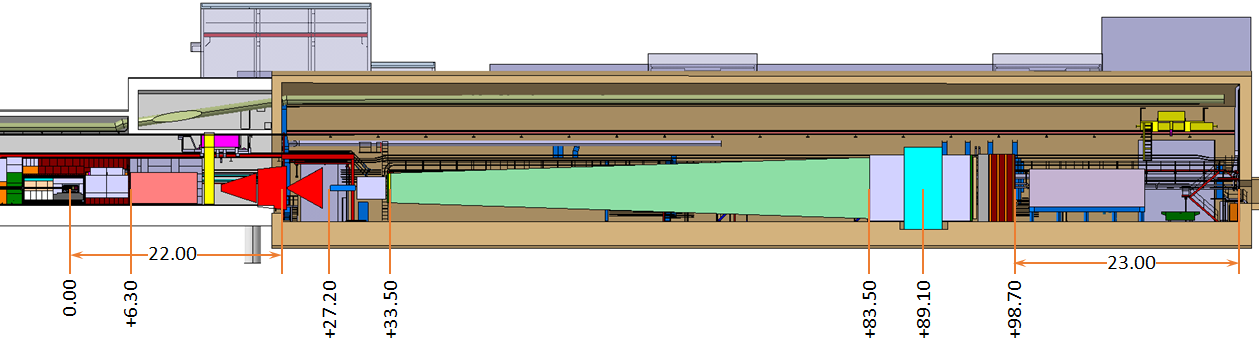}
\caption{\small \label{fig:SHiP_geometry}  Space reservation and location of the SHiP detector components with respect to the centre of the BDF/SHiP proton target.}
\end{figure}

The five metres long hadron absorber stops hadrons and electromagnetic radiation emerging from the proton target. It is equipped with a coil which magnetises the iron shielding blocks~\cite{Kershaw_2018} to serve as the first section of the active muon shield. The rest of the muon shield consists of free-standing magnets. The configuration presented in~\cite{bib:BDFSHIP_PROPOSAL_2023}, shown in Figure~\ref{fig:SHiP_geometry}, consists of a first SC section followed by a NC section. The target complex and part of the free-standing muon shield is located at the end of the TCC8 target hall, while the subsequent muon shield magnets are located in the taller ECN3 experimental hall.

The implementation of BDF/SHiP in ECN3 has undergone a series of radiation protection studies with nominal beam operation of $4\times 10^{19}$ PoT per year and 15 years of operation ~\cite{RP-BDF-0, RP-BDF-1, RP-BDF-2}. 
Compared to the original CDS design, it has been possible to significantly reduce the amount of shielding at strategic locations by benefiting from the thick soil layer above TCC8 and ECN3 and already existing activated shielding. Consequently, decommissioning of the facility would also involve less newly produced radioactive waste. Studies of prompt radiation above the target complex and beyond demonstrate that dose rates are well below the limit for an NDA. Furthermore, the doses due to stray radiation at the CERN fence downstream of ECN3 and beyond have been investigated. Results show that the ambient dose equivalent limit for the CERN fence would be met with a substantial margin and that the effective dose to the public would remain well below~\SI{10}{\micro\sievert}/year and is considered as optimized~\cite{SafetyCodeF}.

Residual dose rates in the target area as well as the soil activation were evaluated for the fifteen years of beam operation showing that the target area is well optimized and compatible with the given soil activation design limits~\cite{RP-BDF-0}. 
Studies for air and nitrogen/helium activation occurring inside of the nitrogen/helium target vessel and the surrounding air have further demonstrated that air and nitrogen/helium releases into the environment have a negligible radiological impact on the public~\cite{RP-BDF-0}. In order to further simplify the installation and increase the lifetime of the facility, it is currently considered the option of employing primary vacuum; this will further reduce the radiological impact of the facility, reduce operational costs and increase the capability of the system to run for longer periods (i.e. by reducing the risks of radiation accelerated corrosion).

The radiation to the detector and electronics in ECN3 is expected to be significantly below levels that require special measures with the exception for the first part of the muon shield together with the side of ECN3 along the stream of muons, but in any case not requiring the development or application of radiation tolerant electronics~\cite{Esposito_PBC_ACC_2203203}.

The updated dimensions of the muon shield and the detectors allow integrating SHiP in the existing TCC8/ECN3 hall below the existing bridge cranes (Figure~\ref{fig:SHiP_geometry}). While the distance between the Sal\`eve-side wall and the decay volume in ECN3 is between $\approx$4~--~\SI{2}{\m} (upstream/downstream), the Jura-side wall is at about the same distance of $\approx$9~--~\SI{7}{\m} as in the original CDS design, leaving sufficient space for detector assembly and maintenance.

Limited modifications to the ECN3 floor will be necessary under the spectrometer magnet in the form of a 5$\times$\SI{7}{\meter\squared} pit with a depth of \SI{1}{\m}. A detailed investigation of the impact and reuse of existing services and infrastructure has been performed. The implementation of BDF/SHiP will not interfere with services for other NA facilities and a number of existing detector services may be reused. The current access shaft to TCC8/ECN3 of 4$\times$\SI{8}{\meter\squared} is considered a limiting factor in performing the works associated with both TCC8 and ECN3. An additional shaft of $8\times$\SI{8}{\square\m} at the end of ECN3 would allow separating the activities associated with TCC8 and the target complex, and the detector activities in ECN3. It would reduce interference and significantly ease and simplify the detector installation. In order to build the new shaft, part of the building 918 will be demolished and the existing services will be rerouted. A new access building will be constructed on top of the shaft and equipped with an overhead crane for transport purposes. Access control will be needed. The reduced surface building 918 appears sufficient to host detector electronics, services and computing, and space for operating the detector. 

A new power converter will be required for the hadron absorber, six for the SND muon system and one for the decay spectrometer. Power converters will also be needed for the BDF dilution system magnets. The power converters will be installed in the target service building.

\section{Preliminary schedule and preliminary cost estimate}
\label{sec:ScheduleCost}

\subsection{North Area operation, beamline and infrastructure schedule}
\label{sec:schedule}

The main constraints for the ECN3 HI implementation are the availability of resources, which will be critical given the concurrence with the HL-LHC, ATLAS/CMS Phase-II upgrades, and NA-CONS~Phase~1, as well as the fact that the upgrade of the accelerator infrastructure upstream of TCC8 must be ready for operation after LS3 to avoid impacting other NA users. In addition, given the length of cool-down required in highly radioactive areas, major modifications in TDC2/TCC2 are (most likely) not compatible with the LS3 timeline.

 These constraints can be met if the TCC8/ECN3 upgrade is decoupled from the upstream accelerator infrastructure (access, cooling and ventilation, etc.) by allowing at least 1 year to complete work in TCC8/ECN3 after LS3 and during Run~4, whilst the rest of the NA is operational. This possibility has been confirmed by the team responsible for the access system.
As already mentioned in Section~\ref{sec:Integration} the consolidation of BA82 during LS3 is not feasible because of lack of resources and in the following it is assumed that the power converters for the experimental magnets will be hosted in the new target station service building. 

\begin{figure}[h]
     \centering
     \includegraphics[width=\textwidth]
     {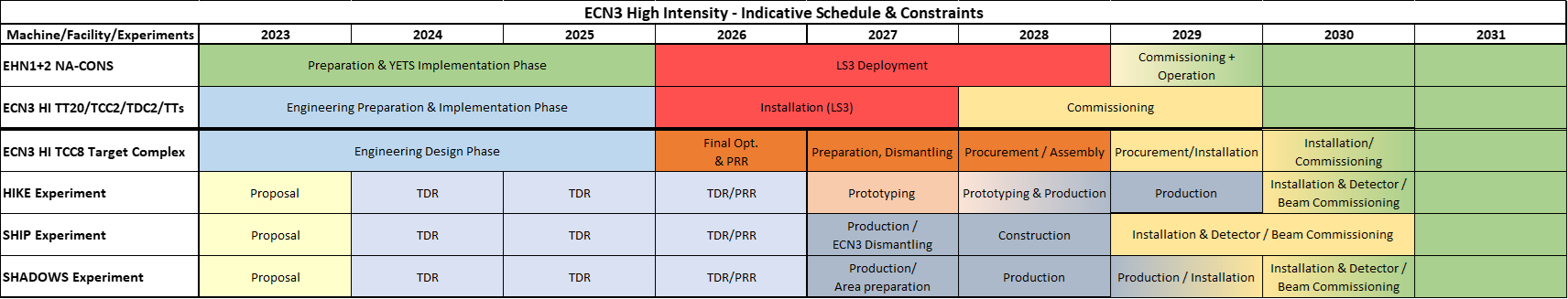}
     \caption{\small Preliminary implementation schedule of the ECN3 High Intensity facility.}
     \label{fig:ecn3_schedule}
\end{figure}

A preliminary implementation timeline for the ECN3 HI facility is shown in Figure~\ref{fig:ecn3_schedule}.

The proposed schedule assumes a decision on the experimental program by the end of 2023 to address the outstanding issues on the experiment-dependent target and secondary beamline design. Engineering studies must be completed before LS3 to keep compatibility with Phase 1 of NA-CONS and execution in LS3. The TDR/Project Readiness Review~(PRR) phases of the intensity upgrade would start immediately in 2024. Note also that the proposed schedule assumes adequate access to test beams by the selected experimental program until 2030 for the development and calibration of the detectors (see Sections \ref{subsec:hikeshadowsschedule} and \ref{subsec:bdfshipschedule}).

\subsection{Beamline and infrastructure cost estimate}

Following the mandate of the ECN3 Beam Delivery TF, the short pre-study primarily focused on an evaluation of the technical feasibility of a future ECN3 HI facility. Consequently, the provided resource estimates~\cite{bib:EDMScostTF} are in several cases based only on group expert estimates without time for an extensive engineering study. The overall uncertainty range for the cost estimate here summarized is not expected to be better than C3 –- C4~\cite{bib:DOEGuide}~\footnote{A Class 3 estimate uncertainty has a lower range between -10\% and -20\% and an upper range between +10\% and +30\%. A Class 4 estimate uncertainty has a lower range between -15\% and -30\% and an upper range between +20\% and +50\%. }.
Two main cost categories have been identified:

\begin{enumerate}
  \item high-intensity beam delivery (including the corresponding engineering phase):
  \begin{enumerate}
    \item a set of additional NA-CONS requirements, beyond the initial baseline and identified during the ECN3 TF evaluation, or as a result of the operational experience in 2021-2022. These consolidation items are not directly linked to the intensity upgrade requirements but to the future reliability of a new facility;
    \item high-intensity upgrade specific beam delivery requirements allowing for maximum beam intensities safely and reliably delivered to ECN3;
  \end{enumerate}
  \item experiment-specific target complex and infrastructure requirements for TCC8 and ECN3. 
\end{enumerate}

With respect to the initial estimate presented in~\cite{bib:EDMScostTF}, the latest information concerning the shielding requirements for HIKE-Phase 1 and SHADOWS and the corresponding adaptation of the civil engineering work implies an increase of $\approx$~4~MCHF of the cost for HIKE-Phase 1 and SHADOWS. 

The updated cost estimates are now equal for both options,  within the uncertainties. They include:

\begin{enumerate}
    \item High-Intensity beam delivery with the corresponding engineering phase: 14~MCHF;
    \item TCC8 target Complex and ECN3 Infrastructure: 50~MCHF.
\end{enumerate}

A detailed list of funding requests has been presented during the NA-CONS CSSR~\cite{bib:NA-CONS_CSSR}. The above cost estimate is based on the following main assumptions:

\begin{itemize}
    \item BA82 consolidation will take place during NA-CONS~Phase~2;
    \item recovery of iron shielding blocks for the TCC8 target station from the CNGS hadron absorber, TT7 dump/absorber, and OPERA;
    \item staging of the beam instrumentation upgrade compatibly with the available resources;
    \item no need of increasing the scope of the electrical infrastructure consolidation beyond that considered by NA-CONS;    
    \item no need of an additional cooling tower beyond the already planned NA-CONS scope.
\end{itemize}

The cost estimates for the TCC8 target complex and ECN3 infrastructure are either derived from the studies carried out in the scope of the BDF CDS~\cite{Ahdida:2703984}, or from updated civil engineering studies performed by an external consultant, and/or reviewed by equipment/service group experts to provide an expert estimate in line with the pre-study requirements~\cite{bib:JointBDFCBWG}.
The experiment requirements summarized in Section~\ref{sec:OPScenarii} and presented in the LoIs~\cite{bib:SHADOWSLOI_2022,bib:BDFSHIPLOI_2022,bib:HIKELOI_2022} were considered in compiling a related infrastructure requirement document~\cite{userrequirements} and iterated together with NA-CONS. Despite the different types of target complex implementations, as well as civil engineering needs for HIKE/SHADOWS or BDF/SHIP, the total implementation cost envelope remains comparable when considering equivalent operation periods.
The cost includes:
\begin{itemize}
    \item civil engineering needed for the target complex including a surface building also housing the required additional power converters and respective Heating, Ventilation and Air Conditioning (HVAC) infrastructure;
    \item a new target complex in the form of a dilution system, target systems and shielding, instrumentation and inertisation system (for BDF/SHIP), or a separated production target and TAX implementation with cooling and longitudinal shielding (for HIKE/SHADOWS);
    \item power converters and DC cables for the experimental magnets and muon shielding;
    \item TCC8/ECN3 general infrastructure, as well as services and support needed for the new detectors;
    \item dismantling and decommissioning of the existing TCC8 target complex (for both implementation scenarios), integration and installation activities, as well as beamline and infrastructure modifications required between HIKE Phase~1 and Phase~2.
\end{itemize}

Spectrometer magnets and muon shields are not included in the cost estimate and they are expected to be primarily covered by the experimental collaborations. No cryogenics services were considered as not requested in the experiment LoIs. 

The operation with~\SI{1.2}{\s}-long~\textit{dedicated ECN3 spills} might have an impact on the specifications of the electrical infrastructure beyond the NA-CONS scope that remains to be evaluated (see Section~\ref{sec:transfer}). 
The beamline design for HIKE-Phase 2 is ongoing and the specifications on the proton beam dump (not included in the initial cost estimate) are being defined. The corresponding radiation protection and integration studies have to be done, with expected implications for the design of the shielding and of the K12 magnets and busbars in the area between the target and the TAX and in general for the services required for HIKE Phase~2~(see Section~\ref{sec:IntegrationHIKESHADOWS}). The requirements for the cryogenic system for the first SC section of the SHiP muon shield have not been specified, yet (see Section~\ref{sec:BDFintegration}). The cost implications of the above items will have to be addressed in the TDR phase.

From the analysis conducted including the outcome of the NA-CONS CSSR, no showstopper for the ECN3 HI implementation according to the schedule proposed in Section~\ref{sec:schedule} has been identified.

\subsection{HIKE/SHADOWS cost and schedule}\label{subsec:hikeshadowsschedule}

The indicative operation timeline of HIKE/SHADOWS as given in Figure~\ref{fig:indicativeschedule} assumes start of nominal operation in 2031. 

The HIKE timeline to first beam is as follows~(see Figure~\ref{fig:ecn3_schedule}):
\begin{itemize}
\item 2024-2025: detector studies  
\item 2026: TDR
\item 2027-2028: prototyping 
\item 2028-2029: production 
\item 2030: installation and possible commissioning with lower-intensity beam
\item 2031: data acquisition starts with high-intensity beam
\end{itemize}

The estimated material cost of the HIKE detector upgrades and new components is estimated to 27.5 MCHF summing both Phase 1 and 2 contributions. It is largely dominated by the cost of Phase 1.

In case of positive approval of the experiment by the end of 2023/beginning of 2024, SHADOWS plans to prepare a TDR by mid of 2026 and to undergo a PRR by the end of 2026. This would allow a timely start of construction in 2027, that could last until mid-2029, followed by one year of installation/commissioning in 2029-2030. The first pilot run could be performed already by the end 2030 or beginning of 2031~(Figure~\ref{fig:ecn3_schedule}). The following expected nominal operation will include 8 years where SHADOWS will operate 50~\% of the time in BD mode together with HIKE, interleaved with long shutdown periods. This operational configuration will allow to consolidate the set-up along the lifetime of the experiment.

The overall cost of the SHADOWS detector is driven by the choice of the detector technologies that are currently under scrutiny. 

The current best estimate is a total material cost of 12.4~M\texteuro{}, out of which 9.6~M\texteuro{} correspond to the main spectrometer and 2.8 ~M\texteuro{} to the NaNu subdetector. The overall uncertainty range for this cost estimate is not expected to be better than C3. 
SHADOWS expects that the MIB system, the dipole magnet of the main spectrometer and the decay vessel will be provided by CERN as Host Laboratory, while the rest of the cost will be shared among the other collaborating institutions.

\subsection{BDF/SHiP cost and schedule} \label{subsec:bdfshipschedule}

Given the extensive studies performed during the Technical Proposal and the CDS phases, it is expected that the TDR phase will require 3--4 years, depending on the subsystem. The construction phase is expected to start in LS3 to allow commissioning the BDF in 2030, with first year of data taking in 2031~(see Figure~\ref{fig:ecn3_schedule}). LS4, currently scheduled for 2033, presents an opportunity for consolidation, if necessary. 

The operational schedule stretches over 15 years, with several opportunities for extensions and upgrades of BDF/SHiP, as discussed in~\cite{bib:BDFSHIP_PROPOSAL_2023}.

The cost estimate of the detector includes the muon shield, the SND and the HSDS detectors, and all associated infrastructure. The estimate initially prepared for the CDS report has been revised according to the new detector configuration and dimensions, and updated with 2023 rates~\cite{bib:BDFSHIP_PROPOSAL_2023}. It amounts to $\approx$~51\,MCHF with an uncertainty at the level of $^{+30\%}_{-10\%}$, making it compatible with a Class 3 cost estimate. The accuracy is derived from the uncertainties associated with each individual component. At the same time the total cost is conservative given that upper estimates have been used and the most expensive options have been included, wherever applicable, e.g. muon shield in the hybrid configuration with a superconducting magnet, SBT with maximum number of compartments, etc~\cite{bib:BDFSHIP_PROPOSAL_2023}. 

\FloatBarrier

\bigskip

\section{Physics potential }
\label{sec:PhysPotential}

In the following the main areas of fundamental physics that could be significantly impacted by the projects proposed at ECN3 are considered. It should be noted that the first two topics, i.e., the physics of feebly-interacting particles (FIPs) and flavour physics, benefit from an extensive body of literature and from many existing and dedicated studies, whereas the third topic, neutrino physics, presents various novel ideas that have not yet been studied at the same level of detail.

\subsection{FIP physics}
\label{sec:FIPs}

This section focuses on the FIP searches of the experiments proposed at ECN3. After a general introduction on the physics motivations and measurement issues, the projects detail their respective strategies and how FIP simulations and background estimates are performed. The international landscape of competing experiments and proposals is then briefly discussed before presenting exemplary FIP sensitivity projections.

\subsubsection{Introduction}

The Standard Model (SM) of particle physics is highly successful in accurately predicting experimental observations across many different processes and energy scales. Nevertheless, there are a number of both theoretical arguments and experimental observations pointing to the need for new physics beyond the Standard Model (BSM). From the theoretical perspective, the parameters of the SM appear finely tuned, in particular the electroweak scale (known as the hierarchy problem) and the CP-violating phase of strong interactions (known as the strong CP problem), and there is no compelling explanation for its flavour structure and its accidental global symmetries. On the experimental side, there is clear evidence for a particle-antiparticle asymmetry in the universe beyond the SM prediction, for non-zero neutrino masses and for the existence of a new form of matter called dark matter.

While these problems are often addressed by postulating new physics at high energies beyond the reach of existing particle colliders, there exist compelling solutions also in terms of light particles, which are kinematically easily accessible but have evaded detection due to their tiny couplings. For example, the hierarchy problem can be solved dynamically through the relaxion mechanism, which introduces a new spin-0 particle (the relaxion), which may have a mass in the MeV range and couple to SM particles via a tiny Higgs mixing~\cite{Flacke:2016szy,Banerjee:2020kww}. The strong-CP problem is commonly solved via the Peccei-Quinn mechanism, which predicts the existence of a new particle (the QCD axion). While QCD axions are usually considered to be extremely light, in certain models they can have a mass in the MeV-GeV range~\cite{Agrawal:2017ksf,Hook:2019qoh}. Particles with similar coupling structures (so-called axion-like particles) furthermore arise naturally in many theories with spontaneously broken global symmetries, such as supersymmetry breaking~\cite{Bellazzini:2017neg} or string theory~\cite{Conlon:2006tq,Svrcek:2006yi,Choi:2009jt,Arvanitaki:2009fg,Acharya:2010zx,Cicoli:2012sz,Halverson:2017deq,Cicoli:2022fzy}.

The problem of neutrino masses can be elegantly solved by introducing three right-handed neutrinos below the electroweak scale~\cite{Asaka:2005pn}. While the lightest of these sterile neutrinos may be a viable dark matter candidate~\cite{Drewes:2016upu}, the two heavier sterile neutrinos (called heavy neutral leptons~\cite{Bondarenko:2018ptm}) may explain the baryon asymmetry of the universe through their decays into Standard Model particles. In this set-up, the lightest sterile neutrino would have a mass in the keV range and such tiny couplings that these particles evade all laboratory searches and never enter into thermal equilibrium in the early universe.

An alternative avenue to address the dark matter puzzle is to postulate the existence of new particles that are in thermal equilibrium with the SM bath at high temperatures, but then decouple as the universe cools down. Due to its insensitivity to initial conditions, this so-called freeze-out mechanism has for many years been the leading paradigm to predict the abundance of dark matter in the present universe. While it has traditionally been assumed that the interactions that keep the dark matter particles in thermal equilibrium are mediated by SM particles (in particular electroweak gauge and Higgs bosons), this possibility has been increasingly constrained by the non-observation of dark matter signals in collider and direct detection experiments~\cite{Arcadi:2017kky}. These constraints have led to a shift of focus towards dark matter models that introduce new interactions, mediated by new BSM particles~\cite{Feng:2008ya}. Such interactions may arise for example from simple gauge extensions of the SM, such as a spontaneously broken $U(1)'$ symmetry. These so-called dark sector models can in principle be realized for dark matter masses anywhere between a few MeV (the lower bound being imposed by the agreement of the Big Bang Nucleosynthesis predictions with observed element abundances~\cite{Sabti:2019mhn}) and hundreds of TeV (the upper bound stemming from the so-called unitarity limit~\cite{Griest:1989wd}). Nevertheless, it is particularly attractive to consider dark matter masses that fall below the energy threshold of direct detection experiments searching for nuclear recoils, which rapidly lose sensitivity for sub-GeV dark matter.

Finally, it should also be mentioned that fundamental extensions of the SM, notably string theory, quite generally have a tendency to contain whole sectors of particles, very weakly coupled to the particles that make up the experiments. Such ``hidden'' or ``dark'' sectors can be coupled to the SM via ``portal'' interactions, e.g. dark photons~\cite{Abel:2008ai, Goodsell:2009pi, Goodsell:2010ie, Cicoli:2011yh} or axion-like particles~\cite{Conlon:2006tq,Svrcek:2006yi,Choi:2009jt,Arvanitaki:2009fg,Acharya:2010zx,Cicoli:2012sz,Halverson:2017deq,Cicoli:2022fzy}.

All of the examples above motivate searches for new particles at the MeV to GeV scale, called FIPs.\footnote{In the following we will always implicitly assume this mass range when referring to FIPs. In general both lighter as well as heavier FIPs can be of interest, cf.~\cite{Agrawal:2021dbo,Antel:2023hkf}.} While the same arguments can in principle be used to predict specific coupling structures, the range of possibilities is so large that it makes sense to combine this top-down approach with a more model-agnostic bottom-up approach, in which we consider coupling structures that resemble interactions known from the SM. For scalar particles this means couplings similar to those of the SM Higgs boson, while for axion-like particles inspiration can be taken from the neutral pions, i.e. the Goldstone bosons of chiral symmetry breaking. GeV-scale heavy neutral leptons would interact through mixing with the active neutrinos of the SM, while the interactions of new gauge bosons (called dark photons) would resemble electromagnetism. This approach leads to a small number of well-defined benchmark scenarios, which have been spelled out explicitly by the Physics Beyond Colliders initiative~\cite{Beacham:2019nyx}.

In principle it is possible to search for such FIPs at the energy frontier, i.e.\ using high-energy proton-proton collisions. A key limitation however arises from the typical detector dimensions, which limit the range of observable decay lengths to a few (tens of) meters. Taking into account the substantial boost factors of light particles produced in high-energy collisions, one immediately concludes that the LHC is not the ideal environment to search for long-lived (neutral) particles with a proper decay lengths above 1 metre. Much higher sensitivities can be achieved by experiments operating with larger detectors at lower centre-of-mass energies. Indeed, many of the leading constraints on FIPs stem from beam-dump experiments carried out several decades ago. Given modern beam intensities and detector technologies, it will be easily possible to surpass the sensitivity of these experiments by orders of magnitude and probe deeply into the unexplored parameter regions of FIPs models.

A typical FIP event in a beam-dump experiment would consist of a FIP being produced in the proton target (with an angular and energy distribution that depends on the specific production mechanism), propagating into the decay volume and then decaying into several SM particles. In the simplest case, the decay produces exactly two charged particles, such that the vertex position and the mass of the decaying particle may be reconstructed. However, in practice many more complicated decay modes are of interest, involving neutral particles (such as photons or neutral pions) in the final state and three-body decays. Detecting and identifying all final-state particles and reconstructing the vertex position and the mass of the decaying particle as accurately as possible is of utmost importance in order to achieve a background-free environment, as well as the possibility of characterising a signal~\cite{Morandini:2023pwj}.

In Beam Dump experiments, the physics backgrounds originate mainly from the three following processes related to muons and neutrinos emerging from the dump, for which the experiments have developed mitigation strategies:
\begin{itemize}
\item{} Muon combinatorial: This type of background arises when two opposite-sign muons within the same proton spill appear to form a vertex and point back to the target. 

\item{} Muon DIS: Muons may interact inelastically in the material of the detector or in the surrounding infrastructure. These DIS interactions produce $V^0$s but also, more importantly, false $V^0$s due to random combinations of tracks from the same DIS interaction. Given the small energy transfer, the DIS interactions lead to energetic products that are aligned with the direction of the incoming muon. Hence, muon DIS background is dominated by those originating in the material in the close vicinity of the fiducial volume.

\item{} Neutrino DIS: Similarly to the muon DIS background, the dominant source of neutrino-induced background comes from neutrino DIS in the material close to fiducial volume. 

\end{itemize}

\bigskip

\subsubsection{HIKE FIP searches}

 Kaon and beam-dump data sets are sensitive to complementary FIP processes and mass ranges. Operation in both kaon and beam-dump modes will allow HIKE to address a uniquely broad range of hidden-sector scenarios covering a mass range spanning from about 10~MeV to a few GeV. Moreover, operation in kaon mode provides excellent sensitivity to non-minimal dark sector scenarios involving short-lived FIPs, which completely evade detection in beam-dump experiments~\cite{Harris:2022vnx,Gori:2022vri}.

Prospects for searches for FIP production in kaon decays, including non-minimal scenarios, have received much attention recently, and are reviewed in~\cite{Goudzovski:2022vbt}. 
HIKE kaon datasets will bring significant sensitivity improvements for dark photon (via the $\pi^0\to\gamma A^\prime$ and possibly $K^+\to\pi^+A^\prime$ decays), dark scalar (via the $K^+\to\pi^+S$ decay), heavy neutral leptons with electron and muon couplings (via the $K^+\to e^+N$, $K^+\to\mu^+N$ and $\pi^+\to e^+N$ decays), and axion-like particles (via the $K^+\to\pi^+a$ decay). Depending on the FIP mass and coupling constant values, the searches at HIKE will consider both invisible final states (via missing mass), and searches for production of FIPs followed by their decays (including prompt and displaced decay vertices). The HIKE projections are detailed in the proposal and, in many cases, based on analyses of the existing NA62 datasets~\cite{NA62:2020xlg,NA62:2021zjw,NA62:2020mcv,NA62:2021bji}. Therefore the projections are robust, and fully account for such factors as background and resolution.

 The HIKE experiment plans to collect a substantially larger sample than NA62 in dump mode: the HIKE sensitivity curves are obtained for a total of $5\times 10^{19}$~PoT, to be compared to the $10^{18}$~PoT expected to be collected in dump mode by NA62 at LS3. The HIKE sensitivity to the FIP benchmarks~\cite{Beacham:2019nyx} has been studied with the data from NA62 and using full Monte Carlo simulations evolved from the NA62 Monte Carlo framework. This framework is a C++, \textsc{Geant4}-based code, containing a detailed description of the subdetectors and the K12 beamline. 
The NA62 beam-dump datasets~\cite{Dobrich:2023dkm} have been used to extrapolate the expected background level, and to quantify additional possible improvements due to upgraded or extra detectors. 
The overall background expected is: $< 0.01$, $<0.8$, $<0.07$ and $<0.1$ events for $\mu^+ \mu^-$, $e^+ e^-$, $\pi^+\pi^- (\gamma)$, $\ell^\pm \pi^\mp$ final states, respectively.

\subsubsection{SHADOWS FIP searches}

The SHADOWS sensitivity to different benchmarks has been studied with the use of the full Monte Carlo simulation.
As for HIKE, the SHADOWS full Monte Carlo is part of the general NA62 Monte Carlo framework (see above) able to simulate the interactions of the particles with the detector elements. The detailed geometry of the detector and the technologies chosen for each sub-detector has been included in the Monte Carlo along with a detailed description of the K12 beam line, the muon sweeping system, and the experimental hall.

\vskip 2mm
The signals are generated with \textsc{PYTHIA 8.32} and the background with the \textsc{Geant4}-based \textsc{Beam Delivery Simulation} or \textsc{BDSIM} package~\cite{Nevay:2018zhp}. 
The output of \textsc{BDSIM} package and of \textsc{PYTHIA 8.32}~\cite{Sjostrand:2007gs} is then handed over to the SHADOWS full Monte Carlo where the simulation of interactions of the  particles with the detector material and SHADOWS magnetic elements is performed. The inelastic interactions of neutrinos and muons with the detector material are simulated using the \textsc{GENIE}~\cite{Andreopoulos:2009rq} and \textsc{PYTHIA6}~\cite{Sjostrand:2006za} generators, respectively, where the interactions are forced to occur to enhance the sample and a weight representing their probability is stored together with the event.
The  muon inelastic interactions have been  studied also using the physics lists contained in \textsc{Geant4}, and without forcing the interactions to occur. The results obtained with \textsc{Geant4} agree within a factor of 2 with those obtained with \textsc{PYTHIA6}. 
The impact of this difference on the final result is negligible, as the background arising from muon inelastic interactions in SHADOWS is very low.

\vskip 2mm
A detailed discussion of the backgrounds samples can be found in the Proposal \cite{bib:SHADOWS_PROPOSAL_2023} together with techniques and methods used to mitigate them. A brief summary of the findings is reported here.

\vskip 2mm
The proton interactions with the dump give rise to a copious direct production of short-lived resonances, pions and kaons. While the TAX length is sufficient to absorb the hadrons and the electromagnetic radiation produced in the proton interactions, the decays of pions, kaons and short-lived resonances result in a large flux of muons and neutrinos. Muons and neutrinos emerging from the dump are the two major sources of background for FIP searches in SHADOWS.
The muon flux predicted by simulation has been validated with two campaigns of measurements performed in ECN3 both on-axis and off-axis, when the K12 beam line was operated in beam dump mode, in November 2021 and June 2023. The measurement of the muon flux off-axis has been a major achievement with respect to the LoI. Two full size modules of the muon system and several telescopes with different technologies (silicon pixels, scintillating tiles, and micromegas) have been funded and built on purpose for that.

\vskip 2mm
The three main backgrounds discussed in the introduction are evaluated in the SHADOWS baseline setup, featuring 1 mbar of pressure in the decay volume instrumented with both the upstream background veto and the lateral background veto on its full length. Before any analysis step, the momentum and directions of every charged track are smeared to account for the detector resolutions after full reconstruction. Signal selection is based on two tracks pointing to a common vertex in the target area. Background rejection makes full use of the precise tracking and timing capabilities of the spectrometer and its background vetos. Table~\ref{tab:all_bkg} summarises the overall background in SHADOWS that can mimic a signal final state in the full SHADOWS dataset of $5\times 10^{19}$ PoT.

\vskip 2mm
It should be reminded that prior to any suppression technique the rates of these background components in the geometric acceptance of SHADOWS are significantly lower than in an on-axis setup, especially for neutrino-induced background. This is a direct consequence of the specific kinematics of these components that favor small polar angles and therefore emission mostly in the forward direction.
In addition, the lower momentum of muons and neutrinos emitted off-axis significantly reduces the probability of inelastic interactions with respect to an on-axis setup, since the inelastic cross-section raises with the momentum of the involved particles.

\begin{table}[t]

\centering
\begin{tabular}{lcc} 
\hline \hline
   background type  & fully reconstructed events & partially reconstructed events\\ \hline
  Muon combinatorial  & $10^{-3}$ & 0.7\\
   Muon DIS & $<2.5\cdot 10^{-2}$ & $<0.90$ \\  
  Neutrino DIS & $< 0.01$ & $< 0.01 $ \\     
\hline\hline
\end{tabular}
\caption{\small Estimated background in SHADOWS for $5 \times 10^{19}$~PoT. Partially reconstructed events correspond to FIP channels with invisible particles such as neutrinos in their decay.} 
\label{tab:all_bkg}
\end{table}

\subsubsection{SHiP FIP searches}

BDF/SHiP's expected physics performance in ECN3 has been studied in detail with the help of the full \textsc{Geant}-based Monte-Carlo framework that was developed for the original proposal. The software framework is based on the \textsc{FairRoot} package~\cite{FairRoot} and is called \textsc{FairShip}. The framework incorporates \textsc{Geant4}~\cite{Agostinelli:2002hh,geant4b} to simulate the particles through the target and the experimental setup, \textsc{PYTHIA8}~\cite{Sjostrand:2007gs} for the primary proton fixed-target interaction, \textsc{PYTHIA6}~\cite{Sjostrand:2006za} for muon DIS and cascade production of charm and beauty~\cite{CASCADE}, and \textsc{GENIE}~\cite{Andreopoulos:2009rq} for interactions of neutrinos. The production and decays of various types of FIPs have been implemented in \textsc{FairShip}. Mainly \textsc{PYTHIA8} is used to generate the different signal processes. 

The validity of the \textsc{FairShip} prediction of the beam-induced particle fluxes has been verified by comparing to the data from the CHARM beam-dump experiment at CERN~\cite{Dorenbosch:164101}. The most realistic cross-check of \textsc{FairShip} has been performed in summer 2018 in a dedicated experiment at the CERN SPS~\cite{vanHerwijnen:2267770}. It has directly measured the rate and momentum of muons produced by 400\,GeV protons dumped on a replica of the BDF/SHiP target, and found a very good agreement between the prediction by the simulation and the measured spectrum~\cite{Ahdida:2020doo}.

The background simulations have been performed with strongly enhanced muon production from the relevant processes, such as resonance decays and gamma conversion. These have been found to produce rare but difficult background events. Dedicated samples of charm and beauty hadrons have been produced. These are both a source of signals and challenging backgrounds. The effect of cascade production of charm and beauty from secondary hadrons are also accounted for in both signal and background.

The SHiP detector response and resolution has been taken into account based on measurements done in test beams with prototypes of all subdetectors during the CDS phase.  For the implementation in ECN3, the \textsc{Geant4} simulation has been updated with the complete geometry of the underground complex, the revised muon shield, and the detectors.

Extensive simulations of the three main background components discussed in the introduction have been done in the SHiP setup.
In order to get large statistics for the background studies of muon and neutrino DIS, the fluxes obtained from the simulation of the minimum bias, and the charm and beauty production, were used to produce DIS events using \textsc{PYTHIA6} for muons and \textsc{GENIE} for neutrinos, and boosting the interaction cross-sections such that every muon/neutrino interacts according to the material distribution of the experimental setup. 

With the use of the upstream vessel wall background tagger~(UBT) and the surrounding wall background tagger (SBT), coincidence timing, and a simple and common set of selection criteria~\cite{bib:BDFSHIPLOI_2022} based on reconstructed quantities, the resulting expected background levels are shown in  Table~\ref{Tab:bkgs}. They do not differ significantly from the CDS results~\cite{Ahdida:2704147}. The adaptation to ECN3 and the results of the background studies bear witness of the redundancy built in to the combined performance of the suppression of beam-induced particle rates and the detector.
The selection above is entirely inclusive with respect to different types of long-lived particle decays in the fiducial volume. This ensures maximum sensitivity in the FIP searches, while remaining generic to new models that may be proposed in the future. It preserves close to 100~\% of the signal efficiency in fully reconstructed modes, while in general, the efficiency for partially reconstructed modes is around 70~\%, obtained by simulating the signals with the full simulation. It has also been verified that the probability that an actual signal candidate is wrongly vetoed by an uncorrelated hit in the SBT remains insignificant. With the simple regional veto that requires the SBT hit to be upstream of the signal candidate vertex and within a time window of $3\times \sigma_{\mathrm{SBT}}$ (time resolution $\sigma_{\mathrm{SBT}} \approx$\,ns) the probability is roughly a percent.

\begin{table}[t]

\centering
\begin{tabular}{lcc} 
\hline \hline
   background type  & fully reconstructed & partially reconstructed \\ \hline
  Muon combinatorial  & \multicolumn{2}{c}{$ (1.3\pm 2.1)\times 10^{-4}$} \\
  Muon DIS&  $<\,5\times 10^{-3}$ & $<\,0.2$ \\  
  Neutrino DIS & $<0.1$ & $<\,0.3$ \\     
\hline\hline
\end{tabular}
\caption{Expected background in BDF/SHiP in the search for FIP decays at 90~\%\,CL for $6\times 10^{20}$ protons on target after applying the pre-selection, the timing, and the UBT and SBT veto. The neutrino- and muon-induced backgrounds are given separately for the set of criteria corresponding to the fully and partially reconstructed signal modes.} 
\label{Tab:bkgs}
\end{table}

To avoid irreducible neutrino DIS background from neutrinos interacting with the air molecules inside the vessel, a level of vacuum below $10^{-2}$ bar is sufficient. The background from cosmics can be reduced to negligible levels using the SBT~\cite{Anelli:2007512}.

For the sensitivity to LDM scattering, the principal background comes from neutrino events with only one reconstructed outgoing electron at the primary vertex constitute, mimicking the signature $\chi e^{-}\to \chi e^{-}$. The \textsc{GENIE} Monte-Carlo generator~\cite{Andreopoulos:2009rq}, interfaced with \textsc{FairShip}, has been employed for a full simulation to provide an estimate of the expected background for $6\times 10^{20}$~PoT. After imposing a selection optimised for the signal, the total residual neutrino background amounts to $ \approx$600 events~\cite{bib:BDFSHIP_PROPOSAL_2023}. The dominant background contribution arises from $\nu_e$ quasi-elastic scattering $\nu_e n\to e^{-}p$, where the soft proton remains unidentified, and from topologically irreducible sources, i.e., $\nu_{e}(\bar{\nu}_{e})$ elastic and $\bar{\nu}_{e}$ quasi-elastic scattering ($\bar{\nu}_e\,p\to e^{+}n$).

LDM signal events have been simulated with the help of the MadDump software~\cite{Buonocore:2018xjk}, and assuming pair-production ($\chi\bar{\chi}$) in the prompt decays of dark photons. In the considered dark photon mass range of $M_{V} \approx{\mathcal O}$(1)\,GeV/c$^2$, only contributions from the decay of light mesons ($\pi,\,\eta,\,\omega$) and proton bremsstrahlung have been included. Prompt-QCD and heavier DY-like production mechanisms have been proven to be negligible.

\subsubsection{International landscape}

\begin{figure}[t!]
\centering
\includegraphics[width=0.55\textwidth]{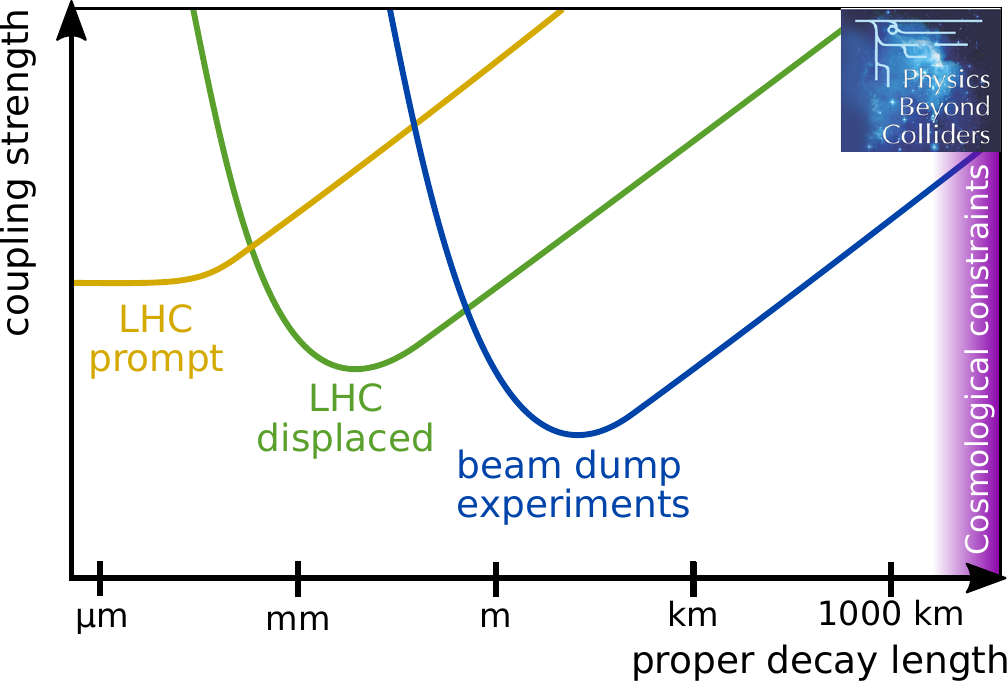}
\caption{\small
Schematic illustration of the complementarity of searches for FIPs at existing LHC experiments and at the proposed ECN3 beam-dump experiments. Searches for promptly decaying particles at the LHC typically lose sensitivity as soon as the boosted decay length $\beta \gamma c \tau$ exceeds the vertex resolution (of the order of $0.1$ mm). Searches for displaced vertices are most sensitive to boosted decay lengths between millimetres and metres, corresponding to decays in the inner parts of the detectors. Beam-dump experiments, on the other hand, can detect particles with boosted decay lengths of up to tens of meters. Since the beam energy is smaller than at the LHC, for fixed mass particles the typical boost factor $\gamma$ is usually smaller. The difference in terms of proper decay length $c \tau$ is even larger. The larger total number of proton collisions furthermore implies that beam-dump experiments can achieve an unparalleled sensitivity in terms of the underlying coupling strength. For $\tau > 0.1 \, \mathrm{s}$, constraints from cosmology, in particular Big Bang Nucleosynthesis, become relevant.
\label{fig:FIPs_sketch}}
\end{figure}

Broadly speaking, constraints on FIPs stem from two types of experiments: fixed-target experiments searching for the scattering or decay of FIPs in a downstream detector and collider experiments searching for displaced signatures. Fixed-target experiments operate at much higher effective luminosity (i.e.\ much larger number of collisions) but lower centre-of-mass energy. They achieve very low backgrounds due to the long distance between the interaction point and the decay/scattering volume. As a result, both proton and electron beam dump experiments have unique sensitivity to FIPs with tiny couplings and decay lengths greater than $1\,\mathrm{m}$. For shorter decay lengths, there are strong constraints from existing collider experiments, such as Belle II (see~\cite{Belle-II:2018jsg}) and LHCb (see~\cite{Alimena:2019zri}), which are projected to substantially improve their reach with increasing integrated luminosity in the coming decade. The complementarity of fixed-target experiments and collider experiments is illustrated in Figure~\ref{fig:FIPs_sketch}.

To extend the reach of the LHC to longer lifetimes, various new experiments have been proposed. These can be divided into detectors placed in the forward direction and detectors placed at large angle. The FASER experiment~\cite{FASER:2018eoc} provides a proof of principle of an experiment from the former category, and much more sensitive experiments could be built at a dedicated Forward Physics Facility (FPF)~\cite{Feng:2022inv}. In the following, FASER2 will be taken as a representative proposed forward experiment, noting that similar sensitivities may be achieved by competing proposals such as FACET~\cite{Cerci:2021nlb}. Among the various proposed experiments at large angle are CODEX-b~\cite{Aielli:2019ivi}, ANUBIS~\cite{Bauer:2019vqk} and MATHUSLA~\cite{Curtin:2018mvb}. Out of these, MATHUSLA is most ambitious, requiring significant civil engineering, whereas the first two may be accommodated within existing facilities. In the following CODEX-b will therefore be considered as a representative of LHC large angle (LHC-LA) experiments, noting that similar sensitivities may be achieved by ANUBIS and significantly better sensitivities by MATHUSLA~\cite{Antel:2023hkf}.

The two on-going CERN fixed-target experiments most relevant in the context of FIPs physics are NA62 (see section~\ref{sec:current_status}), which takes proton beam data both in kaon mode and in beam-dump mode, and NA64~\cite{Banerjee:2019pds,NA64:2023wbi,NA64SPSC}, which searches for missing energy in an active dump using either a $(100-150)\,\mathrm{GeV}$ electron beam or a $160\,\mathrm{GeV}$ muon beam~\cite{Gninenko:2640930,NA64SPSC}. Both experiments already achieve world-leading sensitivity to certain FIPs models and will continue taking and analysing data in coming years. The two most advanced proposals for fixed-target experiments outside of CERN are DarkQuest~\cite{Apyan:2022tsd}, which is a proposed upgrade of the running SpinQuest experiment using a 120 GeV proton beam at Fermilab, and LDMX~\cite{LDMX:2018cma}, which will initially operate with a 4 GeV electron beam at SLAC, with a subsequent upgrade to 8 GeV beam energy discussed in Ref.~\cite{Akesson:2022vza}. While the former would be able to probe similar models as NA62 and the various proposed ECN3 experiments, the latter resembles NA64 and focuses primarily on missing energy signatures from (meta)stable FIPs. However, the lower beam energy of DarkQuest means that it will be unable to achieve the same sensitivity as the ECN3 experiments to FIPs above the GeV scale, as well as to FIPs produced dominantly in $B$ meson decays.

Finally, particles that couple primarily to photons can also be produced at the European XFEL at DESY. An experiment to detect such particles using an optical dump, called LUXE-NPOD, has been proposed in~\cite{Bai:2021gbm}.

\subsubsection{Results}

To illustrate the sensitivity that can be achieved by the ECN3 experiments, a number of benchmark physics cases (BCs) are considered. They have been first proposed in~\cite{Beacham:2019nyx}, further refined in~\cite{Agrawal:2021dbo,Antel:2023hkf}, and have since become a community standard. Specifically, they include models of dark photons with visible (BC1) and invisible (BC2) decays, dark scalars (BC4), heavy neutral leptons with electron (BC6) and tau lepton (BC8) couplings, and axion-like particles with photon (BC9) and fermion (BC10) couplings. This subset of benchmarks has been selected to capture the full range of production modes and final states that are relevant for the ECN3 experiments in order to highlight the unique opportunities and facilitate the comparison between the proposals. Some further benchmarks can be found in~\cite{Antel:2023hkf}. 

While in many cases there exist sizeable theoretical uncertainties regarding the production and decay modes of these particles, considerable effort has been made in the past years to reduce these uncertainties and define a common framework for all experiments. 
That said, there still remain some differences between the experiments both in the underlying assumptions and in the concrete numerical implementations.
This should be kept in mind when interpreting the sensitivity projections.

\begin{figure}[t!]
\begin{center}
\includegraphics[width=0.6\textwidth]{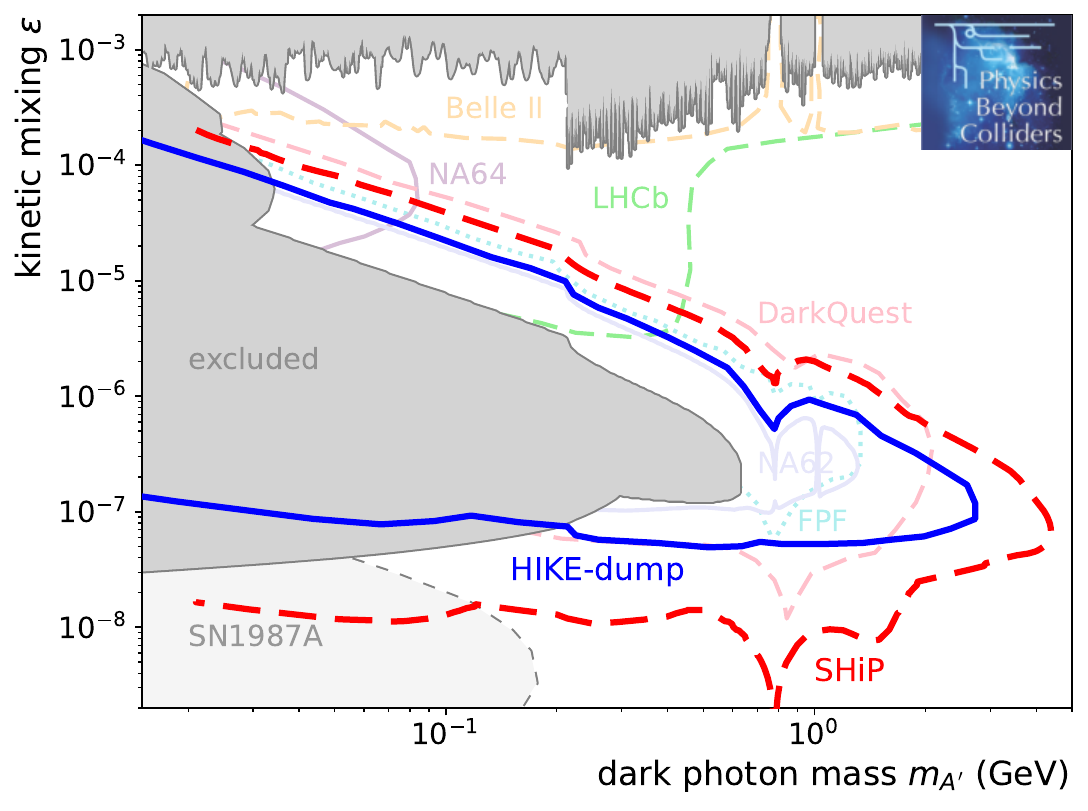}
\caption{\small
Sensitivity projections for the FIP Physics Centre~(FPC) benchmark BC1 (dark photons with kinetic mixing).
\label{fig:BC1}}
\end{center}
\end{figure}

\begin{figure}[t!]
\begin{center}
\includegraphics[width=0.6\textwidth]{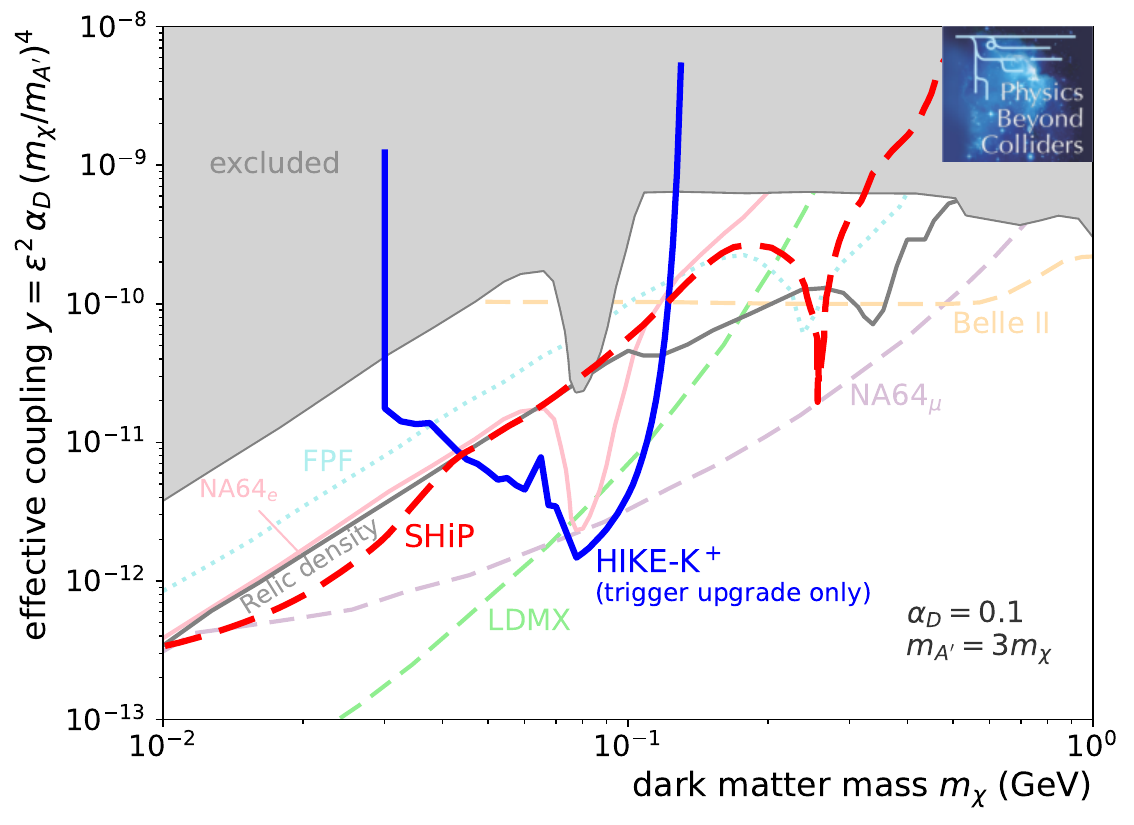}
\caption{\small
Sensitivity projections for the FPC benchmark BC2 (dark photons decaying into stable dark fermions).  SHiP searches for the scattering of the dark fermions off the electrons of the SND target, while HIKE searches for $K^+ \to \pi^+ \pi^0$ followed by $\pi^0 \to A^\prime(\to \text{invisible}) \gamma$. Existing constraints and projections from~\cite{Antel:2023hkf} have been updated to include new NA64 results from~\cite{NA64:2023wbi,NA64SPSC}. The curve labelled FPF corresponds to the FLArE proposal~\cite{Kling:2022ykt}. Note that there is no NA62 projection due to the lack of a dedicated trigger. The HIKE curve is extrapolated from NA62 data assuming the HIKE trigger upgrade only. Further potential improvements in HIKE sensitivity from background reduction thanks to detector upgrades are under study~\cite{bib:HIKE_PROPOSAL_2023}. 
\label{fig:BC2}}
\end{center}
\end{figure}

\begin{figure}[t!]
\begin{center}
\includegraphics[width=0.6\textwidth]{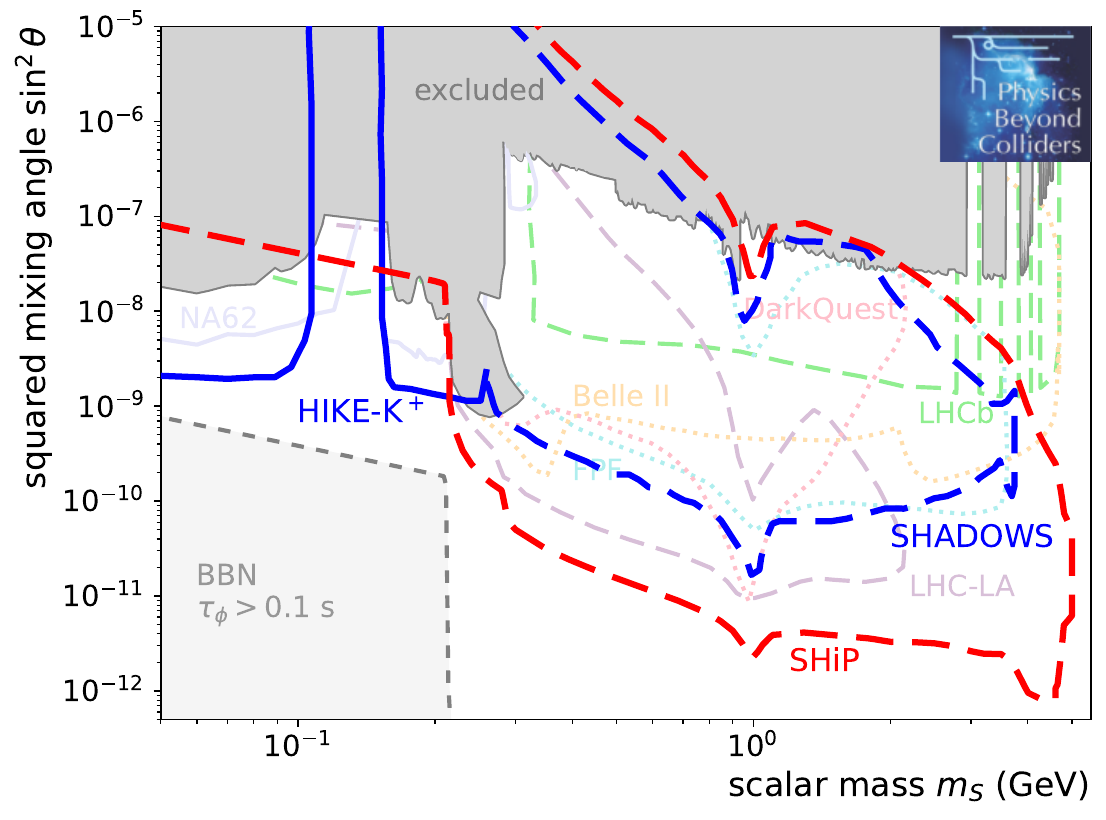}
\caption{\small
Sensitivity projections for the FPC benchmark BC4 (dark scalars with Higgs mixing). The inclusive branching ratio for $B \to S + X$ is calculated following the appendix of~\cite{Winkler:2018qyg}, while the decay kinematics are taken from the decay $B^+ \to K^+ S$.
\label{fig:BC4}}
\end{center}
\end{figure}

\begin{figure}[t!]
\begin{center}
\includegraphics[width=0.6\textwidth]{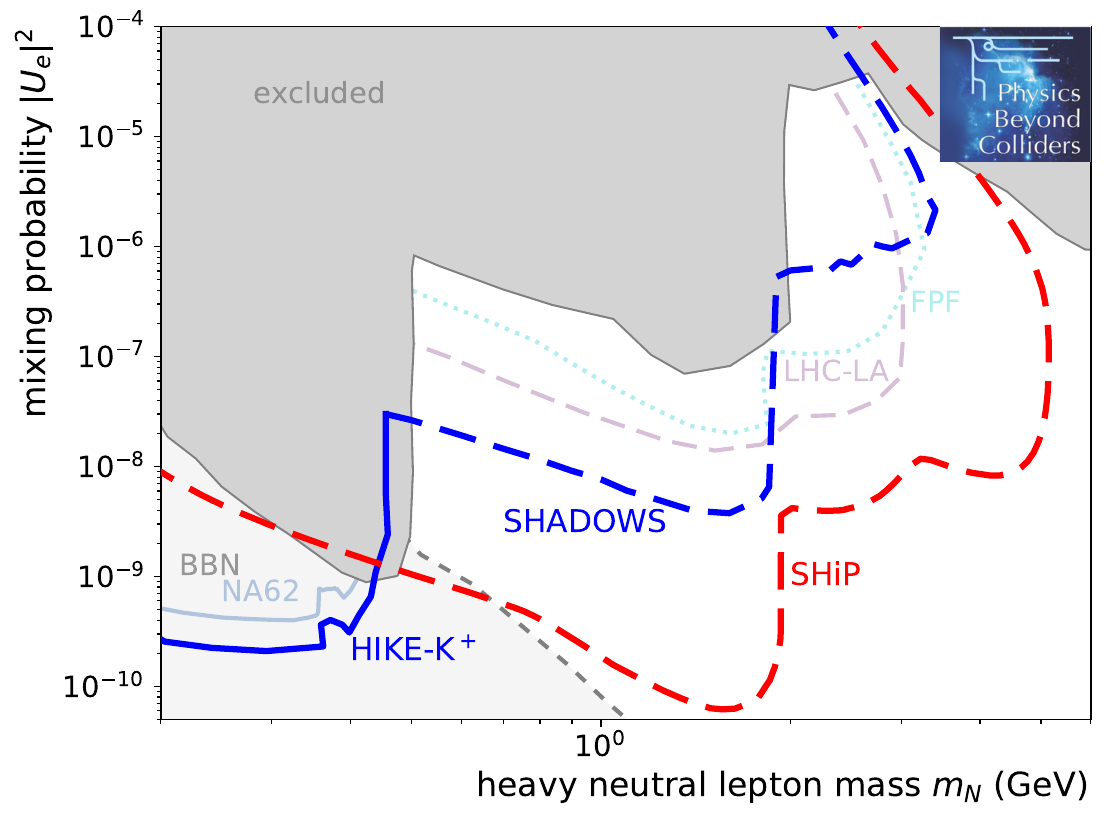}
\caption{\small
Sensitivity projections for the FPC benchmark BC6 (heavy neutral leptons with electron mixing).
\label{fig:BC6}}
\end{center}
\end{figure}

\begin{figure}[t!]
\begin{center}
\includegraphics[width=0.6\textwidth]{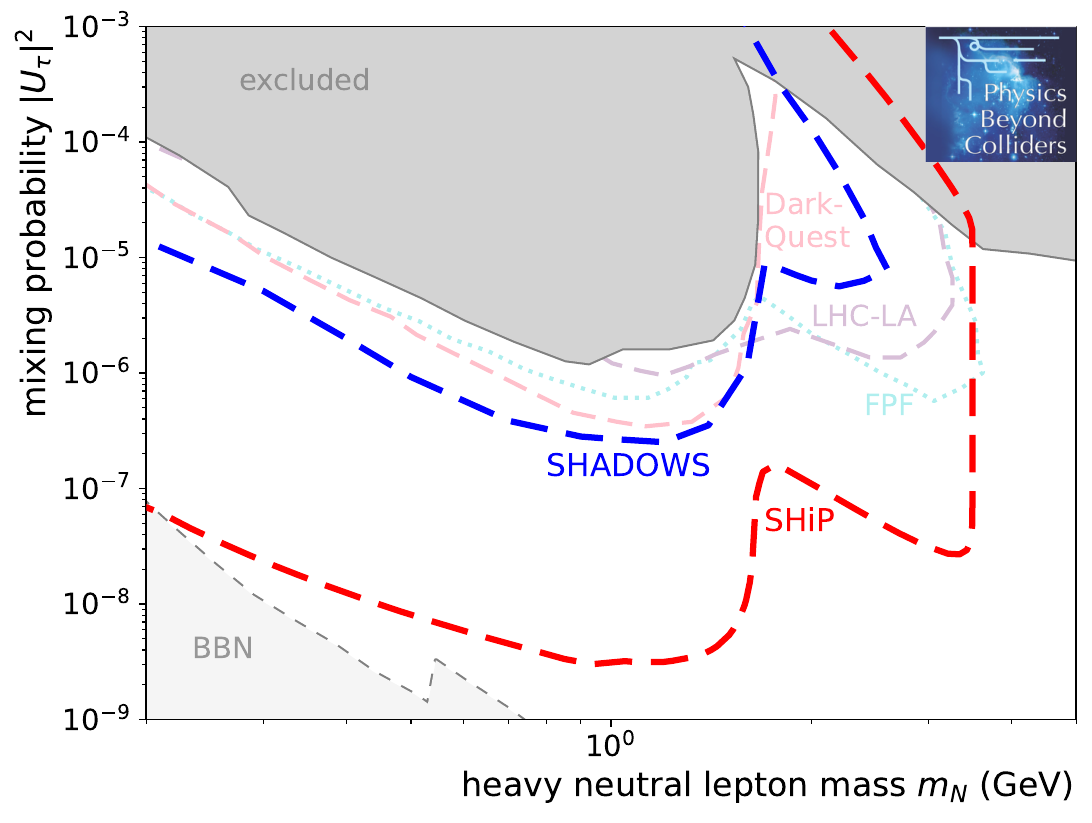}
\caption{\small
Sensitivity projections for the FPC benchmark BC8 (heavy neutral leptons with tau mixing).
\label{fig:BC8}}
\end{center}
\end{figure}

\begin{figure}[t!]
\begin{center}
\includegraphics[width=0.6\textwidth]{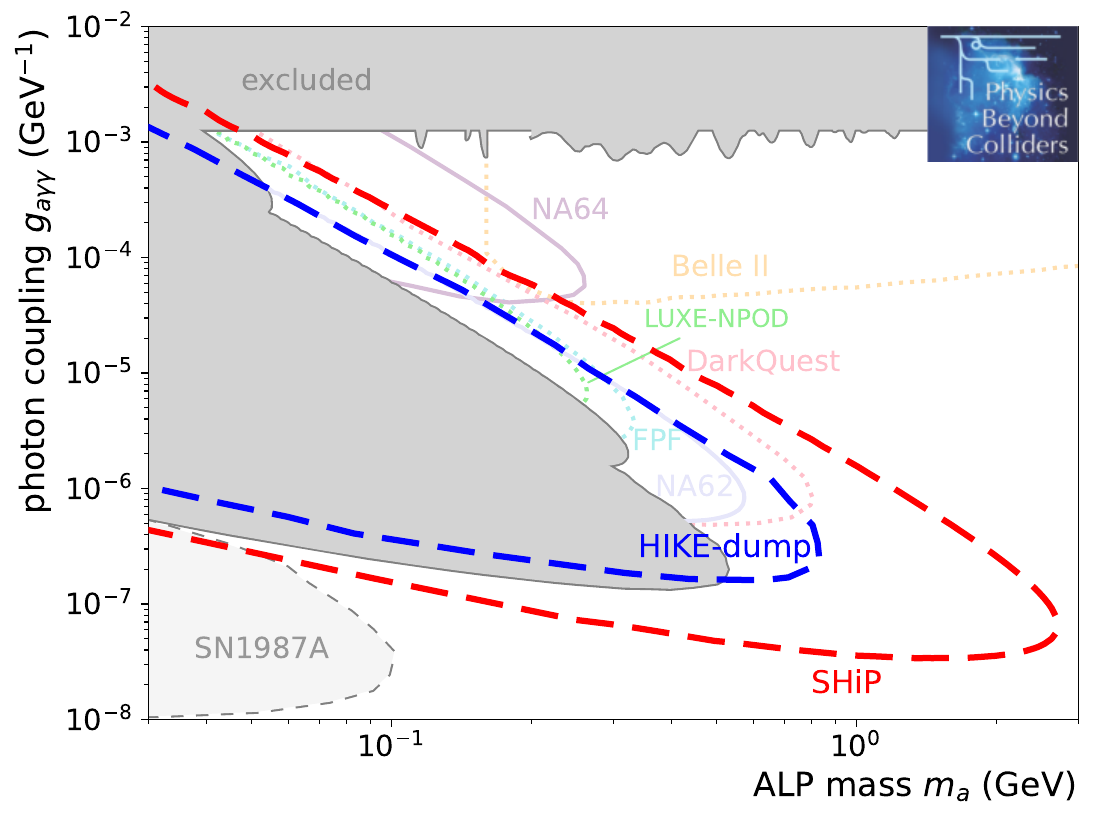}
\caption{\small
Sensitivity projections for the FPC benchmark BC9 (axion-like particles with photon couplings). The SN1987A constraint is taken from Ref.~\cite{Ertas:2020xcc}.
\label{fig:BC9}}
\end{center}
\end{figure}

\begin{figure}[t!]
\begin{center}
\includegraphics[width=0.6\textwidth]{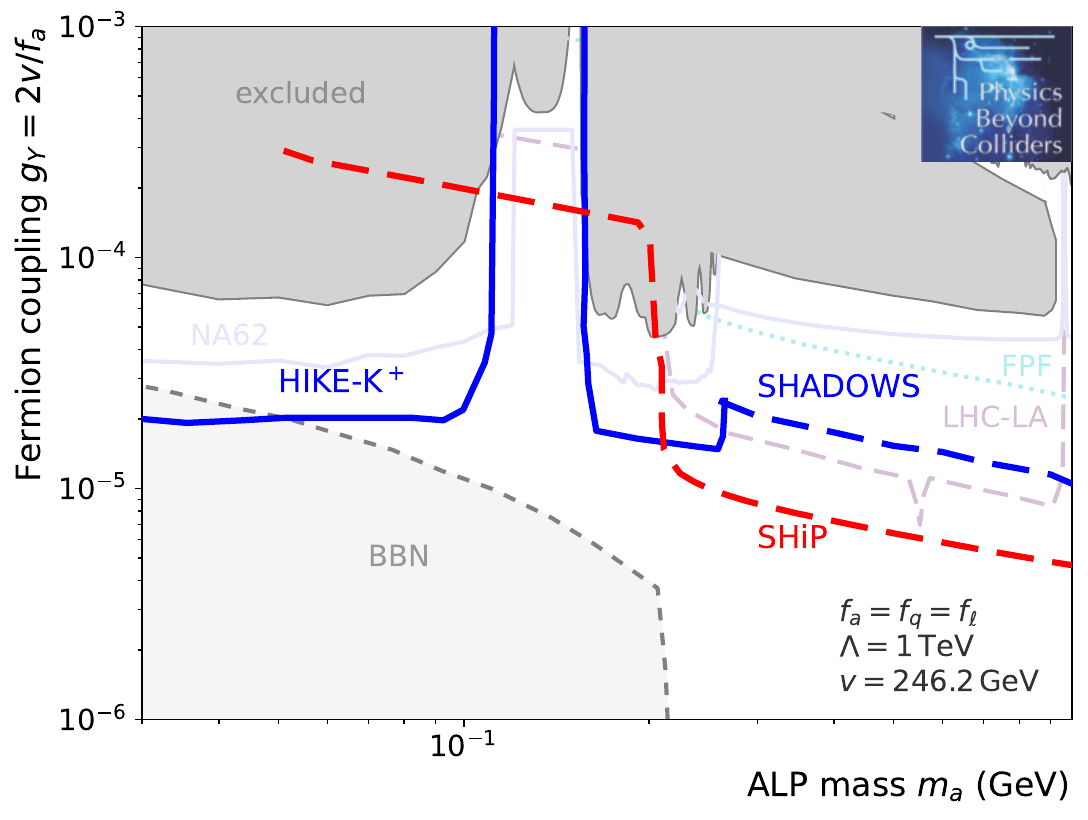}
\caption{\small
Sensitivity projections for the FPC benchmark BC10 (axion-like particles with fermion couplings). The plot range is restricted to $m_a < \eta^\prime$ to avoid large theoretical uncertainties at higher masses.
\label{fig:BC10}}
\end{center}
\end{figure}

The results are shown in Figures~\ref{fig:BC1}--\ref{fig:BC10} with a layout adapted from~\cite{Ferber:2023iso}. All curves are exclusion limits at 90\% confidence level. For the various projections, the line style reflects the maturity of the background estimates: solid lines correspond to background estimates based on the extrapolation of existing data sets, dashed lines indicate background estimates based on
full Monte Carlo simulations, and dotted lines represent projections based on toy Monte Carlo simulations or on the assumption that backgrounds are negligible. In these figures, existing constraints and projections for non-ECN3 experiments are (unless mentioned otherwise) taken from~\cite{Antel:2023hkf}\footnote{Projections for LHC experiments are shown for the HL-LHC era (FPF integrated luminosity of $3 \, \mathrm{ab^{-1}}$), for Belle II assuming $50\,\mathrm{ab^{-1}}$ (except for BC1, where $20\,\mathrm{fb^{-1}}$ are assumed), for NA62 assuming $10^{18}$ PoT in dump mode and $10^{19}$ PoT in kaon mode, for NA64 assuming $1 \times 10^{13}$ electrons and $2 \times 10^{13}$ muons on target, for DarkQuest assuming $10^{20}$ PoT, for LDMX assuming $1.6 \times 10^{15}$ electrons on target in a 8 GeV beam and for LUXE-NPOD assuming a 40 TW laser operating for one year.}, see there for additional details and references. The ECN3 sensitivity projections are for
the baseline 
detector designs and reference integrated intensities given in Section~\ref{sec:OPScenarii}. When alternative designs are considered by the experiments, the corresponding sensitivity curves are given in the 
proposals.

For all benchmarks under consideration, the proposed ECN3 high-intensity facility has the potential to make CERN the world leader in the search for MeV-GeV FIPs, improving existing sensitivities by orders of magnitude and offering unparalleled opportunities for discovery.

\FloatBarrier

\subsection{Flavour physics}\label{subsec:flavor}
The flavour sector of the SM has a puzzling and rich flavour structure featuring significant intrinsic hierarchies and the only known source of CP violation.
Rare decays of kaons provide a vital and powerful test of this sector. Notably, high precision measurements of processes that are strongly suppressed in the SM -- often related to an underlying  symmetry structure -- enable to probe scales far in excess of the direct reach of existing colliders, cf., e.g.,~\cite{Aebischer:2022vky}. 

Currently, ECN3 hosts the NA62 experiment, a world-leading multi-purpose kaon experiment whose primary goal is the measurement of ultra-rare $K^+$ decays, most notably $K^+ \to \pi^+ \nu \bar{\nu}$ for which clean theory predictions are available. The most recent measurement of the $K^+\to\pi^+\nu \bar\nu$ decay rate based on the NA62 Run~1 (2016--18) dataset~\cite{NA62:2021zjw} achieves a precision of $ \approx~\!\!40~\%$, expected to be improved to $ \approx~\!\!15-20~\%$ until LS3. Furthermore, the NA62 experiment pursues a broad programme of rare $K^+$ and $\pi^0$ decay measurements, precision tests of low-energy QCD, precision tests of lepton universality, searches of lepton flavour/number violation, and searches for production and decays of hidden-sector mediators in $K^+$ and $\pi^0$ decays and in beam-dump mode. 

The HIKE project would bring the above rare $K^+$ and $\pi^0$ decay programme to a new level of precision with respect to NA62, improving for example the precision of $\text{BR}(K^+ \to \pi^+ \nu \bar{\nu})$ by a factor of $\approx 3-4$.
In addition, HIKE will accomplish a similarly broad rare $K_L$ decay programme. This section details the unique opportunities of the combined $K^+$ and $K_L$ programme to probe BSM physics and the potential impact on fundamental physics.

The overview starts with a discussion of the state-of-the-art at both the theoretical and the experimental front. It identifies two scenarios of particular interest: violation of unitarity of the Cabibbo-Kobayashi-Maskawa (CKM) matrix and violation of lepton flavour universality (LFU), for which the sensitivities and potential impact of HIKE are evaluated. Further science goals are then considered in the international landscape.

\subsubsection{State of the art}

\subsubsection*{Recent developments in SM predictions for rare kaon decays}

The amplitudes for flavour-changing neutral current (FCNC) kaon decays
receive contributions from physics at several energy scales. Within
the SM, the relevant scales are the electroweak scale, the charm-quark
threshold, and the hadronic scale of the order of the kaon
mass. Effective field theory (EFT) techniques are used to factorize the
amplitudes into Wilson coefficients (typically calculated in
renormalization-group improved perturbation theory) and matrix
elements of effective operators. For decays that are mediated (or at
least dominated) by a $Z$ penguin diagram, the Glashow--Iliopoulos--Maiani (GIM) mechanism suppresses the low-energy contributions, and perturbative uncertainties are well under control.

If precise knowledge on the hadronic matrix is available, the corresponding decay is ``clean''. The prime examples are the rare decays $K_L \to \pi^0 \nu \bar\nu$ and $K^+ \to \pi^+ \nu \bar\nu$. The SM predictions for their branching ratios are exceptionally clean since the requisite hadronic matrix elements can be extracted from the well measured $K \to \pi \ell \nu_\ell$ modes, including higher-order chiral corrections~\cite{Mescia:2007kn}. Correspondingly, the next-to-leading logarithmic QCD and the next-to-leading logarithmic QED corrections have been calculated, resulting in a residual (non-parametric) theory uncertainty at the percent
level~\cite{Buchalla:1992zm, Misiak:1999yg, Buras:2005gr,
  Buras:2006gb, Brod:2008ss, Brod:2010hi}.  The rare decays $K_L \to \pi^0 \ell^+ \ell^-$ are less clean, due to the contributions of the photon penguin, but provide important probes of non-vectorial
contributions of BSM physics~\cite{Mescia:2006jd}.

The rare decays $K \to \mu^+\mu^-$ are dominated by long-distance (LD) contributions which makes their use as a precision probe of BSM physics challenging. 
Expected progress in the lattice determination of the dominant two-photon intermediate state might change this picture in the future~\cite{Christ:2020bzb}. 
Interestingly, it has been pointed out recently that the direct CP-violating, short-distance contribution to $K_S \to \mu^+ \mu^-$ can, in principle, be extracted experimentally using $K_L - K_S$ interference data~\cite{DAmbrosio:2017klp,  Dery:2021mct}. Including the effects of indirect CP violation~\cite{Brod:2022khx} and recently obtained information on a relative strong phase~\cite{Dery:2022yqc}, the corresponding branching
ratio is now also predicted with a residual theory uncertainty at the
percent level.

\subsubsection*{CKM unitarity measurements and the Cabbibo angle anomaly}

The main (semi)leptonic decay modes of relevance for this topic are $K\to\pi\ell\nu$ ($K_{\ell 3}$) and $K\to\ell\nu$ ($K_{\ell 2}$). Thanks to a global effort involving several experiments, lattice QCD simulations and analytical QCD calculations, an impressive precision has been achieved for these modes, typically below the percent level~\cite{Bryman:2021teu,FlavourLatticeAveragingGroupFLAG:2021npn,Seng:2021nar,Cirigliano:2022yyo}. 

In the context of the SM, these results offer the possibility of the most precise extractions of the $V_{us}$ CKM element using $\Gamma_{K_{\ell 3}}$ rate and the $\Gamma_{K_{\mu 2}}/\Gamma_{\pi_{\mu 2}}$ ratio, as well as a clean window to interesting QCD physics~\cite{FlaviaNetWorkingGrouponKaonDecays:2008hpm,FlaviaNetWorkingGrouponKaonDecays:2010lot}. 
On the other hand, precision measurements of these decays represent an important BSM physics probe, e.g., in scenarios with SM-like flavour and CP structure. 
As with rare decays, EFTs represent a very useful setup that covers a vast variety of BSM physics models. A particularly simple and interesting case is the $U(3)^5$-symmetric one\footnote{Here $U(3)^5$ refers to the flavour symmetry of the gauge part of the SM Lagrangian. Each $U(3)$ factor refers to a rotation in generation space of the gauge fermion multiplets ($q_L,u_R,d_R,\ell_L,e_R$).}, where all BSM effects are absorbed in the phenomenological CKM elements~\cite{Cirigliano:2009wk}. Thus the only BSM probe is a CKM unitarity test: $|V_{ud}|^2 + |V_{us}|^2 + |V_{ub}|^2 = 1$, where the last term can be neglected in practice. If the $U(3)^5$ symmetry is not imposed, a rich variety of effects take place, such as Lepton Flavour Violation~(LFV) effects or non-standard currents, which would affect differently each decay mode.

Until a few years ago, there was a good agreement between the $V_{us}$ values obtained from $\Gamma_{K_{\ell 3}}$, $\Gamma_{K_{\mu 2}}/\Gamma_{\pi_{\mu 2}}$, and with the CKM unitarity prediction (using the $\beta$-decay $V_{ud}$ value). This overall agreement entailed strong constraints on BSM physics, corresponding to effective TeV scales, with an interesting synergy with LHC direct searches~\cite{FlaviaNetWorkingGrouponKaonDecays:2010lot,Gonzalez-Alonso:2016etj}. 
However, recent theoretical and experimental improvements in kaon and beta-decay physics moved apart the various $V_{us}$ determinations~\cite{Bryman:2021teu,FlavourLatticeAveragingGroupFLAG:2021npn,Seng:2018yzq,Seng:2018qru,Gorchtein:2018fxl}, yielding an interesting yet unclear situation, known as the Cabibbo angle anomaly~\cite{Grossman:2019bzp}. This intriguing situation has sparked an intense activity in model building, EFT studies and the reevaluation of the SM contributions, see, e.g.,~\cite{Coutinho:2019aiy,Cirigliano:2021yto,Seng:2022wcw}.

\begin{figure}[t!]
\centering
\includegraphics[width=0.90\textwidth]{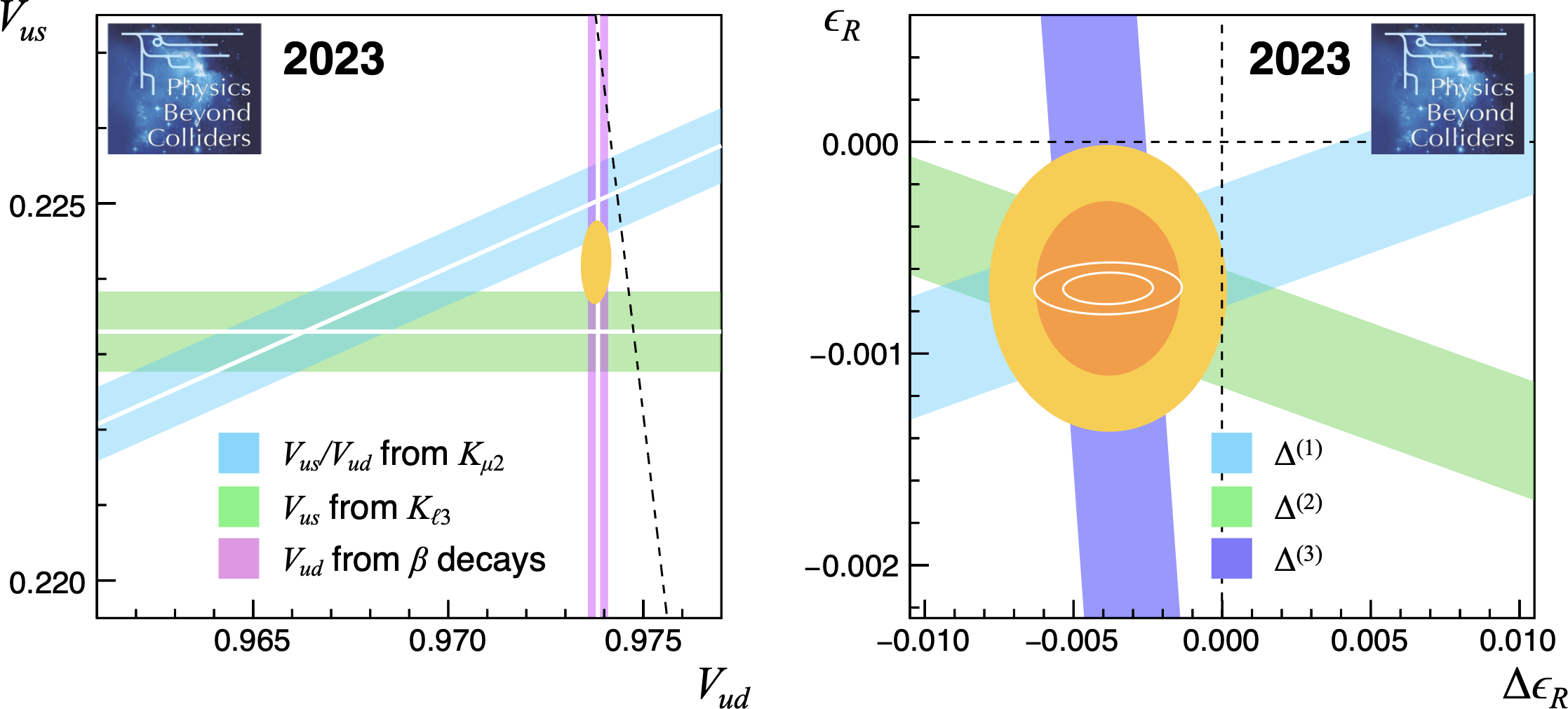}
\caption{\small
Status of first-row CKM unitarity in 2023. Left: Measurements of $V_{us}$, $V_{us}/V_{ud}$, and $V_{ud}$ and relation to CKM unitarity. The yellow ellipse represents the 68.27~\% CL confidence interval from a fit for the best values of $V_{us}$ and $V_{ud}$. The unitarity curve is illustrated by the dashed line.
Right: Constraints on right-handed currents from observed unitarity deficits. The yellow and orange ellipses illustrate the 95.45~\% CL and 68.27~\% CL confidence intervals from a fit.
The white ellipses indicate the much smaller confidence intervals that would be obtained if the only source of uncertainty were from the kaon decay measurements.
} 
\label{fig:ckm_2023}
\end{figure}

Figure~\ref{fig:ckm_2023}, left, shows the current experimental constraints in the $V_{us}$-$V_{ud}$ plane.
The tension between the values of $V_{us}$ from $K_{\mu2}$ and $K_{\ell3}$ decays is seen in the fact that corresponding bands do not intersect at a common point with the band for $V_{ud}$ from nuclear and neutron beta decays. 
The right panel of Figure~\ref{fig:ckm_2023} illustrates the constraints from CKM unitarity on the contributions to the leptonic and semileptonic kaon decay amplitudes from right-handed quark currents, following the analysis of~\cite{Cirigliano:2022yyo}. Specifically, denoting by $\epsilon_R$ the contributions of right-handed currents to the decays of non-strange quarks and by $\epsilon_R^{(s)}$ those to the decays of strange quarks, the following relations to the unitarity deficits can be written:
\begin{align}
\Delta_{\rm CKM}^{(1)} &\equiv |V_{ud}^{\beta}|^2 + |V_{us}^{K_{\ell3}}|^2 - 1 = 2\epsilon_R + 2\Delta\epsilon_R V_{us}^2,\\
\Delta_{\rm CKM}^{(2)} &\equiv |V_{ud}^\beta |^2\, \left[1 + (|V_{us}/V_{ud}|^{K_{\mu2}})^2\right] - 1 = 2\epsilon_R - 2\Delta\epsilon_R V_{us}^2,\\
\Delta_{\rm CKM}^{(3)} &\equiv |V_{us}^{K_{\ell3}}|^2\, \left[(|V_{us}/V_{ud}|^{K_{\mu2}})^{-2} + 1\right] - 1 
= 2\epsilon_R - 2\Delta\epsilon_R (2 - V_{us}^2),
\end{align}
with $\Delta\epsilon_R \equiv \epsilon_R - \epsilon_R^{(s)}$. 
The colored bands in the plot show the constraints from the different constructions of the unitarity deficit in the plane of $\epsilon_R$ vs.\ $\Delta\epsilon_R$; note that the bands intersect by construction.

\subsubsection*{Lepton flavour universality violation in rare kaon decays}

In the SM the three lepton flavours ($e$, $\mu$ and $\tau$) have exactly the same gauge interactions and are distinguished only through their couplings to the Higgs field and hence the charged lepton masses. Models of BSM physics, on the other hand, do not necessarily conform to the LFU hypothesis and may thereby induce subtle differences between the different generations that cannot be attributed to the different masses. Among the most sensitive probes of these differences are rare kaon decays with electrons, muons or neutrinos in the final state.

The FCNC decay $s\to d$ can be described with the effective Hamiltonian
\begin{equation}\label{eq:Heff}
\mathcal{H}_{\rm eff}=-\frac{4G_F}{\sqrt{2}}V_{td}V_{ts}^*\frac{\alpha_e}{4\pi}\sum_k C_k^{\ell}O_k^{\ell}\,,
\end{equation}
where $G_F$ denotes Fermi's constant, $\alpha_e$ the fine-structure constant and the Wilson coefficients $C_k^\ell$ multiply the effective operators $O_k^\ell$. The present discussion can be limited to the following sub-set of effective operators motivated by various anomalies in $B$ physics~\cite{Alguero:2023jeh}:
\begin{align}
&{O}_9^{\ell} = (\bar{s} \gamma_\mu P_L d)\,(\bar{\ell}\gamma^\mu \ell)\,,
&&{O}_{10}^{\ell} = (\bar{s} \gamma_\mu P_L d)\,(\bar{\ell}\gamma^\mu\gamma_5 \ell)\,,&&
  {O}_L^{\ell} = (\bar{s} \gamma_\mu P_L d)\,(\bar{\nu}_\ell\,\gamma^\mu(1-\gamma_5)\, \nu_\ell)\,.
\end{align}
For the study of BSM physics contributions to $\delta C_k^\ell$ it is possible to reduce the set of operators further by considering only scenarios where the neutral and charged leptons are related by SU(2)$_L$ gauge symmetry, such that $\delta C_{L}^{\ell} \equiv \delta C_9^{\ell} = - \delta C_{10}^{\ell}$.


\begin{figure}[t!]
\centering
\includegraphics[width=0.5\textwidth]{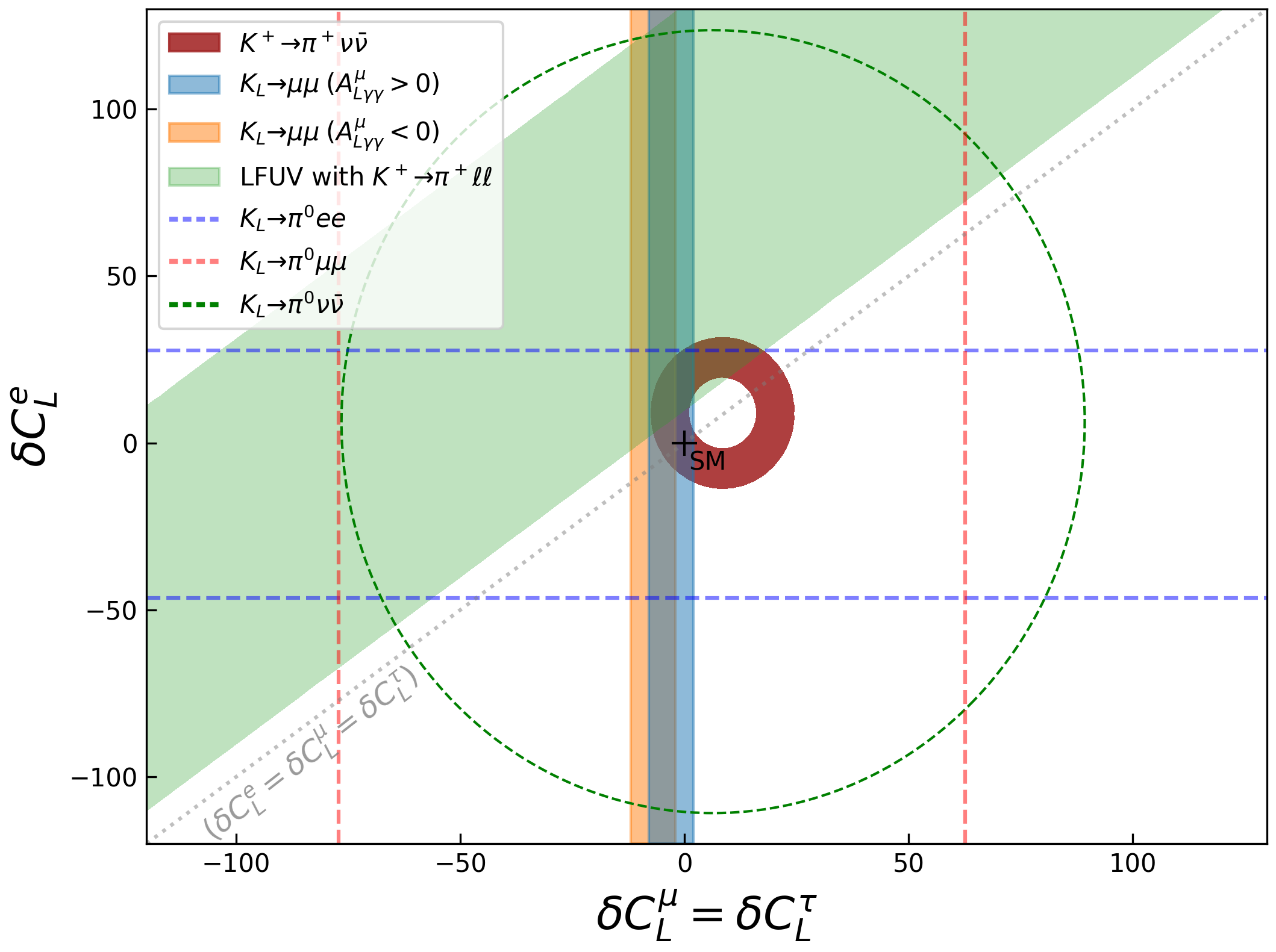}%
\includegraphics[width=0.5\textwidth]{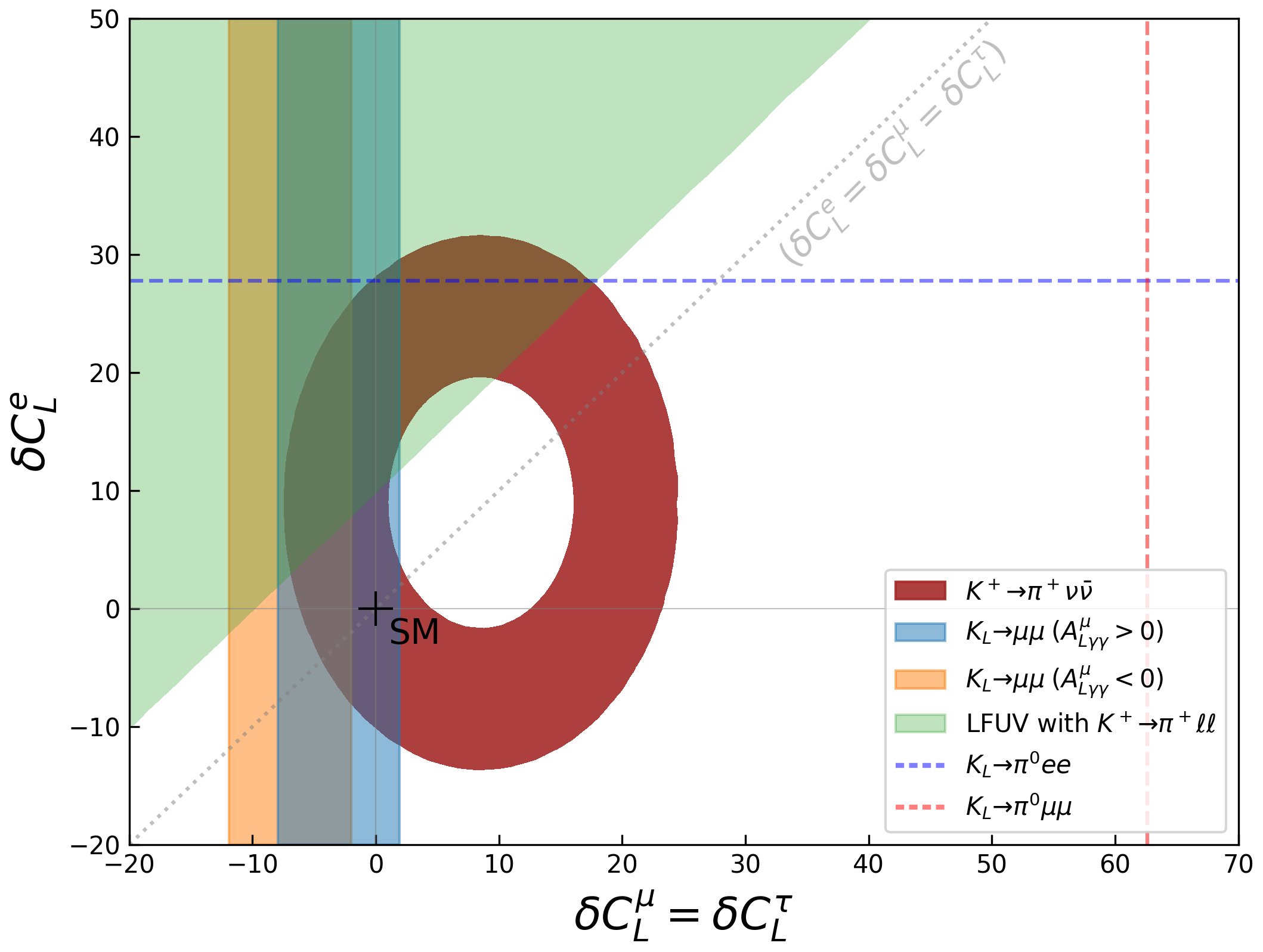}
\caption{\small
Current bounds on the BSM contributions $\delta C$ to the Wilson coefficients $C_L^e$ and $C_L^\mu = C_L^\tau$ from individual kaon observables. The right panel is a zoomed version of the left panel. See Figure~7 in~\cite{DAmbrosio:2022kvb} (from where the figure is taken) for further information. 
It can be clearly seen that in particular the decays $K^{+}\to\pi^{+}\nu\bar{\nu}$ and $K_{L}\to \mu\mu$ provide significant sensitivity to these BSM effects. Figure~\ref{fig:fitcomparison} shows the (combined) improvements possible with HIKE. 
\label{fig:all_obs_individually}}
\end{figure}

The individual constraints on $\delta C_L^e$ and $\delta C_L^\mu = \delta C_L^\tau$ are shown in Figure~\ref{fig:all_obs_individually} (taken from~\cite{DAmbrosio:2022kvb}), where it is readily seen that the main constraining observables are BR($K^+\to \pi^+ \nu\bar\nu$) and BR($K_L\to\mu\bar\mu$),  where for the latter the unknown LD sign plays an important role. For theories with LFU New Physics effects, such that $\delta C_L^e = \delta C_L^\mu = \delta C_L^\tau$, the NA62 measurement of $K^+ \to \pi^+ \nu \bar \nu$ already puts rather strong constraints on possible lepton-flavour universal BSM effects. However, these constraints are relaxed considerably if LFU-violating BSM effects are allowed.

\subsubsection{HIKE sensitivity}

The potential impact of HIKE on the physics landscape described above is discussed in details in the HIKE proposal \cite{bib:HIKE_PROPOSAL_2023} and shortly summarized below. 

\subsubsection*{CKM unitarity}

Given that the dominant contribution to the uncertainty on the measurement of the first-row unitarity deficit is from the determination of $V_{ud}$ from nuclear beta decays, the fact that experimental and theoretical sources contribute approximately equally to the current overall uncertainty on $V_{us}$, and the substantial set of kaon decay measurements in world data,
HIKE can contribute to the understanding of the anomaly mainly by providing experimental confirmation of the leptonic and semileptonic branching ratio values, to help to exclude an experimental origin. Indeed, the experimental situation is complex, with a few measurements of the branching ratios playing an outsize role in the overall determination of $V_{us}$. 

For charged kaon decays, HIKE Phase 1 is poised to make a significant impact. Not only does the value of $V_{us}/V_{ud}$ used in the unitarity analysis derive from a single measurement of ${\rm BR}(K_{\mu2})$ with a 0.27~\% total uncertainty~\cite{KLOE:2005xes}; this measurement also impacts the normalization of all other branching ratio measurements in the $K^+$ decay rate fit to world data, e.g., by the PDG or the analysis of~\cite{FlaviaNetWorkingGrouponKaonDecays:2010lot}. The importance of the measurement of the ratio ${\rm BR}(K_{\mu3})/{\rm BR}(K_{\mu2})$ to settle this question is discussed in~\cite{Cirigliano:2022yyo}. HIKE could also make a very precise measurement of ${\rm BR}(K_{\mu3})/{\rm BR}(K_{e3})$, an important test of LFU, as well as of other important ratios amenable to measurement with good precision, such as ${\rm BR}(K_{e3})/{\rm BR}(K_{\pi2})$, ${\rm BR}(K_{\mu3})/{\rm BR}(K_{\pi2})$, and ${\rm BR}(K_{\pi2})/{\rm BR}(K_{\mu2})$, possibly with a unified analysis.
With the ratios between the widths for four of the six main $K^+$ decay modes thus determined, current world data on the branching ratios for $K_{\mu2}$, $K_{\pi2}$, $K_{e3}$, and $K_{\mu3}$ can be omitted from the $K^+$ rate fit, allowing HIKE to make a nearly independent determination of the $K_{\mu2}$ and $K_{\ell3}$ branching ratios.

The limiting systematic uncertainties are difficult to predict, but the HIKE sensitivity can be estimated on the basis of past experience. NA48/2 measured the ratios
${\rm BR}(K_{e3})/{\rm BR}(K_{\pi2})$, 
${\rm BR}(K_{\mu3})/{\rm BR}(K_{\pi2})$, and 
${\rm BR}(K_{e3})/{\rm BR}(K_{\mu3})$ at the level of 0.4~\%~\cite{NA482:2006vnw}. 
It should be easy for HIKE to match or exceed this precision, especially for the ratios 
${\rm BR}(K_{\mu3})/{\rm BR}(K_{\mu2})$ and ${\rm BR}(K_{\mu3})/{\rm BR}(K_{e3})$, for which significant cancellations of systematic uncertainties are expected. 
The HIKE Phase-1 sensitivity estimate
assumes 0.2~\% total uncertainty for the measurements of these two ratios, and 0.4~\% total uncertainty for the measurements of ${\rm BR}(K_{e3})/{\rm BR}(K_{\pi2})$, ${\rm BR}(K_{\mu3})/{\rm BR}(K_{\pi2})$, and ${\rm BR}(K_{\pi2})/{\rm BR}(K_{\mu2})$.

The potential improvements to the knowledge of the semileptonic branching ratios for $K_L$ decays from HIKE Phase 2 are more challenging to evaluate. 
One possible set of HIKE Phase-2 measurements that could be added to the current $K_L$ world data set to improve the precision of the $K_{\ell3}$ branching ratios consists of high-precision measurements of ${\rm BR}(K_{e3})/{\rm BR}(K_{\mu3})$ and ${\rm BR}(\pi^+\pi^-)/{\rm BR}(K_{e3})$, as well as a good measurement of ${\rm BR}(\pi^+\pi^-)/{\rm BR}(\pi^+\pi^-\pi^0)$ with less stringent precision requirements, to assist in normalization via the global fit. The corresponding Phase-2 sensitivity estimates assume total uncertainties of 0.3~\%, 0.4~\%, and 0.6~\%, respectively. These are consistent with or slightly more conservative than the assumptions for HIKE Phase 1. In particular, NA48 made a statistically dominated measurement of ${\rm BR}(\pi^+\pi^-)/{\rm BR}(K_{e3})$ with a systematic uncertainty of 0.3~\%~\cite{NA48:2006jeq}. 

\begin{figure}[t!]
\centering
\includegraphics[width=0.9\textwidth]{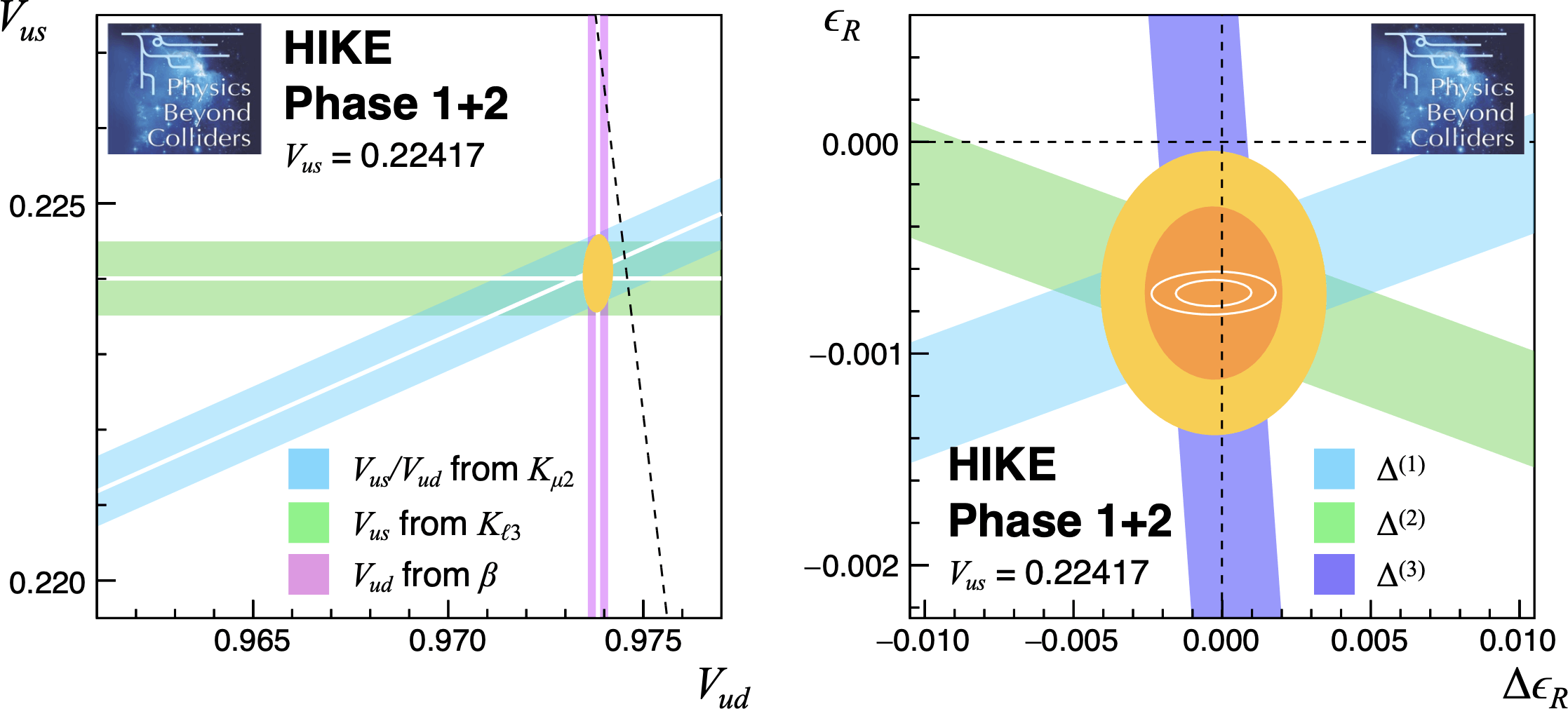}
\caption{\small
Status of first-row CKM unitarity in future scenario with measurements from HIKE Phases 1 and 2 confirming $V_{us} = 0.22417$.
Left: Measurements of $V_{us}$, $V_{us}/V_{ud}$, and $V_{ud}$ and relation to CKM unitarity.
Right: Constraints on right-handed currents from observed unitarity deficits.}
\label{fig:ckm_phase2}
\end{figure}

The impact of adding the HIKE measurements from both phases to the global fit in this scenario can be seen in Figure~\ref{fig:ckm_phase2}.
The increase in sensitivity from the reduced uncertainties for the branching ratios can be appreciated in the smaller size of the white ellipses.
Under the assumption that consistent results are obtained for $K_{\mu2}$ and $K_{\ell3}$, the values obtained for $V_{us}$ are perfectly consistent, indicating that if the unitarity deficit is attributed to right-handed currents, they must be SU(3) flavour universal. 
The level of exclusion of the point $\epsilon_R = \Delta\epsilon_R = 0$ is greatly decreased: the current $3.1\sigma$ evidence for right-handed currents is reduced to a mere $2.2\sigma$ curiosity.
In this scenario, while the kaon measurements are consistent, the unitarity deficit remains; the precision obtained in the kaon sector strongly motivates further progress on the determination of $V_{ud}$, especially in the theoretical calculation of the radiative corrections.

\subsubsection*{LFU violation}

\begin{table}[t]
\vspace{-5mm}
\renewcommand{\arraystretch}{1.39}
\begin{center}
\setlength\extrarowheight{1pt}
\scalebox{0.75}{
\hspace*{-2mm}
\begin{tabular}{llllc}\hline\hline
\bf{Observable} & \bf{SM prediction}& \bf{Experimental results} & \bf{Reference}& \bf{HIKE projections} \\ \hline
BR$(K^+\to \pi^+\nu\bar\nu)$    & $(7.86 \pm 0.61)\times 10^{-11}$  & $(10.6^{+4.0}_{-3.5} \pm 0.9 ) \times 10^{-11}$ & \cite{NA62:2021zjw}& 5\% (Phase 1) \\
LFUV($a_+^{\mu\mu}-a_+^{ee}$)&\multicolumn{1}{c}{0}&$-0.031\pm 0.017$&\cite{DAmbrosio:2018ytt,NA62:2022qes}&$\pm0.007$ (Phase~1)\\
BR$(K_L\to \mu\mu)$ ($+$)   & $(6.82^{+0.77}_{-0.29})\times 10^{-9}$    & \multirow{2}{*}{$(6.84\pm0.11)\times 10^{-9}$} & \multirow{2}{*}{\cite{ParticleDataGroup:2020ssz}} &  \multirow{2}{*}{1\% (Phase~2)}\\
BR$(K_L\to \mu\mu)$ ($-$)   &  $ (8.04^{+1.47}_{-0.98})\times 10^{-9}$    &  & & 
\\
BR$(K_S\to \mu\mu)$         & $(5.15\pm1.50)\times 10^{-12}$    & $ < 2.1(2.4)\times 10^{-10}$ @$90(95)\%$ CL & \cite{LHCb:2020ycd} & {\small Upper bound kept to current value}\\
BR$(K_L\to \pi^0 ee)(+)$         & $(3.46^{+0.92}_{-0.80})\times 10^{-11}$    & \multirow{2}{*}{$ < 28\times 10^{-11}$ @$90\%$ CL} & \multirow{2}{*}{\cite{KTeV:2003sls}}&  \multirow{2}{*}{20\% (Phase~2)}\\
BR$(K_L\to \pi^0 ee)(-)$         & $(1.55^{+0.60}_{-0.48})\times 10^{-11}$    &  &  & \\
BR$(K_L\to \pi^0 \mu\mu)(+)$         & $(1.38^{+0.27}_{-0.25})\times 10^{-11}$    & \multirow{2}{*}{$ < 38\times 10^{-11}$ @$90\%$ CL} & \multirow{2}{*}{\cite{KTEV:2000ngj}} &  \multirow{2}{*}{20\% (Phase~2)} \\
BR$(K_L\to \pi^0 \mu\mu)(-)$         & $(0.94^{+0.21}_{-0.20})\times 10^{-11}$    &  &  &  \\
\hline \hline
\end{tabular}}
\caption{\small
The SM predictions, current experimental status and the expected HIKE sensitivity for the different observables, that are the inputs for the global fit theory plots. The ``($+$)'' and ``($-$)'' signs in the first column correspond to constructive and destructive interference of the amplitudes.
\label{tab:data}}
\end{center}
\end{table}

Table~\ref{tab:data} summarises the SM predictions for various (semi)leptonic and rare kaon decays from~\cite{DAmbrosio:2022kvb}, the current experimental status and the HIKE sensitivities. These values are the inputs for the theory analysis described below. The HIKE Phase 1 sensitivity is estimated thanks to the extensive experience of the NA62 experiment, corresponding to a factor four increase in PoT and kaon decays. HIKE, with new or upgraded detectors and readouts to profit the most from the increased beam intensity, will improve the acceptance of kaon decays and keep the random veto under control at much higher intensity. Improved upstream detectors will be used to control the dominant background modes. 

The sensitivity to the $K_L\to\pi^0\ell^+\ell^-$ decays at Phase~2 is determined primarily by the irreducible Greenlee background $K_L\to\gamma\gamma\ell^+\ell^-$~\cite{Greenlee:1990qy,Husek:2015sma}. This background is suppressed exploiting the reconstructed mass of the di-photon system (which peaks at the $\pi^0$ mass for the signal), photon energy asymmetry in the kaon frame (which has a flat distribution for the signal and peaks at $\pm1$ for the background), and the minimal angle between any of the photons and any of the leptons in the kaon frame (which is on average higher for the signal than for the radiative Greenlee process). The expected numbers of SM signal ($N_S$) and Greenlee background ($N_B$) events in five years of HIKE Phase~2 operation, evaluated using a full \textsc{Geant4} simulation, reconstruction and analysis chain, are summarised in Table~\ref{tab:KL_decays}. The $K^+\to\pi^0\pi^+\pi^-$ background with pion decays in flight to the $K_L\to\pi^0\mu^+\mu^-$ decay is found to be sub-dominant using a full simulation. HIKE is expected to provide the first observation (above $5\sigma$) and measurement of both $K_L\to\pi^0\ell^+\ell^-$ decay modes, making it possible to determine the corresponding branching ratios with a precision of 20~\%.

\begin{table}[t]

\centering
\begin{tabular}{cccc}
\hline
Mode & $N_S$ & $N_B$ & $N_S/\sqrt{N_S+N_B}$ \\
\hline
$K_L\to\pi^0 e^+e^-$ & 70 & 83 & 5.7\\
$K_L\to\pi^0\mu^+\mu^-$ & 100 & 53 & 8.1\\
\hline
\end{tabular}
\caption{\small\label{tab:KL_decays} Expected number of signal ($N_S$) and background ($N_B$) events at HIKE Phase 2, as well as the resulting "signal-to-background" ratio for different $K_L$ decay modes.}
\end{table}

\begin{figure}[t]
\centering
\includegraphics[width=0.48\textwidth]{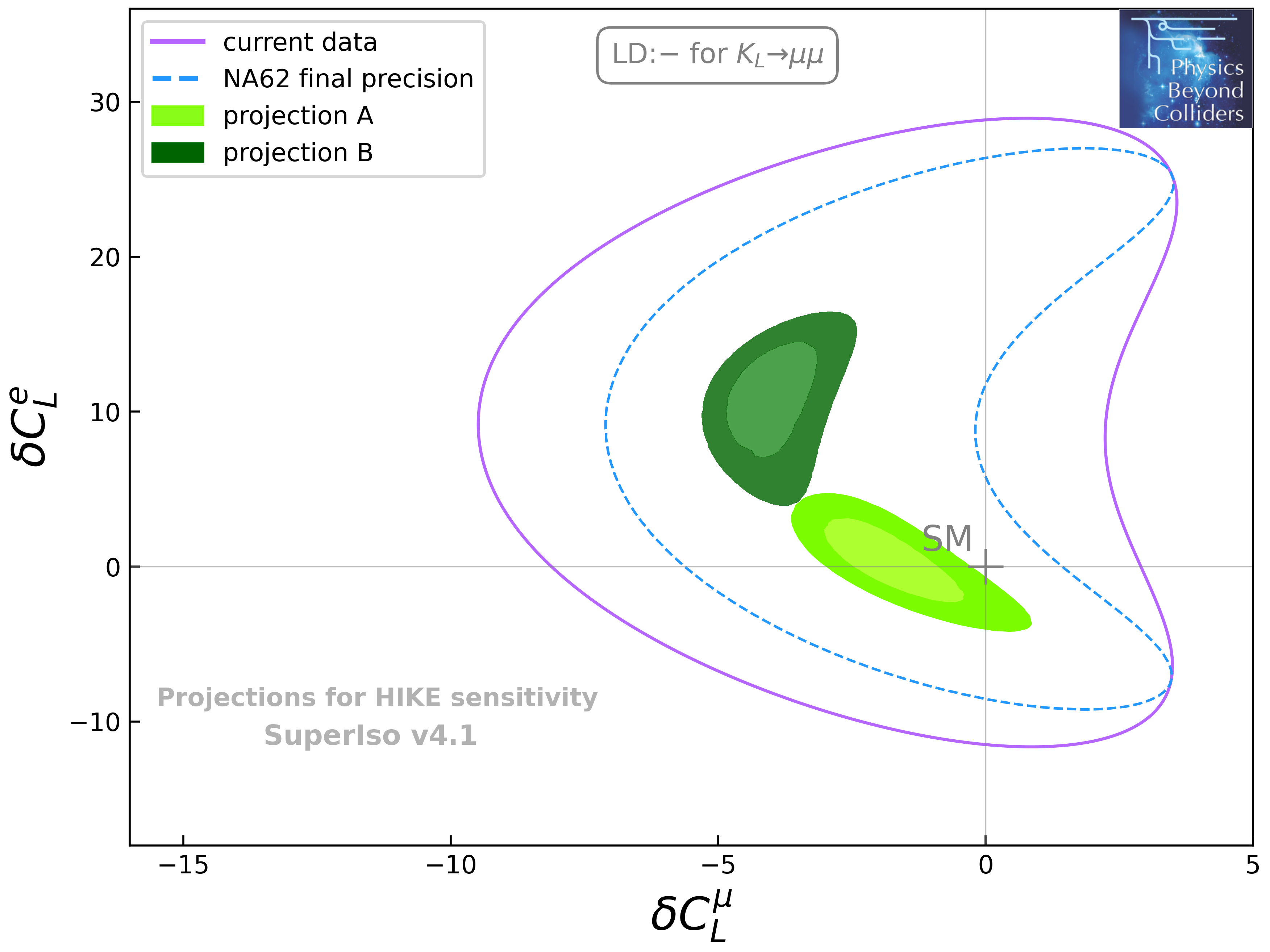}
\includegraphics[width=0.48\textwidth]{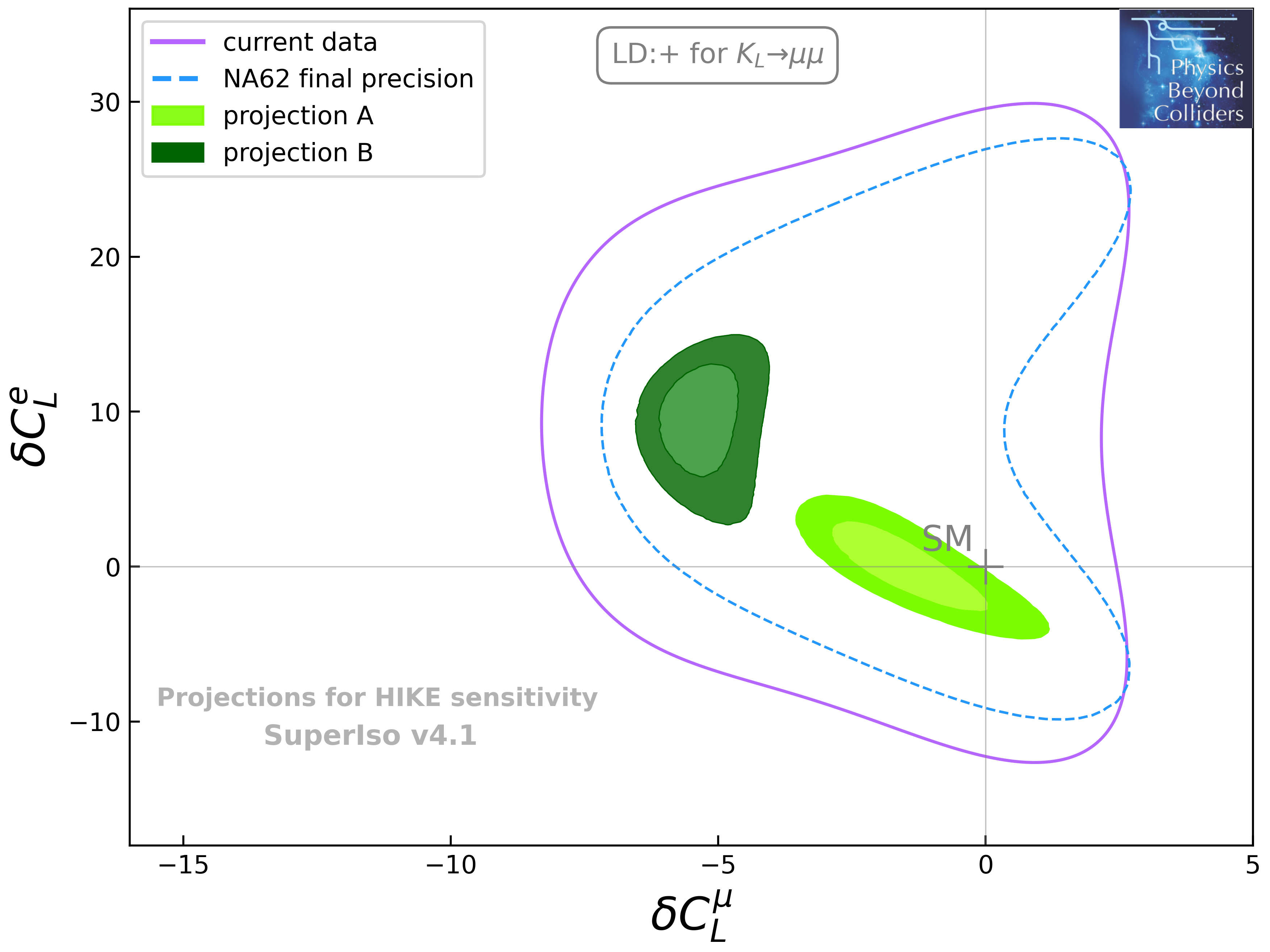}
\caption{\small
Global fits in the $\{ \delta C_L^{e}, \delta C_L^\mu(=\delta C_L^{\tau})\}$ plane with current data (2$\sigma$ confidence level, solid purple contour), the projected sensitivity for NA62 (2$\sigma$ confidence level, dashed blue contour) and the projected scenarios with the sensitivities for HIKE as in Table~\ref{tab:KL_decays} (green regions corresponding to 1$\sigma$ and $2\sigma$ confidence level). The two panels correspond to the different signs for the LD contribution to $K_L\to \mu\bar\mu$ (left: negative, right: positive), which affects the SM prediction and hence the interpretation of the experimental result (see Table~\ref{tab:data} and Figure~\ref{fig:all_obs_individually}). Figure updated from~\cite{DAmbrosio:2022kvb}. 
\label{fig:fitcomparison}}
\end{figure}

Following the strategy of~\cite{DAmbrosio:2022kvb}, projection fits of Wilson coefficients of equation 5 are made (using SuperIso v4.1~\cite{Mahmoudi:2008tp,Neshatpour:2022fak}) for the future kaon measurements that will become possible with the HIKE program. The projection fits require both the possible future measured values as well as the experimental precision. For the latter the expected HIKE sensitivities are taken from Table~\ref{tab:data}, while for the projected central values two scenarios are assumed:
\begin{itemize}
 \item projection A: the predicted central
values for those observables with only an upper bound is projected to be the same as the
SM prediction while for the measured ones the current central values are taken;
 \item projection B: the central values for all of the observables are projected with the best-fit points obtained from the fits with the existing data.
\end{itemize}
Both projections do not assume any improvement in the theoretical precision. The projected fits of the two scenarios are shown in Figure~\ref{fig:fitcomparison} where the 68 and 95~\% CL regions are shown with the two shades of light-green for projection A and the two shades of dark-green for projection B. The two panels correspond to the two possible signs of the LD contributions to $K_L\to \mu\bar\mu$.

The two scenarios give quite different results with projection A indicating overall consistency with SM at the level of $3\sigma$ while projection B clearly departs from the SM at more than $3\sigma$, especially for positive LD. The sign of the LD contributions to $K_L\to \mu\bar\mu$ has a clear impact on how precisely BSM can be probed and although currently the theory uncertainty overshadows the experimental error, in case of future improvement of theory prediction, decrease in experimental uncertainty will be relevant 
for extracting information on BSM physics as well as identifying the correct sign of $A_{L\gamma\gamma}^\mu$.

\subsubsection{Further science goals}

\subsubsection*{Lepton flavour violation}

Individual lepton flavours -- electron, muon, and tau number -- are conserved in the SM but known to be violated in nature, as evidenced from neutrino oscillations. No LFV has yet been observed in the charged-lepton sector, but is generically expected in many extensions of the SM, notably those that aim to generate neutrino masses~\cite{Calibbi:2017uvl,Davidson:2022jai}. An observation would provide groundbreaking indirect evidence for new elementary particles, e.g.~heavy neutrinos~\cite{Atre:2009rg}, additional Higgs bosons, or leptoquarks~\cite{Shanker:1981mj,Mandal:2019gff}.
The absence of model-independent predictions leads to explore a wide variety of LFV signatures~\cite{Davidson:2022jai}. HIKE will be able to search for LFV in kaon and $\pi^0$ decays, reaching the sensitivity to branching fractions down to ${\cal O}(10^{-13})$. Recent results from NA62 include limits on the decays $K^+\to \pi^+e^+\mu^-$~\cite{NA62:2021zxl}, $\pi^0\to e^+\mu^-$~\cite{NA62:2021zxl}, and $K^+\to \mu^- e^+e^+ \nu$~\cite{NA62:2022exp}, with analogous charge-flipped final states currently only constrained by older experiments~\cite{ParticleDataGroup:2022pth}. Phase~1 of HIKE will improve on the processes listed above, as well as other modes including $K^+\to e^-\mu^+\mu^+\nu$, $K^+\to\pi^+\pi^0e^+\mu^-$, $K^+\to\pi^+(\pi^0)e^-\mu^+$, $\pi^0\to e^+\mu^-$.
Phase 2 of HIKE will study $K_L$ decays and is likely to improve limits on LFV decays such as $K_L\to e^\pm \mu^\mp (\pi^0)(\pi^0)$~\cite{BNL:1998apv,KTeV:2007cvy} and $K_L\to e^\pm e^\pm \mu^\mp \mu^\mp$~\cite{KTeV:2002kut} that still stem from the BNL-E871 and KTeV experiments.
The LFV signatures above implicitly assume heavy new physics, but HIKE will also be sensitive to several LFV channels mediated by FIPs, involving displaced vertices.

\subsubsection*{Lepton number violation}

While the individual lepton flavours are without a doubt broken in nature, the same is not known for total lepton number: no lepton-\emph{number}-violating process has ever been observed, in agreement with the SM prediction~\cite{FileviezPerez:2022ypk}. An observation would again provide evidence for additional particles beyond the SM and have wide-ranging consequences for our understanding of fundamental physics and even cosmology, since lepton number violation could be the reason for the observed dominance of matter over antimatter~\cite{Davidson:2008bu}. Neutrino masses can serve as motivation for these violations too: if neutrinos are Majorana particles then lepton number is broken and corresponding signatures are expected, the  most sensitive of which arguably being neutrinoless double beta decay $(A,Z)\to (A,Z+2)+2 e^-$~\cite{Furry:1939qr,Rodejohann:2011mu}. Meson decays provide a complementary probe that is sensitive to different flavour structures~\cite{Littenberg:1991ek}. Phase 1 of HIKE will be able to improve on the bounds recently set by NA62 in the channels 
$K^+\to \pi^-\mu^+\mu^+$~\cite{NA62:2019eax},
$K^+\to \pi^-e^+\mu^+$~\cite{NA62:2021zxl}, and
$K^+\to\pi^-(\pi^0)e^+e^+$~\cite{NA62:2022tte}, reaching the sensitivity to branching fractions down to ${\cal O}(10^{-13})$. Searches for processes with displaced vertices involving emission and decay of a heavy Majorana neutrino, such as $K^+\to\mu^+N$, $N\to\pi^-\mu^+$, are also of interest.

\subsubsection*{Precision tests of low-energy QCD}

Most kaon decays are governed by long-distance physics and are described by chiral perturbation theory (ChPT), the low-energy EFT of QCD. 
Kaon decay amplitudes are evaluated in the ChPT framework using the so-called low-energy constants determined from experimental data. Comprehensive measurements of kaon decay rates and form factors represent both essential tests of the ChPT predictions and crucial inputs to the theory. A complete overview of kaon decays in relation to the ChPT can be found in~\cite{Cirigliano:2011ny}. 
The HIKE dataset will provide a unique opportunity to perform a wide range of precision measurements of rare and radiative decays of both $K^+$ and $K_L$ mesons:
\begin{itemize}
\item Precision measurements of $K^+\to\pi^+\ell^+\ell^-$  allow for the determination of the sign of the form-factor $a_S$, since different combinations of ChiPT parameters enter the ${\cal O}(p^4)$  
chiral Lagrangian~\cite{DAmbrosio:1998gur}.
\item Precise measurements of $K^+\to\pi^+\gamma\gamma$, $K^+\to\pi^+\gamma\ell^+\ell^-$ provide interesting chiral tests, including determination of the ${\cal O}(p^4)$ weak chiral Lagrangian and relations among low-energy observables~\cite{Gabbiani:1998tj}.
\item The decays $K^+\to\pi^+\pi^0\gamma$, $K^+\to\pi^+\pi^0\ell^+\ell^-$ are interesting to determine the weak chiral Lagrangian~\cite{Cappiello:2017ilv} and to study CP asymmetries.
\item A measurement of $K^+\to e^+\nu\gamma$ aiming at ${\cal O}(p^6)$ will be very interesting since the ChPT Lagrangian terms here are not known from other data, and a recent measurement from J-PARC~\cite{J-PARCE36:2022wfk} departs from the $O(p^4)$ theory result.
\item The radiative decay $K^+\to\pi^0 e^+\nu\gamma$ has been accurately studied  theoretically, and can limit to 1~\% the novel structure-dependent contributions of new physics~\cite{Bijnens:1992en,NA62:2023lnp}.
\item Measurements of the principal kaon decay modes $K\to 2\pi$ and $K\to 3\pi$ provide overall information on all isospin amplitudes, $\pi\pi$ phase shifts, the $\delta I =1/2$ rule, and a test of the weak chiral Lagrangian~\cite{Cirigliano:2011ny,DAmbrosio:2022jmd}.
\end{itemize}

The recent NA62 $K^+\to\pi^+\mu^+\mu^-$ experimental measurements has already improved the theoretical determination of the form-factors~\cite{NA62:2022qes}.
HIKE expects to collect background-free samples of several times $10^5$ events of both $K^+\to\pi^+e^+e^-$ and $K^+\to\pi^+\mu^+\mu^-$ decays, allowing for crucial improvements in the precision of the extracted form factors. Measurements of the branching ratios of decays $K^{+}\to e^{+}\nu\gamma$, $K^{+}\to \pi^{0}e^{+}\nu\gamma$, and $K^{+}\to\pi^{+}\gamma\gamma$ are expected to reach a relative precision of a few per mille. The decays $K^{+}\to\pi^+\gamma e^+e^-$, $K^{+}\to\pi^+\pi^0 \gamma$ and $K^{+}\to\pi^+\pi^0 e^+e^-$ are expected to be measured with a few per cent relative precision.
Studies of the $K\to 2\pi$ and $K\to 3\pi$ decays will provide important inputs to ChPT parameter fits. 

\subsubsection{International landscape}

This overview is concluded with a brief discussion of how the physics potential of HIKE compares with other ongoing or planned experimental efforts. 

\subsubsection*{Kaon facilities}

A central player in the field of kaon physics is the KOTO experiment at J-PARC, whose physics programme is entirely focused on the decay $K_L \to \pi^0 \nu \bar \nu$. The Grossman-Nir (GN) bound~\cite{Grossman:1997sk} states that under mild assumptions the partial decay width for this process must be smaller than the one for $K^+ \to \pi^+ \nu \bar \nu$, which translates to $\mathcal{B}(K_L \to \pi^0 \nu \bar \nu) < 4.3 \mathcal{B}(K^+ \to \pi^+ \nu \bar \nu)$. Given the KOTO upper bound $\mathcal{B}(K_L \to \pi^0 \nu \bar \nu) < 3.0 \times 10^{-9}$ at 90~\% CL~\cite{KOTO:2018dsc} and the NA62 measurement of $\mathcal{B}(K^+ \to \pi^+ \nu \bar \nu)$, it can be concluded that KOTO is currently not sensitive to models of BSM physics that satisfy the GN bound. 

Nevertheless, KOTO provides valuable tests of BSM models that circumvent the GN bound. This can happen for example through the direct production of a new long-lived particle $X$ via $K\to \pi + X$~\cite{Kitahara:2019lws,Afik:2023mhj}. Naively, the production of such new particles is also subject to the GN bound, such that the leading sensitivity should stem from $K^+\to\pi^+ + X$ decays.  
However, there are several differences 
such as experimental acceptances~\cite{Kitahara:2019lws} and  violation of flavour symmetries~\cite{Gori:2020xvq,Pospelov:talk,Ziegler:2020ize}. In the context of such models, KOTO (including its future upgrade KOTO-II) and HIKE are highly complementary.

\subsubsection*{Other probes of flavour physics}

\begin{figure}[t]
    \centering
    \includegraphics[width=0.55\textwidth]{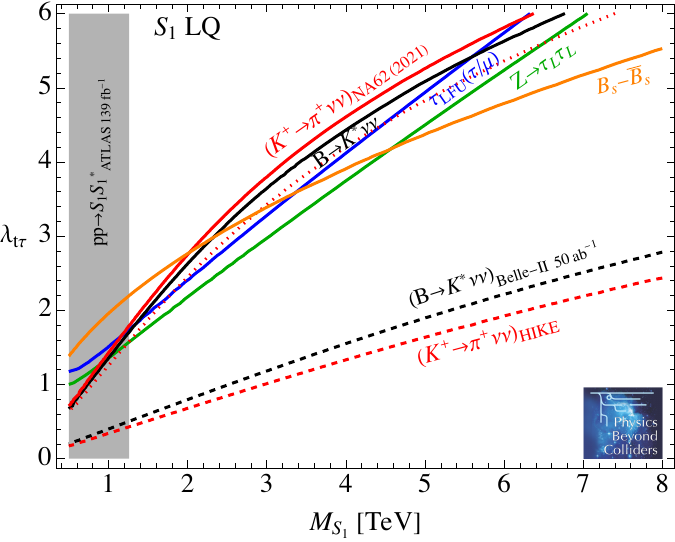}
    \caption{\small Constraints on the $\lambda_{t\tau}$ coupling of the $S_1$ leptoquark from flavour and electroweak observables, as function of the leptoquark mass $M_{S_1}$. Constraints are shown from NA62 (red), Belle \cite{Belle:2017oht} (black), lepton-flavour universality in $\tau$ decays \cite{Pich:2013lsa} (blue), $Z$ boson couplings to tau leptons \cite{ALEPH:2005ab} (green), and from $B_s - \bar{B}_s$ mixing \cite{UTfit:2007eik} (orange) (other $\Delta F=2$ transitions provide similar but slightly weaker constraints). The shaded gray region is excluded by ATLAS from pair-production searches \cite{ATLAS:2021jyv}. Also shown are the projected sensitivity for NA62 (dotted red) and HIKE (dashed red), and for $B \to K^* \nu \bar{\nu}$ from Belle-II with 50ab$^{-1}$ of luminosity \cite{Belle-II:2018jsg} (dashed black). The constraints are derived using the complete one-loop matching of this leptoquark to the SMEFT derived in~\cite{Gherardi:2020det}, following the phenomenological analysis of Refs.~\cite{Gherardi:2020qhc,Marzocca:2021miv}.}
    \label{fig:scalar_leptoquark}
\end{figure}

\begin{figure}[t!]
\centering
\hspace*{2.4cm}\includegraphics[width=0.68\textwidth]{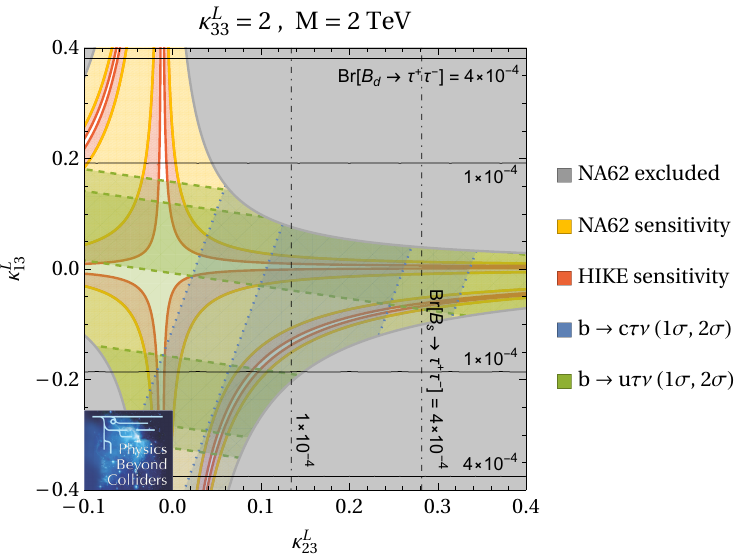}
\caption{\small
Constraints on a vector leptoquark $SU(2)$ singlet with mass $M = 2\,\mathrm{TeV}$ and dominant coupling to the third generation ($\kappa_{33}^L$ = 2) as a function of the (real) flavour-changing couplings $\kappa_{13}^L$ and $\kappa_{23}^L$, see~\cite{Crivellin:2018yvo} for details. The grey shaded region is excluded by NA62, the blue (green) shaded regions are preferred by $b \to c(u) \tau \nu$ data~\cite{Charles:2004jd,HFLAV:2016hnz}. The NA62 (HIKE) sensitivity is indicated by  yellow (red) shading. Black lines represent the branching ratios for $B_s \to \tau^+ \tau^-$ and $B_d \to \tau^+ \tau^-$, which can be constrained by LHCb and Belle II.
\label{fig:vector_leptoquark}}
\vspace{2mm}
\centering
\includegraphics[width=0.52\textwidth]{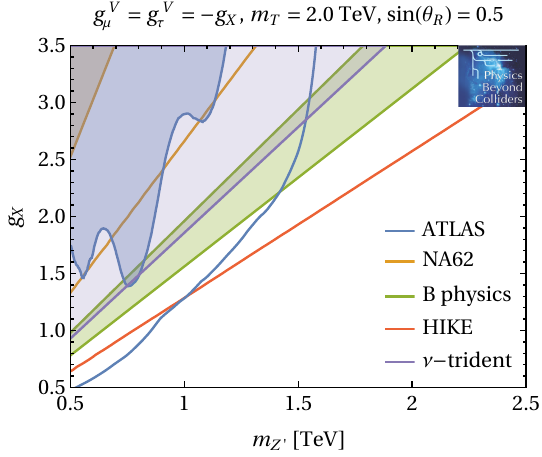}
\caption{\small
Constraints on a top-philic $Z'$ with mass $m_{Z'}$ and gauge coupling $g_X$, see~\cite{Kamenik:2017tnu,Fox:2018ldq} for details. Couplings to top quarks are assumed to be induced via mixing with a vector-like quark with mass $m_T = 2 \, \mathrm{TeV}$ and mixing angle $\sin \theta_R = 0.5$, and vector couplings to muons and tau leptons are assumed to be $g_\mu^V = g_\tau^V = - g_X$, such that various anomalies in $b \to s$ transitions can be fitted (green shaded region), see~\cite{Alguero:2023jeh}. The purple shaded region is excluded by neutrino trident production~\cite{Altmannshofer:2014pba}. Blue shaded regions (blue lines) indicate the current exclusion with 139 fb$^{-1}$ (projection for 3 ab$^{-1}$) for ATLAS~\cite{ATLAS:2019erb}, while orange shaded regions (orange lines) indicate the current NA62 exclusion limit (projected sensitivity). The red line corresponds to the HIKE projected sensitivity.
\label{fig:Zprime}}
\end{figure}

In the context of LFU violation, rare $B$ meson decays have received substantial interest in recent years. While the hints for LFU violation in $b\to s\ell^+\ell^-$ transitions have disappeared and $\mathcal{B}(B_s\to \mu^+\mu^-)$ is in good agreement with the SM prediction, there is still strong tension in observables such as $\mathcal{B}(B\to K\mu^+\mu^-)$, $\mathcal{B}(B_s\to \phi\mu^+\mu^-)$ as well as angular observables in $B\to K\mu^+\mu^-$ and $B_s\to \phi\mu^+\mu^-$~\cite{HeavyFlavorAveragingGroup:2022wzx}. Together, they point towards LFU-violating BSM in the Wilson coefficient $C_9$ within a global fit~\cite{Alguero:2023jeh,Gubernari:2022hxn}. Furthermore, the measurements of $R(D)$ and $R(D^*)$ point towards LFU violation in charged currents~\cite{HeavyFlavorAveragingGroup:2022wzx}. While the former anomalies might lead to an enhancement of $K_{L,S}\to\mu^+\mu^-$~\cite{Crivellin:2016vjc}, the latter are particularly relevant for $K\to\pi\nu\nu$~\cite{Crivellin:2018yvo,Crivellin:2019dwb,Marzocca:2021miv} since left-handed tau leptons are linked to tau neutrinos via $SU(2)_L$ invariance and the neutrino flavour in $K\to\pi\nu\nu$ is not detected.

To illustrate the interplay of different constraints and future experiments, three specific models of BSM physics are considered. The first, discussed in~\cite{Gherardi:2020det,Gherardi:2020qhc}, introduces a scalar leptoquark $S_1 \sim ({\bf \bar{3}}, {\bf 1})_{+1/3}$ coupled only to the third generation of quark and lepton $SU(2)_L$ doublets:
\begin{equation}
    \mathcal{L} \supset \lambda_{t \tau} \bar{q}_3^c l_3 S_1 + \text{h.c.}~,
\end{equation}
where $q_3 = (t_L, \, V_{t d_j} d_L^j)$, $l_3 = (\nu_{\tau}, \, \tau_L)$. In this up-quark basis, the coupling to left-handed down quark $d^i_L$ is proportional to the corresponding $V_{t d_i}$ CKM element. The second, discussed in~\cite{Crivellin:2018yvo}, considers a vector leptoquark $SU(2)$ singlet with hypercharge -4/3 and dominant couplings to third-generation leptons:
\begin{equation}
    \mathcal{L} \supset (\kappa_{fi}^L \overline{Q_f} \gamma_\mu L_i + \kappa_{fi}^R \overline{d_f} \gamma_\mu e_i) V^{\mu\dagger}_1 + \text{h.c.}~.
\end{equation}
Finally, the third model is based on the  top-philic $Z'$ proposed in Refs.~\cite{Kamenik:2017tnu,Fox:2018ldq}. In contrast to the model set-up considered there, vector couplings to both muons and tau leptons are considered, giving rise to an interesting interplay between the LHC (which gives the dominant constraints for small $Z'$ masses) and flavour physics (which achieves leading sensitivity for large $Z'$ masses).

Exclusion regions, interesting parameter regions and sensitivity projections for the three models are shown in Figures~\ref{fig:scalar_leptoquark}--\ref{fig:Zprime}. In all plots, the NA62 exclusion limit corresponds to $\text{BR}(K^+ \to \pi^+ \nu \bar{\nu}) < 0.42 \times \text{SM}$ and $\text{BR}(K^+ \to \pi^+ \nu \bar{\nu}) > 2.04 \times \text{SM}$~\cite{NA62:2021zjw}, while the NA62 (HIKE) sensitivity projections assume that the SM value for $\text{BR}(K^+ \to \pi^+ \nu \bar{\nu})$ will be confirmed with 20~\% (5~\%) uncertainty. We emphasize that in these models there is non-trivial interference between the SM and the new physics contribution to $\text{BR}(K^+ \to \pi^+ \nu \bar{\nu})$. As a result, the deviation from the SM does not simply scale proportionally to the coupling strength squared, and it is possible for the branching ratio to become smaller than in the SM. In such a case the sensitivity improvement in terms of the underlying couplings achievable by HIKE may differ from the naive expectation based on the improvement in precision. 

\FloatBarrier

\subsection{Neutrino Physics}

Collisions induced by the high-intensity and high-energy proton beam extracted from the SPS and reaching the ECN3 experimental hall will produce copious amounts of neutrinos, which---despite faint interaction rates---will enable a comprehensive neutrino-physics program at ECN3.
In particular, the use of emulsion detectors (cf.~Sections~\ref{sec:SHADOWS_exp} and \ref{sec:SHiP_exp}) allows for various measurements with so-far poorly explored \(\nu_\tau\) neutrinos.
An obvious highlight would be the high-significance observation of the \(\bar{\nu}_\tau\) in concurrence with LHC-neutrino experiments.

In general, precise and reliable theoretical predictions for the scattering rates of (anti-)neutrinos on proton and
nuclear targets constitute a central ingredient for the interpretation of
a wide variety of ongoing and future neutrino experiments.
In turn, measurements of these neutrino scattering rates can provide valuable probes
of the partonic structure of nucleons and nuclei as well as of fundamental SM parameters.
Both the $\nu$-ECN3 experiments and the Forward Physics Facility (FPF~\cite{Feng:2022inv}) proposed at the LHC will be able to carry out at least some of these studies including with tau neutrinos. 

After a general overview of the potential physics topics that could be addressed by the projects, specific information on the expected performance of the experiments is given in
Sections~\ref{sec:nu_SHADOW}, \ref{sec:nu_SHiP} and, for common issues, \ref{sec:nu_commonSNDandNaNu}. A summary and comparison of the physics reach in the international landscape, including the  FPF proposed at CERN, is provided in \ref{sec:nu_FPFetc}.

\subsubsection{Physics case}
\label{sec:nu_for_SMandBSM}

An overview of the expected fluxes of neutrino interactions and reconstructed events is given in Table~\ref{tab:nuFluxes}, complemented by the number of charmed particles expected to be detected by SHiP SND.

\begin{table}[t]
\begin{center}
    \footnotesize
	
	\begin{tabular}{c|l|c|c|c|c|c|c}
		\hline
 & Neutrino flavour   & $\nu_e$ & $\bar \nu_e$ & $\nu_\mu$ & $\bar \nu_\mu$ & $\nu_\tau$ & $\bar \nu_\tau$ \\
		\hline
\multirow{3}{*}{SHADOWS NaNu} 
 & Expected  $\nu$ interactions   & $4.1\cdot10^3$ & $1.0\cdot 10^3$ & $4.0\cdot10^4$ & $9.0\cdot 10^3$ & 120 & 70 \\
 & Reconstr. $\nu$ interactions   & $2.5\cdot10^3$ & $0.6\cdot 10^3$ & $2.5\cdot10^4$ & $5.0\cdot 10^3$ & 60 & 40 \\	
 & With charge-ID       & --- & --- & $2.5 \cdot 10^4$ & $5.0\cdot 10^3$ & 10 & 7 \\ \hline
\multirow{4}{*}{SHiP SND} 
 & Expected  $\nu$ interactions    & $2.7\cdot10^6$ & $0.6\cdot 10^6$ & $8.0\cdot10^6$ & $1.8\cdot 10^6$ & $8.8\cdot10^4$ & $6.1\cdot10^4$\\
 & Reconstr. CC $\nu$ interactions & $2.4\cdot10^6$ & $0.5\cdot 10^6$ & $6.8\cdot10^6$ & $1.5\cdot 10^6$ & $7.4\cdot10^4$ & $5.2\cdot10^4$\\	
 & With charge ID    & --- & --- & $4.3\cdot10^6$ & $1.0\cdot10^6$ & 3800 & 2900\\	
 & charmed particles detected     & $1.3\cdot10^5$ & $2.3\cdot10^4$ & $2.6\cdot10^5$ & $5.2 \cdot10^4$ & --- & --- \\\hline
	\end{tabular}
 \caption{\small Expected yields of reconstructed neutrino interactions at SHADOWS NaNu for a collected
            data set of $5\times10^{19}$ PoT as well as of reconstructed charge-current neutrino
            interactions and of charmed hadrons produced in neutrino interactions at SHiP SND for 
            $6 \times 10^{20}$ PoT.
            \label{tab:nuFluxes}}
\end{center}
\end{table}

A number of potentially interesting physics topics are listed below. 
It should be emphasized that decisive quantitative evaluations are not available for all topics, and that some of them are listed to serve as inspiration for further feasibility studies.
\begin{itemize}

\item{}\textbf{\(\bar{\nu}_\tau\) observation:}
The \(\bar{\nu}_\tau\) is the only particle in the SM of particle physics that remains to be experimentally observed.

\item{}\textbf{Lepton-flavour Universality (LFU):}
The ECN3 and FPF neutrino experiments are able to simultaneously measure the $\nu q \to \ell q'$ charged-current~(CC) $\nu$-scattering cross sections for all three neutrino flavours. This helps to reduce associated systematic uncertainties and allows for a more precise comparison of those cross sections, e.g., to search for hints of BSM physics. Strong constraints however already exist  \cite{Falkowski:2021bkq} since similar contact operators or diagrams would also contribute to meson decays via $q \to q' \ell \nu$ and to LHC scattering via $q q' \to \ell \nu$.

\item{}\textbf{DIS structure functions \(F_4\) and \(F_5\):}
Deep-inelastic scattering (DIS) of a neutrino on a nucleon was defined in the generic (model independent) way assuming intermediate vector boson (IVB) exchange between lepton and hadron currents 
with five independent structure functions~\cite{Llewellyn:1972, Albright:1974ts, Kretzer:2003iu}.
The most commonly studied ones, \(F_1\) and \(F_2\), only require single-photon exchange. Electroweak effects give rise to \(F_3\). These three have all been measured quite precisely for the proton. \(F_4\) and \(F_5\) are suppressed for small lepton masses. This makes tau leptons the only viable mean of accessing these so far unmeasured structure functions via CC tau-neutrino DIS.

\item{}\textbf{Parton distribution functions (PDFs):}
PDFs are fundamental quantities for describing, e.g., nucleons. Many processes are employed (cf.~discussion in Section \ref{sec:nu_FPFetc}) to extract precise PDFs. Sea quarks are generally more difficult to access, which makes electroweak processes, such as CC $\nu$-DIS, a usually fruitful tool for singling out their distributions.
In practice, $\nu$-DIS data have been taken on nuclear targets and not free nucleons. It is well established that parton distributions get modified when the parent hadron is embedded inside a nucleus. On one hand, this requires careful studies when using such data for measuring nucleon PDFs. On the other hand, $\nu$-DIS off nuclear targets sheds further light on nuclear PDFs, complementary to charged-lepton scattering by nuclei, even more so when different target materials are employed. Significant impact on strangeness PDFs, including different systematics compared to other approaches, can be expected here thanks to the unique possibility of direct tagging of charm quark production in emulsion. 

\item{}\textbf{Charm production in neutrino interaction:}
Neutrino-induced interactions in emulsion detectors can be used to investigate inelastic, quasi-elastic, as well as exclusive charm production, with the added benefit of reconstructing the charm-decay chain. Alternatively, charm can be identified via their muonic decay channels, without the need of emulsion detectors. Such data sets are beneficial to both study charm fragmentation, including charmed baryons and pentaquarks or doubly-charmed hadrons in a specific kinematic regime~\cite{2016SHiPPhysicsCase},  as well as charm-hadron decays, e.g., new decay channels.    

\end{itemize}

Additional interesting aspects that merit but also require more detailed studies are:

\begin{itemize}
\item{}\textbf{\(\nu_\tau\) magnetic moment:}
While the SM predicts very small neutrino magnetic moments to arise at one loop level ($\mu_\nu \sim 10^{-19} \mu_B \times (m_\nu/\text{eV})$ in the Dirac neutrino case), BSM physics could potentially lead to larger magnetic moments~\cite{Giunti:2014ixa}. Solar neutrino measurements have provided the most stringent constraints on the magnetic moment of tau neutrinos, yielding a limit of $\mu_\nu < 1.3\cdot 10^{-11} \mu_B$~\cite{A:2022acy}. Independently, the neutrinos magnetic moments can be probed at accelerator experiments by searching for neutrino-electron scattering events with low-energy recoils. The currently strongest purely laboratory-based bound on the $\nu_\tau$ magnetic moment, $\mu_\nu < 3.9 \cdot 10^{-7} \mu_B$, comes from DONUT~\cite{DONUT:2001zvi}.
While measurements at ECN3 are expected to improve on this (see, e.g., the SHiP study in \cite{bib:BDFSHIP_PROPOSAL_2023}), they will not reach a level comparable to the astrophysical constraints. They therefore provide a welcome independent laboratory confirmation but are likely to be sensitive only to exotic models.

\item{}\textbf{Study of neutral currents:}
The study of the neutral-current (NC) $\nu$-scattering rate, or equivalently the ratios of NC-to-CC cross sections, provide sensitivity to a variety of phenomena. These include, e.g., the weak mixing angle (same for all flavours; requires high precision) or non-standard neutrino interactions (possible flavour or energy dependence). A comparison with measurements at other experiments with different flavour and energy distributions, such as NuTeV, is expected to provide further input to phenomenological studies. Significant deviations of NC results from the SM could indicate interactions of new FIP particles. Studies on NC-related phenomena have been performed by SHiP SND (see section 7.2.5 in~\cite{2016SHiPPhysicsCase}) and FPF (see sections 7.3.2,  7.5.3, and 7.5.8 in~\cite{Feng:2022inv} and~\cite{Abraham:2023psg}).

\item{}\textbf{Sterile neutrino/HNL oscillations:}
Since SM neutrino oscillations are negligible for the ECN3 and FPF experiments, any sign of an oscillation signal would hint toward the existence of an additional eV-scale sterile neutrino. 
Taking into account existing constraints, a possible eV-scale sterile neutrino oscillation signal would cause up to percent-level deviations, which would be experimentally challenging to observe and would require a precise understanding of the expected flux.
\begin{figure}[t!]
\centering
 \includegraphics[width=0.5\textwidth]{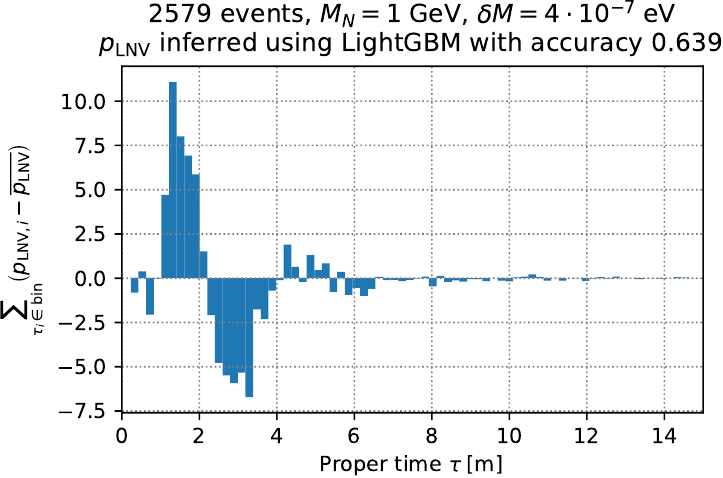}
 \caption{\small Reconstructed oscillations between the lepton number conserving and violating event rates as a function of the proper time for a heavy sterile neutrino with the parameters $m_N$ = 1 GeV, $|U_{\mu}|^2 = 2 \times 10^{-8}$ and mass splitting of $\Delta m = 4 \times 10^{-7}$ eV. Figure taken from~\cite{Tastet:2020}.
 \label{fig:reconstructed_oscillations}}
\end{figure}
While a study was performed for FPF (see Section 7.5.9 in~\cite{Feng:2022inv}), both SHADOWS NaNu and SHiP SND  need to evaluate their sensitivities to eV-scale sterile neutrino oscillations from known neutrino flavours.
SHiP experiment is sensitive to oscillations of GeV-scale heavy sterile neutrino between lepton number conserving and violating states with $m_N \sim$ 1 GeV and $\Delta m \sim 10^{-6}$ eV (see Figure~\ref{fig:reconstructed_oscillations} taken from~\cite{Tastet:2020})). SHiP SND detector is just in front of the SHiP's HSDS decay volume closer to beam dump, so it will be interesting to study how it can improve this result at small values of proper time by adding events with heavy neutrino decays in the magnet or even in the target region.
\end{itemize}

On a more exploratory note, data from the ECN3 neutrino experiments potentially help in validating MC simulation for neutrino oscillation and astroparticle experiments, but more studies would be needed to quantify this physics case.

\subsubsection{SHADOWS neutrino measurements}
\label{sec:nu_SHADOW}

The baseline concept of NaNu, the SHADOWS neutrino detector, foresees two separate detector components: one active detector component closer to the beam-line targeting the study of muon neutrino interactions, and a partly passive detector component based on emulsion, aiming for tau-neutrino physics. 
A detailed description of the experimental setup and possible extensions are discussed in~\cite{Neuhaus:2022hji}. \textsc{Pythia8} was used to estimate the neutrino kinematics at the off-axis location of SHADOWS NaNu, while \textsc{GENIEv3} and \textsc{Geant4} were used to simulate the neutrino interactions and their subsequent decay products in the detector. In the following, only the physics reach of this baseline detector system assuming four full years of operation and a collected data set of $5\times10^{19}$ PoT is summarized.

An overview of the expected neutrino interactions and reconstructed event yields%
\footnote{\(\nu_\mu/\bar{\nu}_\mu\) rates in the NaNu detector include the events in the Tungsten-Micromegas part of the detector without emulsions and closer to beam dump axis.}
is given in Table \ref{tab:nuFluxes}.
The expected number of reconstructed $\nu_\mu$ and $\bar \nu_\mu$ interactions is obtained by requiring a minimal muon momentum of 5 GeV. They are dominantly reconstructed by the active detector component and to a smaller extent by the emulsion detector. Assuming additionally a minimal hadronic energy of the recoil system of 10 GeV, to allow for a sufficiently precise reconstruction of the full event kinematics, the numbers reduce by another 40~\%. The hadronic energy is measured using scintillator plates that are interleaved between the passive tungsten plates. An energy resolution of $200\%/\sqrt{E [\text{GeV}]}$ is expected. 
Differential cross-sections in a two-dimensional binning of $5\times5$ bins in~Bjorken $x$ and squared momentum transfer $Q^2$ can be measured with statistical uncertainties in the range between 5~\% and 10~\% for $\nu_\mu$ and $\bar \nu_\mu$ interactions, respectively (cf. Figure~\ref{fig:ECN3-neut-2} top left).
Those measurements would provide a consistency test of existing neutrino data in the context of global fits of PDFs. 
The muon neutrino measurements at SHADOWS NaNu are expected to be limited by systematic uncertainties, which are expected to be on the order of 2--4~\%, as observed in previous cross-section measurements of muon neutrinos (e.g.~in~\cite{MINERvA:2017ozn}).

Charm-meson production in neutrino events can either be identified in the emulsion target or via the muonic charm decay channels. In the latter case, the full reconstruction can be performed using the active detector components and no reconstruction within the emulsion is required. Taking acceptance and reconstruction efficiencies as well as minimal momentum requirements into account,  about 150 identified charm-meson candidates in a di-muon final state can be reconstructed.

The number of identified  $\nu_\tau$ and $\bar \nu_\tau$ interactions exceeds the currently available statistics by a factor ten, allowing in principle for first $\bar \nu_\tau$ candidates during the first year of operation 
within the baseline experimental setup. While the signal over background ratio for $\nu_\tau$ is very high, we expect background contributions from charm-induced processes for the reconstruction of $\bar \nu_\tau$. Those can be distinguished by their decay signatures, yielding a background estimate for the $\bar \nu_\tau$ channel below 2 events, allowing for a first experimental evidence of $\bar \nu_\tau$ at the end of data taking. 

The inclusive cross section of $\nu_\tau$ 
interactions can be measured with a statistical precision of 10~\% and cross-section measurement of $\nu_\tau$ and $\bar \nu_\tau$ interactions can be used to test the combined effect of $F_4$ and $F_5$ structure functions \cite{Albright:1974ts} on the $\nu_\tau$ cross section for the first time, in particular if it is as large as about 30~\% at $E_\nu$=20 GeV (even larger for $\bar \nu_\tau$ interactions, and decreasing for higher energies), as predicted by available QCD analyses. 
Given that the expected $\nu_\tau$ energies (Figure~\ref{fig:neutrino_energy_NaNu_SHiP}) are in the range where the effect is expected to be maximal,
first constraint on $F_4$ and $F_5$ 
could be possible with SHADOWS NaNu.

Similar to the determination of the upper limit for the $\nu_\tau$ magnetic moment by DONUT~\cite{DONUT:2000fbd,DONUT:2001zvi}, a study on the $\nu_\tau$ magnetic moment can be performed at SHADOWS NaNu. It is reasonable to assume similar systematic uncertainties with improved statistical precision.

SHADOWS NaNu could probe LFU at the $\mathcal{O}(10\%)$ level, the precision driven by the statistical one on the tau-neutrino interaction cross section (the electron and muon neutrino cross sections are systematically limited as pointed out before). One may note, though, that LFU of such size is currently not plausible for these $\nu$ energies~\cite{Falkowski:2021bkq}.

The integration of the neutrino detector system, in particular its active components, into the main SHADOWS experiment would allow SHADOWS to extend the search for long-lived particles. Moreover, the emulsion detector can be used for the direct search of signatures of light bosonic dark matter. Detailed studies are still ongoing.

\subsubsection{SHiP neutrino measurements}
\label{sec:nu_SHiP}

The Scattering and Neutrino Detector in SHiP, SND, consists of three elements: the neutrino target and vertex detector, the target tracker stations, and a muon spectrometer (cf.~Section~\ref{sec:SHiP_exp}).  

The muon spectrometer is meant to measure the charge and momentum of the muons, in combination with the SHiP muon spectrometer of the hidden sector. Given the correlation between the emission angle and momentum, muons with high momentum will be detected in the SHiP decay spectrometer and therefore the SHiP SND spectrometer magnet will focus mostly on those with lower momentum, thus loosening the requirements in terms of field strength times length of the spectrometer.  

\begin{figure}[t!]
\centering
 \includegraphics[width=0.45\textwidth]{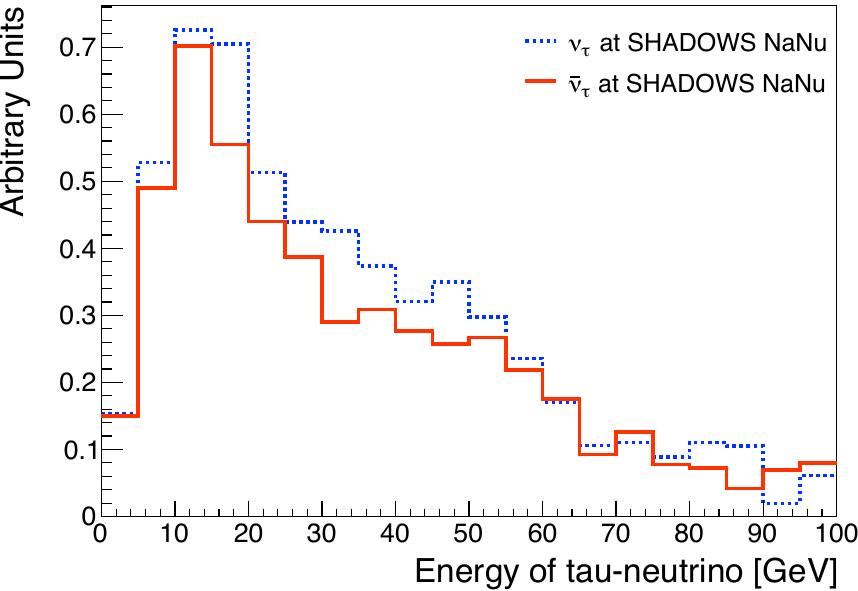}
 \includegraphics[width=0.48\textwidth]{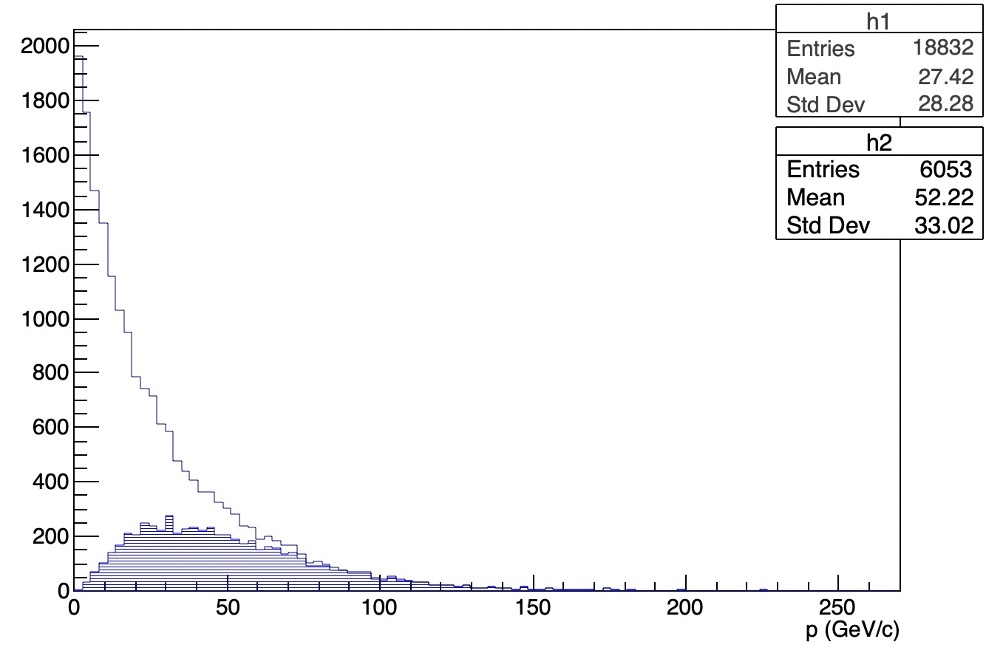}
 \caption{\small Left: Energy distribution of interacting tau neutrinos in SHADOWS NaNu. Right: Muon momentum distribution at SHiP SND, also highlighting the portion of the spectrum measured by the SHiP decay spectrometer.
 \label{fig:neutrino_energy_NaNu_SHiP}}
\end{figure}

The right plot of Figure~\ref{fig:neutrino_energy_NaNu_SHiP} shows the muon momentum spectrum: the portion detectable in the Hidden Sector spectrometer is highlighted as a shaded area and it amounts to about one third. Muons with momentum below 50 GeV will have to be detected in the SHiP SND spectrometer. 
This makes the design of an air core magnet less demanding and more compact. Three tracking stations are foreseen in the spectrometer, one in front, one in the middle, and the third one downstream. The role of the intermediate station is to detect low-energy muons that will not be in the acceptance of the most downstream station. 
The field strength and length is being optimised, currently assumed to provide 3Tm.

The \textsc{Pythia} event generator was used to simulate proton interactions with the target and obtain the neutrino flux. This includes a dedicated simulation of the cascade effect \cite{CERN-SHiP-NOTE-2015-009}. Neutrino interactions are described using the GENIE event generator while the description of the detector response is based on \textsc{Geant4}. The expected rates of reconstructed events of all six neutrino types are given in Table~\ref{tab:nuFluxes} together with the expected number of produced charmed particles. These high rates of charmed particles will allow a rich program of charm physics~\cite{2016SHiPPhysicsCase,delellis}.
Figure \ref{fig:ECN3-neut-2} top middle 
shows the number of muon neutrino CC DIS events in each bin of the probed 2D region in $x$ \& $Q^2$
for $6 \times 10^{20}$ PoT.

The leading systematic uncertainty for an accurate cross-section measurement is the uncertainty on the neutrino flux. This is particularly true for tau neutrinos that are produced via the $D_s$ decay. Charm production in p+p collisions at 400 GeV was measured with an accuracy better than 10~\%  by the NA27 experiment \cite{LEBC-EHS:1988oic}. A dedicated measurement of the $D_s$ production with the identification of the subsequent $D_s \rightarrow \tau$ decay is being carried out by the NA65 experiment \cite{DsTau:2019wjb}. They expect to reconstruct about 1000 $D_s \rightarrow \tau$ decays  in $2.3 \times 10^8$ proton
interactions with a tungsten target \cite{DsTau:2019wjb}. The data, which will become available in the coming years, will narrow down the uncertainty on the tau neutrino flux.

An important aspect is that in a thick target such as the one used for the BDF, charmed hadrons are also produced in the hadron cascade: the relevant process is proton quasi-elastic scattering followed downstream by the same proton inelastic scattering with charm production on a target nucleus. Simulations show that the charm yield increases by more than a factor two due to this effect. In 2018, the SHiP Collaboration successfully conduced a feasibility test of the charm production measurement, including the cascade effect, using the 400 GeV SPS proton beam impinged on a replica of the SHiP target \cite{SHiP:2020hyy}. 
The success of this feasibility test
\cite{SHiP:2021kxh,DiCrescenzo:2743204} paves the way for an extensive measurement campaign.
Ongoing and planned measurements should hence permit to reduce systematic uncertainties on the tau neutrino flux to the percent level.
On the other hand, the high statistics accumulated by the experiment will allow to define different control samples where detection efficiency will be evaluated with data-driven procedures. It is expected that this procedure will lead to an uncertainty on the detection efficiencies at a similar level as that reached on the tau neutrino flux. 

It is worth pointing out that measurements of relative processes, such as the charm production in CC neutrino interactions and the corresponding studies of the strange-quark content of the nucleons, are much less affected by the uncertainty on the absolute flux.

\subsubsection{Common experimental issues}
\label{sec:nu_commonSNDandNaNu}

The important experimental aspects of the proposed experiments at ECN3 are the expected muon fluxes, maximum tolerable track fluxes for emulsion detectors, and corresponding frequency of the emulsion exchange.
Table~\ref{tab:muon-fluxes} summarises the assumed parameters for the neutrino experiments at ECN3
as well as the LHC. 
It is interesting to note that the expected muon flux in SHADOWS NaNu and SHiP SND are close to the experimental muon fluxes at the running FASER$\nu$ and SND@LHC experiments.
So the performance of emulsions in ECN3 neutrino detectors looks feasible both technically and financially.

\begin{table}[h!]
  \begin{center}
    \small
    \renewcommand{\arraystretch}{1.20}
    \begin{tabular}{l|c|c|c|c}
     \toprule
    Experiment               & SHADOWS NaNu & SHiP SND & FASER$\nu$ & SND@LHC \\
          \midrule
    Expected fluxes [$\mu$ tracks/cm$^{2}$/year]   & $\approx 2 \times 10^{6}$      &  $1.0 \times 10^{6}$ & $1.05 \times 10^{6}$   & $1.4 \times 10^{6}$ \\
   Maximum allowed fluxes [$\mu$ tracks/cm$^{2}$] & $10^{6}$ & $10^{6}$      & $4.5 \times 10^{5}$  & $4 \times 10^{5}$ \\
    Number of sets of emulsions per year  & 2      & 1$\div$2            &  3   & 3$\div$4   \\
        \bottomrule
    \end{tabular}
    \caption{\small Comparison of muon background fluxes, maximum allowed muon fluxes, and expected frequency of the emulsion exchanges in the ECN3 (assuming $4 \times 10^{19}$ PoT/year for SHiP SND, $1.2 \times 10^{19}$ PoT/year for SHADOWS NaNu) and LHC (assuming integrated luminosity 70 fb$^{-1}$/year) experiments. The charged particle flux of LHC experiments includes a substantial (up to 50~\%) fraction of non-muon particles (electrons, positrons, pions)
    from the radiation and DIS of muons. At the LHC, signal muon tracks show very similar emission angles as background, thus limiting the tracking capabilities in emulsion. 
    } \label{tab:muon-fluxes}
   \end{center}
\end{table}

\subsubsection{Physics reach in the international landscape}
\label{sec:nu_FPFetc}

\begin{figure}[t]
	\centering
	\includegraphics[width=0.98\textwidth]{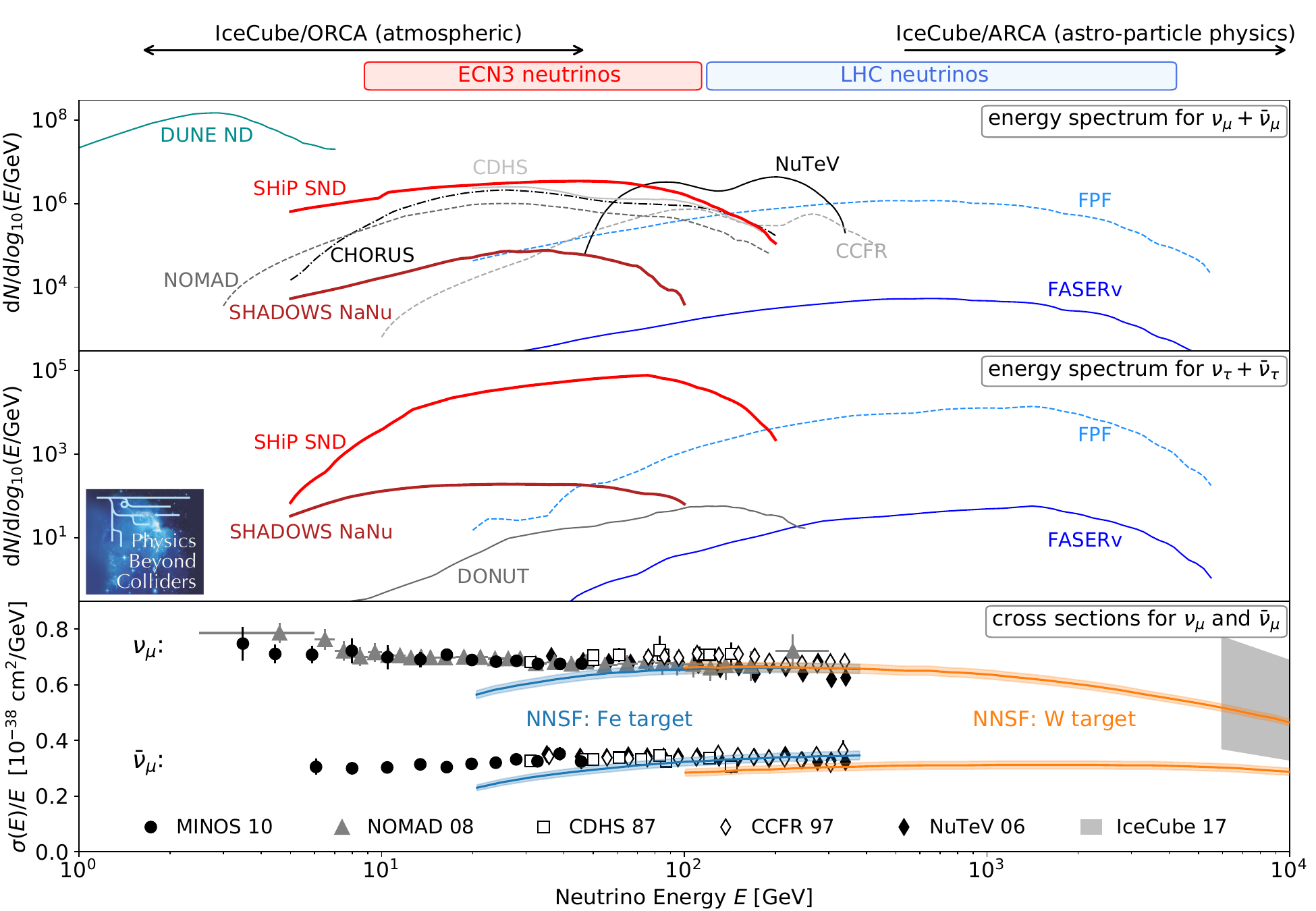}
	\caption{\small Top and Center: Energy spectrum of interacting muon neutrinos (top) and tau neutrinos (center) at SHiP SND and SHADOWS NaNu in comparison to worldwide context including previous accelerator neutrino experiments NuTeV~\cite{NuTeV:2003kth}, CCFR~\cite{CCFRNuTeV:1996vbm}, CDHS~\cite{Berge:1989hr}, CHORUS~\cite{CHORUS:2005cpn}, NOMAD~\cite{NOMAD:2003owt}, DONUT~\cite{DONuT:2007bsg}; the LHC neutrino experiments FASER$\nu$ and those at the proposed FPF (combined statistics of FASER$\nu$2, AdvSND and FLArE); as well as DUNE~\cite{DUNE:2021tad}. The SHiP SND, SHADOWS NaNu and FPF curves are normalized to the event rate given in Table~\ref{tab:neutrino-opportunities}. Bottom: Measurements of the neutrino nucleus interaction cross section at MINOS~\cite{MINOS:2009ugl}, NOMAD~\cite{NOMAD:2007krq}, CDHS~\cite{Berge:1987zw}, CCFR~\cite{Seligman:1997fe}, NuTeV~\cite{NuTeV:2005wsg} and  IceCube~\cite{IceCube:2017roe} in comparison to recent theoretical calculations for the neutrino-nucleus inclusive cross-section~\cite{Candido:2023utz}}
	\label{fig:neutrino-energy}
\end{figure}

The ECN3 experiments are exposed to neutrino beams with energies in the range $10 - 100$~GeV. The corresponding energy spectra of interacting muon neutrino events and tau neutrino events at SHiP SND and SHADOWS NaNu are compared to the worldwide context in Figure~\ref{fig:neutrino-energy}.
A variety of historical neutrino experiments has operated in a similar energy range. They predominantly include experiments placed in beams of  muon neutrinos or anti-neutrinos, such as CDHSW~\cite{Berge:1989hr}, CHARM~\cite{CHARM:1987}, CHARM II~\cite{CHARM_II:1994}, and CHORUS~\cite{CHORUS:2005cpn} at CERN as well as CCFR~\cite{Oltman:1992pq} and NuTeV~\cite{NuTeV:2005wsg} at Fermilab. 
In particular, the Fermilab experiments have collected up to $10^6$ muon neutrinos, which is comparable to the expected rates at SHiP SND.

The data collected by these experiments still provide the most precise data set of high-energy neutrino scattering and are used as input for most proton~\cite{NNPDF:2021njg, Bailey:2020ooq, Hou:2019efy} and nuclear~\cite{Muzakka:2022wey, Eskola:2021nhw, AbdulKhalek:2022fyi} PDF determinations, in particular to probe antiquarks and strangeness~\cite{Faura:2020oom}, to measure the weak mixing angle, or to constrain models of new physics such as Non-Standard Interactions (NSI). 
Figure~\ref{fig:neutrino-energy} also shows previously obtained measurements of the muon-neutrino--nuclei interaction cross sections
as well as their predictions from recent theoretical calculations.
Due to the expected number of muon neutrino interactions
and kinematic coverage, SHiP SND and SHADOWS NaNu will provide complementary input to validate and improve those measurements. 

In addition, there has been another class of accelerator neutrino experiments that were able to detect $\nu_\tau$. This includes DONUT~\cite{DONuT:2007bsg}, which observed 9 directly produced $\nu_\tau$, and OPERA~\cite{OPERA:2018nar}, which observed 10 $\nu_\tau$ produced in oscillations. 
In ECN3, about $1.5\times 10^5$ and $2\times 10^2$ tau neutrinos are expected to undergo CC interaction in the SHiP SND and SHADOWS NaNu detectors, respectively. 
This would significantly increase the number of observed $\nu_\tau$ events compared to DONUT and OPERA, and---thanks to the employed magnets---separate detection of $\nu_\tau$ and $\bar{\nu}_\tau$ events will be possible.
Moreover, at least SHiP SND (and the FPF, see below) should be able to significantly improve DONUT's laboratory-based bound on the $\nu_\tau$ magnetic moment (see, e.g.,~\cite{2016SHiPPhysicsCase}).

Laboratory neutrinos with even higher energies are produced only at the LHC. 
Two experiments, FASER~\cite{FASER:2019dxq, FASER:2020gpr, FASER:2022hcn} and SND@LHC~\cite{SNDLHC:2022ihg}, have recently started their operation and have both reported their first observation of neutrinos~\cite{FASER:2023zcr, SNDLHC:2023pun}.   FASER is an on-axis detector consisting of an emulsion target followed by a magnetized spectrometer. SND@LHC is a slightly off-axis detector consisting of an emulsion target followed by a hadronic calorimeter and muon system. 
These experiments are expected to observe $\sim 10^3$, $\sim 10^4$, and $\sim 10^2$ electron, muon, and tau neutrinos, respectively, 
during the LHC Run-3 data-taking period. The neutrinos have average energies of about a TeV~\cite{Kling:2021gos}, as shown in Figure~\ref{fig:neutrino-energy}.

The currently operating FASER experiment can search for  \(\bar{\nu}_\tau\) during LHC Run-3. However, only $\mathcal{O}(1)$ \(\bar{\nu}_\tau\) CC interaction events with measured muon charge are expected, making a high-significance observation very challenging.

The far-forward LHC neutrinos allow to i) measure neutrino interaction cross sections at TeV energies for the first time and perform tests of LFU, ii) study NC and test NSI~\cite{Ismail:2020yqc}, and iii) provide input for global proton and nuclear PDF fits, including studies of intrinsic charm~\cite{Ball:2022qks,Guzzi:2022rca}. 
In addition, these experiments will provide unique constraints on forward particle production at the high LHC collision energy, which are not accessible by the ECN3 experiments.
Specifically, FASER and SND@LHC will allow to test explanations of the cosmic ray muon puzzle~\cite{Anchordoqui:2022fpn} and study QCD in an otherwise inaccessible regime with $x\sim 4m_c^2/s \sim 10^{-7}$ where novel phenomena such as BFKL
dynamics~\cite{Ball:2017otu,Silvetti:2022hyc} and gluon saturation~\cite{Maciula:2020dxv} are expected, and therefore provide valuable input for astro-particle physics~\cite{Bhattacharya:2023zei,Bertone:2018dse,Gauld:2015yia,Laha:2016dri} including
a direct calibration of the prompt neutrino flux. 

An extension of the LHC neutrino program with significantly increased rates is envisioned during the HL-LHC in the context of the FPF~\cite{Anchordoqui:2022fpn, Feng:2022inv}, 
located approximately 620~m downstream of ATLAS
and directly in the LHC's TeV-energy neutrino beam.
Three experiments with neutrino-detection capabilities are foreseen: FLArE, FASER$\nu$2, and AdvSND. 
The event rates expected at all three experiments together are
shown as blue dashed lines in Figure~\ref{fig:neutrino-energy}.
Notably, the FASER$\nu$2 emulsion detector in conjunction with the FASER2 spectrometer will have the capacity to identify approximately 830 $\nu_\tau$ and 430 $\bar{\nu}_\tau$ separately~\cite{FASER:2019dxq}.
The FPF would be able to constrain tau-neutrino magnetic moments to $\mu_{\nu_\tau} < 6.6\times 10^{-8} \mu_B$~\cite{Abraham:2023psg}, as well as measure the tau-neutrino cross section and probe LFU at the percent level.
The FPF will also be able to constrain the NC/CC ratio at sub-percent precision and search for sterile-neutrino oscillations~\cite{Feng:2022inv}.

\begin{figure}[t]
	\centering
 \includegraphics[width=0.32\textwidth]{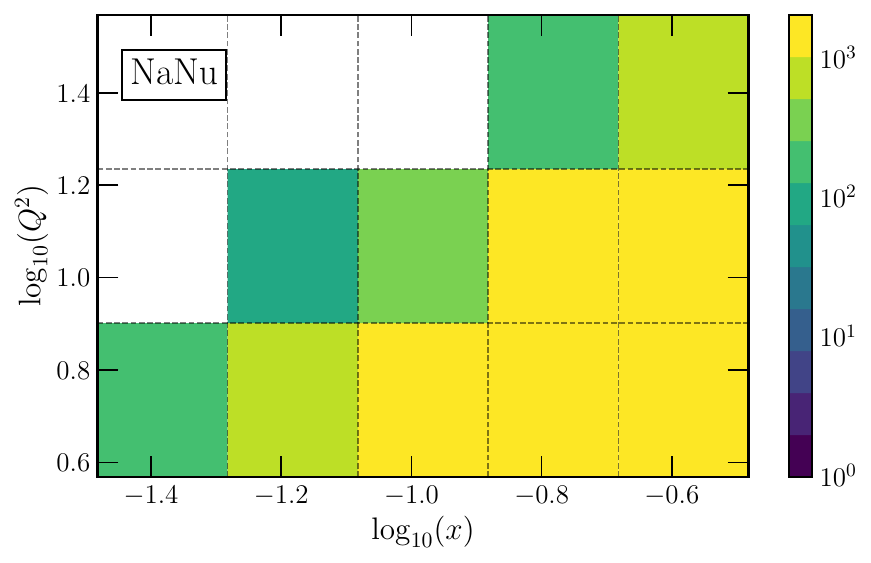}
 \includegraphics[width=0.32\textwidth]{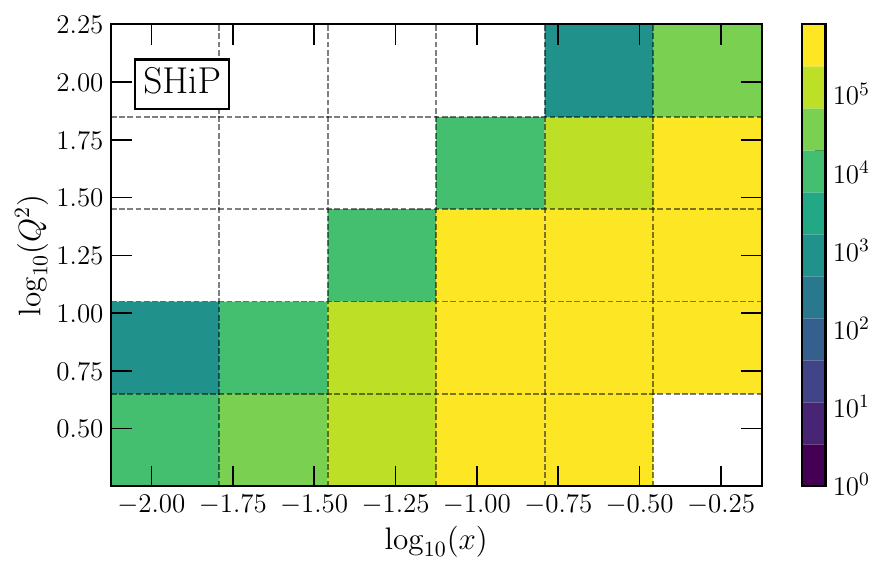}
 \includegraphics[width=0.32\textwidth]{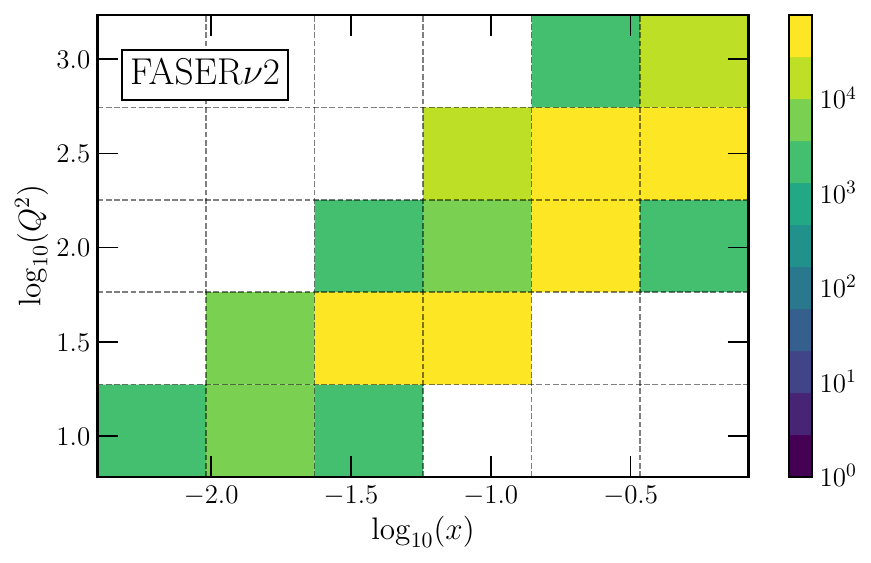}
 \includegraphics[width=0.49\textwidth]{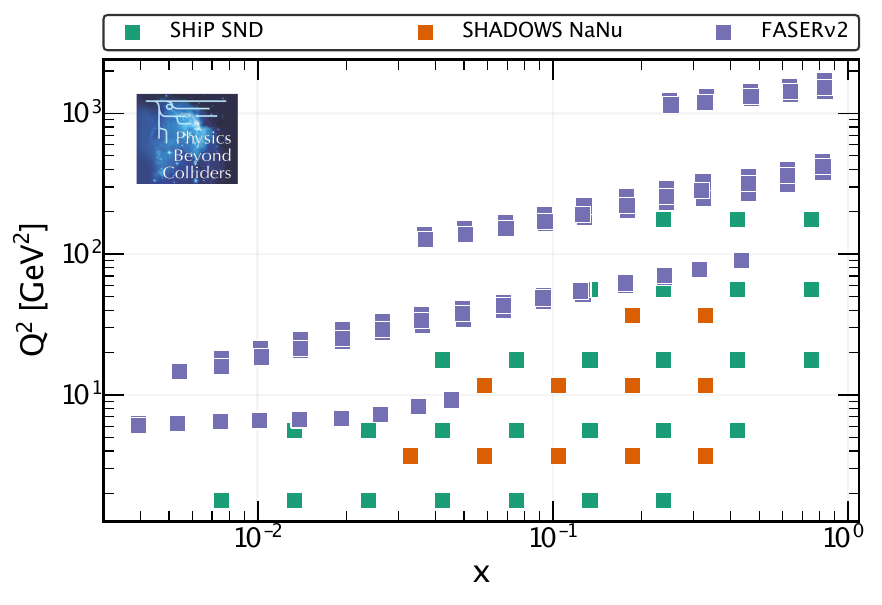}
  \includegraphics[width=0.49\textwidth]{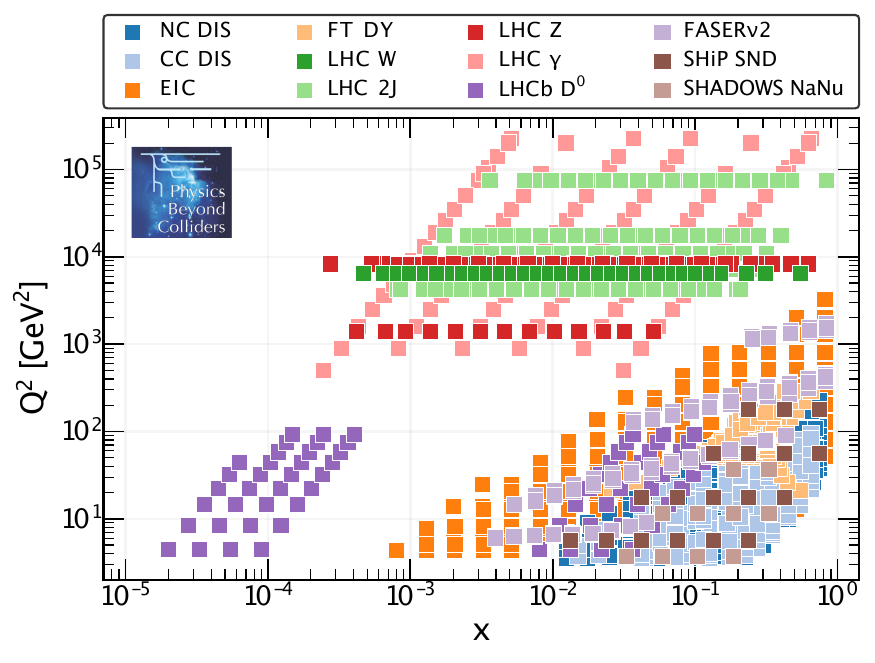}
	\caption{\small 
 Top panels: number of reconstructed
 muon neutrino events within detector
 acceptance with the SHADOWS NaNu (left), SHiP SND
 (center) and FASER$\nu$2 at the FPF (right) detectors in bins of $x$ and $Q^2$. 
 To restrict the comparison to the DIS region, $Q^2\ge 3$ GeV$^2$
 and $W^2\ge 4$ GeV$^2$ are imposed. 
 15 and 4 years of data taking are assumed for SHiP SND and SHADOWS NaNu, respectively, and an integrated luminosity of $\mathcal{L}_{\rm int}=3$ ab$^{-1}$ is assumed for FASER$\nu$2.
 Bottom panels: comparison of the kinematic
 coverage in the $(x,Q^2)$ plane of SHADOWS NaNu, SHiP SND,
 and FASER$\nu$2@FPF between themselves (left) and on an enlarged scale to the worldwide context (right). 
Only bins containing more than 30
 events are retained.
        }
	\label{fig:ECN3-neut-2}
\end{figure}

In order to compare the reach of future CERN neutrino experiments concerning the measurement of DIS inclusive structure functions, the upper panels of Figure~\ref{fig:ECN3-neut-2} display the number of reconstructed muon neutrino events within detector acceptance at SHADOWS NaNu, SHiP SND, and FASER$\nu$2@FPF in different bins of $x$ and $Q^2$. 
Kinematic requirements of $Q^2\ge 3$ GeV$^2$ and $W^2\ge 4$ GeV$^2$ are imposed to restrict the comparison to the DIS region.
The bottom right panels compare the kinematic coverage in the $(x,Q^2)$ plane of SHADOWS NaNu, SHiP SND, and FASER$\nu$2 with each other and in the global context.
In the QCD perturbative region, SHADOWS NaNu, SHiP SND, and FASER$\nu$2 respectively cover $x\ge 0.03$, $x\ge 0.007$, and $x\ge 0.003$,  reaching up to $Q^2\sim 40$ GeV$^2$, $Q^2\sim 200$ GeV$^2$, and $Q^2\sim 2000$ GeV$^2$, respectively. 
Figure~\ref{fig:ECN3-neut-2} indicates that the expected event rates should lead to structure functions measurements with statistical uncertainties at the few percent level or better, which hence are likely to be ultimately limited by systematic uncertainties.
Neutrino DIS measurements of sufficient precision in these regions can be used to inform future global proton and nuclear PDF fits, potentially benefiting searches for BSM physics at the HL-LHC, e.g., via the high-mass Drell-Yan~(DY) process~\cite{Ball:2022qtp}, and reduce theory systematics in key SM measurements such as the $W$-boson mass.

The lower right panel of Figure~\ref{fig:ECN3-neut-2} compares the presented experiments to the world data on hard-scattering processes involving nuclear projectiles or targets.
In particular, this comparison displays the coverage of existing measurements of neutrino DIS on nuclear targets (labelled ``CC DIS''), as well as the expected coverage of electron-nucleus scattering at the Electron-Ion Collider (EIC)~\cite{AbdulKhalek:2021gbh, Khalek:2021ulf}.
While SHiP SND and SHADOWS NaNu overlap with previous neutrino DIS experiments, FASER$\nu$2@FPF covers an uncharted region for CC scattering on nuclear targets and complements the NC measurements to be carried out at the EIC.
It should also be reminded that CC and NC measurements provide access to different PDF flavour combinations, with the former in particular being close to those relevant for $W^\pm$ production at hadron colliders.
In view of the large overlap in kinematics of the ECN3 neutrino experiments with existing high-statistics measurements, a significant impact on PDFs is mainly expected for strangeness where the tagging of charm production in the emulsion detector will play a crucial role, though quantitative estimates exist presently only for SHiP SND~\cite{2016SHiPPhysicsCase}.

\setcounter{topnumber}{2}
\def\topfraction{1}
\setcounter{bottomnumber}{1}
\def\bottomfraction{1}
\setcounter{totalnumber}{3}
\def\textfraction{0}
\def\floatpagefraction{1}
\setcounter{dbltopnumber}{2}
\def\dbltopfraction{1}
\def\dblfloatpagefraction{1}

\begin{table}[t]
  \begin{center}
    \scriptsize
    
    \renewcommand{\arraystretch}{1.80}
    \begin{tabular}{l|c|c|c|c}
      \toprule
      topic & SHADOWS NaNu & SHiP SND  & FPF & world-wide  \\
    \midrule \midrule
      number of years  & 	4	& 	15	& 10 &  \\
       PoT / integrated luminosity & $5\times10^{19}$ 		& 	$6 \times 10^{20}$	& 3~ab$^{-1}$	&  \\
    \midrule
      energy range (in GeV) & [10, 50] & [20, 110] / [5, 100]  & [10,5000] &   \\
      expected \(\nu_e/\bar{\nu}_e\) interactions & $ \approx 4.1 /  1.0\times 10^3  $ &  $2.7 / 0.6 \times 10^6$ & $2.5 / 1.1 \times 10^5$ &   \\
      expected \(\nu_\mu/\bar{\nu}_\mu\) interactions & $\approx 40 /  9\times 10^3$ & $ 8.0 / 1.8 \times 10^6$ & $10 / 3.5 \times 10^5$ &   \\
      expected \(\nu_\tau/\bar{\nu}_\tau\) interactions & $\approx 0.12 /  0.07\times 10^3  $ & $ 8.8 / 6.1 \times 10^4$ & $8.3 / 4.3 \times 10^3$ &   \\
      identified \(\nu_\tau/\bar{\nu}_\tau\) yields &  \begin{minipage}{22mm} \centering no-charge id: 100 \\ charged id: 10 / 7 \end{minipage}  & 3800 / 2900 & 830 / 430 & \begin{minipage}{25mm} \centering DONUT: 9 \\ ~OPERA: 10 \end{minipage}  \\
    \midrule \midrule
      \(\bar{\nu}_\tau\) observation        	& \checkmark & \checkmark & \checkmark & \\
      lepton-flavour universality           	& $<10\%$  & $\sim 1\%$ & $\sim 1\%$ &  \\
      \
      DIS structure functions \(F_4\) \& \(F_5\)	& evidence $F_4/F_5\neq 0$  & \checkmark  & to be studied & \\
      \(\nu_\tau\) magnetic moment    		& $< 4\times 10^{-7} \mu_B$ & $9 \times 10^{-8}\mu_B$ & $6.6\times 10^{-8} \mu_B$ & \begin{minipage}{23mm} \centering lab: $ 3.9\cdot 10^{-7} \mu_B$\\ solar: $1.3\cdot 10^{-11} \mu_B$ \end{minipage}  \\
sterile-neutrino/HNL oscillations          	& to be studied  & to be studied &  \begin{minipage}{23mm} \centering $\Delta m_{14}^2 \gtrsim 10^3 \text{eV}^2$ \\ $U_{\mu 4} \sim 10^{-2}$ \end{minipage} & SBN, NuStorm\\
\midrule
      proton and nuclear PDFs   & \begin{minipage}{25mm} \centering $x \gtrsim 0.03$ \\ $Q^2<40 ~\text{GeV}^2$ \end{minipage}   & \begin{minipage}{23mm} \centering $x \gtrsim 0.007$ \\ $Q^2<200 ~\text{GeV}^2$ \end{minipage} &  \begin{minipage}{23mm} \centering $x \gtrsim 0.003$ \\ $Q^2<2000 ~\text{GeV}^2$ \end{minipage} & \begin{minipage}{22mm} \centering NuTeV, CDHS,  HERA,  \\LHC, EIC etc.  \end{minipage} \\
      \begin{minipage}{35mm}  charmed-hadron \\ production \& decay   \end{minipage}     & $N_c=150$ & $N_c=6.2\cdot 10^5$ ($\varepsilon_{decay} \sim 50$\%) & $N_c=2.5 \cdot 10^5$ & NA65\\
       neutral currents & to be studied & $< 1 \%$  & $< 1 \%$ & NuTeV \\
    \bottomrule
    \end{tabular}
    \caption{\small Overview of selected physics topics that can be pursued with neutrinos
      at SHADOWS NaNu and SHiP SND, compared to those of
      the FPF and other world-wide projects.
      The upper part indicates basic assumptions on the running and numbers of expected neutrino interactions (before reconstruction) in the respective detector volumes. The lower part lists physics opportunities detailed in the main text, separating those associated with tau neutrinos and those based on all neutrino flavours.
      }
    \label{tab:neutrino-opportunities}
  \end{center}
\end{table}

To conclude the discussion, Table~\ref{tab:neutrino-opportunities} provides a summary of the potential neutrino physics topics and scientific reach with the proposed SHADOWS NaNu and SHiP SND subdetectors at ECN3, compared to those of the FPF at the LHC and of other experiments.

\setcounter{topnumber}{2}
\def\topfraction{1}
\setcounter{bottomnumber}{1}
\def\bottomfraction{1}
\setcounter{totalnumber}{3}
\def\textfraction{0}
\def\floatpagefraction{1}
\setcounter{dbltopnumber}{2}
\def\dbltopfraction{1}
\def\dblfloatpagefraction{1}

\clearpage

\label{sec:PhysicsPotential}

\bigskip

\section*{Acknowledgments}
F.K. acknowledges helpful discussions with David Curtin, Marco Drewes, Torben Ferber, Paddy Fox, Jan Jerhot, Maksym Ovchynnikov and Thomas Schwetz, and support by
the Deutsche \linebreak Forschungsgemeinschaft (DFG) through
the Emmy Noether Grant No. KA 4662/1-2 and grant
396021762~--~TRR~257. 
M.G.A. acknowledges support by the Generalitat Valenciana through the plan GenT program (CIDEGENT/2018/014), and by the Spanish Ministerio de Ciencia e Innovación through grants PID2020-114473GBI00 and CNS2022-135595.
B.D. acknowledges funding through the European Research Council under grant ERC-2018-StG-802836 (AxScale project). G.S. acknowledges support by the State Agency for Research of the Spanish Ministry of Science and Innovation through the grant PID2022-136510NB-C33.
J.J. acknowledges support by the European Union's Horizon 2020 research and innovation programme under the Marie Sk{\l}odowska-Curie grant agreement No~860881-HIDDeN. F.Kl. acknowledges support by the Deutsche Forschungsgemeinschaft under Germany's Excellence Strategy - EXC 2121 Quantum Universe - 390833306. 
\newpage

\bibliography{PBCreport}

\newpage
\appendix
\section{Definition of acronyms}
\label{sec:Acronyms}

\setlength{\parindent}{0pt}

\textbf{ACC-CONS}: Acceleration Consolidation project

\textbf{ALARA}: As Low As Reasonably Achievable

\textbf{ALD}: Atomic Layer Deposition

\textbf{ALP}: Axion-Like Particle

\textbf{ASIC}:  Application Specific Integrated Circuit

\textbf{BA}: Batiment Auxiliaire (Auxiliary surface Building)

\textbf{BC}: Benchmark physics case

\textbf{BD}: Beam dump

\textbf{BDF}: Beam Dump Facility

\textbf{BIS}: Beam Interlock System

\textbf{BLM}: Beam Loss Monitor

\textbf{BSG}: Beam SEM Grid

\textbf{BSI}: Beam SEM Intensity

\textbf{BSM}: Beyond Standard Model

\textbf{CC}: Charged Current

\textbf{CDHSW}: CERN-Dortmund-Heidelberg-Saclay-Warsaw, neutrino detector at CERN West Area

\textbf{CDS}: Comprehensive Design Study

\textbf{CHARM}: CERN-Hamburg-Amsterdam-Rome-Moscow, neutrino detector at CERN West Area

\textbf{CHARM II}: CERN-Hamburg-Amsterdam-Rome-Moscow II, neutrino detector at CERN West Area

\textbf{CHORUS}: neutrino detector at CERN West Area

\textbf{ChPT}: Chiral Perturbation Theory

\textbf{CL}: Confidence Level

\textbf{CKM}: Cabibbo-Kobayashi-Maskawa

\textbf{CNGS}: CERN Neutrinos to Gran Sasso

\textbf{CP}: Charge Parity

\textbf{CSSR}: Cost, Schedule and Scope Review

\textbf{DIS}: Deep-Inelastic Scattering

\textbf{DY}: Drell-Yan

\textbf{ECAL}: electromagnetic calorimeter

\textbf{ECC}: Emulsion Cloud Chamber

\textbf{ECN3}: Experimental Cavern North 3

\textbf{ECN3-TF}: PBC ECN3 Beam Delivery Task Force

\textbf{EFT}: Effective Field Theory

\textbf{EHN}: Experimental Hall North

\textbf{EM}: Electromagnetic

\textbf{EPPSU}: European Particle Physics Strategy Update

\textbf{EYETS}: Extended Year-End Technical Stop

\textbf{FASER}: ForwArd Search ExpeRiment at LHC

\textbf{FCNC}: Flavour-Changing Neutral Current

\textbf{FE}: Front End

\textbf{FIP}: Feably Interacting Particle

\textbf{FIRIA}: Fire-Induced Radiological Integrated Assessment

\textbf{FLArE}: Forward Liquid Argon Experiment at FPF at LHC

\textbf{FPF}: Forward Physics Facility at LHC

\textbf{FPC}: FIPs Physics Centre

\textbf{FT}: Flat-Top

\textbf{GIM}: Glashow–Iliopoulos–Maiani 

\textbf{GN}: Grossman-Nir

\textbf{GTK}: Giga Tracker

\textbf{HL-LHC}: High Luminosity LHC

\textbf{HI}: High Intensity

\textbf{HNL}: Heavy Neutral Lepton

\textbf{HIKE}: High Intensity Kaon Experiment

\textbf{HL-LHC}: High-Luminosity LHC

\textbf{HSDS} Hidden Sector Decay Search

\textbf{HVAC}; Heating, Ventilation and Air Conditioning

\textbf{IT}: Information Technology

\textbf{IVB}: Intermediate Vector Boson

\textbf{LD}: Long Distance

\textbf{LDM}: Light Dark Matter

\textbf{LFU}: Lepton Flavor Universality

\textbf{LFV}: Lepton Flavor Violation

\textbf{LHC}: Large Hadron Collider

\textbf{LoI}: Letter of Intent

\textbf{LS}: Long Shutdown

\textbf{LSS}: Long Straight Section

\textbf{MCP}: Micro-Channel Plate 

\textbf{MD}: Machine Development

\textbf{MIB}: Magnetized Iron Block

\textbf{MRPC}: Multigap Resistive Plate Chambers

\textbf{NA}: North Area

\textbf{NC}: Normal-Conducting / Neutral Current

\textbf{NaNu}: North Area NeUtrino experiment

\textbf{NA-CONS}: North Area Consolidation project

\textbf{NDA}: Non-Designated Area

\textbf{NSI}: Non-Standard Interactions

\textbf{OPERA}: Oscillation Project with Emulsion-tRacking Apparatus

\textbf{PANDA}: antiProton ANihilation at DArmstadt

\textbf{PBC}: Physics Beyond Colliders

\textbf{PCB}: Printed Circuit Board

\textbf{PDF}: Parton Distribution Function

\textbf{PDG}: Particle Data Group

\textbf{PPM}: Pulse-to-Pulse Mode

\textbf{PMT}:  Photon Multiplier

\textbf{PoT}: Protons on Target

\textbf{ppp}: particles (protons) per pulse

\textbf{PRR}: Project Readiness Review

\textbf{PS}: Proton Synchrotron

\textbf{QCD}: Quantum Chromo-Dynamics

\textbf{QED}: Quantum Electro-Dynamics

\textbf{R2E}: radiation-to-electronics

\textbf{R\&D}: Research and Development

\textbf{RICH}: Ring-Imaging CHerenkov

\textbf{RP}: Radiation Protection

\textbf{SBT}: Surrounding walls Background Tagger

\textbf{SC}: Super-Conducting

\textbf{SciFi}: Scintillating Fibre

\textbf{SEM}: Secondary Emission Monitor

\textbf{SHADOWS}: Search for Hidden And Dark Objects With the SPS

\textbf{SHiP}: Search for Hidden Particles

\textbf{SiPM}: Silicon Photo-Multiplier

\textbf{SM}: Standard Model

\textbf{SND}:Scattering and Neutrino Detector

\textbf{SPS}: Super Proton Synchrotron

\textbf{SPSC}: SPS and PS Experiments Committee

\textbf{TAX}: Target Attenuator eXperimental areas

\textbf{TBI}: Target Beam Instrumentation

\textbf{TBID}: Target Beam Instrumentation Downstream

\textbf{TBIU}: Target Beam Instrumentation Upstream

\textbf{TBSE}: Target Beam Stopper Extraction

\textbf{TCC}: Tunnel Caverne Cible = Tunnel Target Cavern 

\textbf{TCSC}: Target Collimator Splitter Copper

\textbf{TCX}: Target Collimator mask eXperimental areas

\textbf{TDC}: Tunnel Divider (splitter) Cavern 

\textbf{TDR}: Technical Design Report

\textbf{TED}: Target External Dump

\textbf{TIDVG}: Target Internal Dump Vertical Graphite (SPS Internal Dump)

\textbf{TT}: Transfer Tunnel

\textbf{TZM}: Titanium Zirconium-doped Molybdenum alloy

\textbf{UBT}: Upstream vessel wall Background Tagger

\textbf{WIC}: Warm magnets Interlock Controller

\textbf{WLS}: Wavelength Shifting

$\mathbf{X_0}$: Radiation Length

\textbf{YETS}: Year-End Technical Stop

\end{document}